\documentclass[iop,twocolappendix]{emulateapj}

\usepackage{natbib}
\usepackage{hyperref}
\usepackage{graphicx}
\usepackage{multirow}
\usepackage{mathtools}

\newcommand{\msun}{\mbox{M}_{\odot}}
\newcommand{\mstar}{$\mbox{M}_*$}

\newcommand{\hii}{H~\textsc{ii}\ }
\newcommand{\ii}{~\textsc{ii}}
\newcommand{\iii}{~\textsc{iii}}
\newcommand{\ttwo}{T$_2$}
\newcommand{\ttwogal}{T$_2^{\text{gal}}$}
\newcommand{\ttwohii}{T$_2^{\text{H\textsc{ii}}}$}
\newcommand{\te}{T$_3$}
\newcommand{\tegal}{T$_3^{\text{gal}}$}

\newcommand{\gal}{$^{\text{gal}}$}
\newcommand{\shii}{$^{\text{HII}}$}
\newcommand{\sigha}{$\Sigma_{\text{H}\alpha}$}
\newcommand{\fdig}{$f_{\text{DIG}}$}
\newcommand{\hiimod}{\textit{hiionly}\ }
\newcommand{\stackmod}{\textit{SDSSstack}\ }
\newcommand{\aurmod}{\textit{auroral}\ }

\shorttitle{Biases in metallicity measurements from global galaxy spectra}
\shortauthors{Sanders et al.}

\begin{document}

\title{Biases in metallicity measurements from global galaxy spectra: the effects of flux-weighting and diffuse ionized gas contamination}

\author{Ryan L. Sanders\altaffilmark{1}} \altaffiltext{1}{Department of Physics \& Astronomy, University of California, Los Angeles, 430 Portola Plaza, Los Angeles, CA 90095, USA}

\author{Alice E. Shapley\altaffilmark{1}}

\author{Kai Zhang\altaffilmark{2}} \altaffiltext{2}{Department of Physics and Astornomy, University of Kentucky, 505 Rose Street, Lexington, KY 40506, USA}

\author{Renbin Yan\altaffilmark{2}}

\email{email: rlsand@astro.ucla.edu}

\begin{abstract}
Galaxy metallicity scaling relations provide a powerful tool for understanding galaxy evolution,
 but obtaining unbiased global galaxy gas-phase oxygen abundances requires proper treatment of
 the various line-emitting sources within spectroscopic apertures.
  We present a model framework that treats galaxies as ensembles of \hii
 and diffuse ionized gas (DIG) regions of varying metallicities.
  These models are based upon empirical relations between line ratios and electron
temperature for \hii regions, and DIG strong-line ratio relations from SDSS-IV MaNGA IFU data.
  Flux-weighting effects and DIG contamination can significantly affect properties inferred from global galaxy
 spectra, biasing metallicity estimates by more than 0.3~dex in some cases.
  We use observationally-motivated inputs to construct a model matched to typical
 local star-forming galaxies, and quantify the biases in strong-line ratios,
 electron temperatures, and direct-method metallicities as inferred from global galaxy spectra
 relative to the median values of the \hii region distributions in each galaxy.
  We also provide a generalized set of models that can be applied to individual galaxies
 or galaxy samples in atypical regions of parameter space.
  We use these models to correct for the effects of flux-weighting and DIG
 contamination in the local direct-method mass-metallicity and fundamental metallicity relations,
 and in the mass-metallicity relation based on strong-line metallicities.
  Future photoionization models of galaxy line emission need to include
 DIG emission and represent galaxies as ensembles of emitting regions with varying metallicity,
 instead of as single \hii regions with effective properties, in order to
 obtain unbiased estimates of key underlying physical properties.
\end{abstract}

\keywords{galaxies: ISM --- galaxies: abundances}

\section{Introduction}\label{sec:intro}

The formation and growth of galaxies over cosmic history are governed by the relationship between
 gas accretion, star formation, and feedback.  Understanding this process, known as the ``cycle of baryons,''
 is of critical importance to gaining a full picture of galaxy growth, but directly observing gas in
 inflow and outflow stages is observationally challenging.  
  In lieu of direct observations, the cycle of baryons
 can be probed indirectly by measuring the chemical abundances of galaxies.  In particular, the scaling of
 gas-phase oxygen abundance, which we refer to in this work as ``metallicity,'' with global galaxy properties
 such as stellar mass (\mstar) and star-formation rate (SFR) can give insight into the interplay between
 inflows, outflows, and star formation.

A monotonic increase in metallicity with increasing stellar mass
 has been observed for local star-forming galaxies, and is known as the ``mass-metallicity relation''
 \citep[MZR; e.g.,][]{tre04,kew08,and13}.  The $z\sim0$ MZR has been found to have a secondary dependence
 on SFR, with the relationship among \mstar, SFR, and metallicity known as the ``fundamental metallicity
 relation'' \citep[FMR; e.g.,][but see \citealt{san13,bar17} for conflicting results using spatially-resolved data]{man10,lar10,and13,sal14}.  Chemical evolution models make predictions for the
 shape and normalization of these metallicity scaling relations under different sets of assumptions
 about the nature of galactic winds and the balance between inflow, outflow, and SFR
 \citep[e.g.,][]{fin08,pee11,zah14,dav17}.  An accurate comparison between chemical evolution models
 and the observed MZR and FMR can elucidate the nature of feedback and cosmological
 accretion.  However, such a comparison depends critically on robust measurements of metallicity
 for observed star-forming galaxy samples, and additionally requires the measurement of a metallicity
 that is compatible with metallicities extracted from cosmological hydrodynamical simulations.
  It is thus of paramount importance to eliminate observational biases in galaxy metallicity estimates.

The gas-phase oxygen abundances of galaxies in the local universe are typically estimated using
 one of two methods.  In the so-called ``direct method,'' the temperature-sensitive
 ratio of the intensities of an auroral emission feature
 (e.g., [O\iii]$\lambda$4363, [O\ii]$\lambda\lambda$7320,7330, [N\ii]$\lambda$5755)
 to strong emission lines from the same ionic species
 (e.g., [O\iii]$\lambda\lambda$4959,5007, [O\ii]$\lambda\lambda$3726,3729, [N\ii]$\lambda\lambda$6548,6584)
 is used to measure the electron temperature of the ionized gas \citep{ost06}.
  The cooling efficiency of ionized gas increases as the metal abundance increases.
  Thus, the gas-phase metallicity can be determined from the equilibrium electron
 temperature, assuming a heating and cooling balance \citep{izo06, pil12}.
  This method is the most accurate method of metallicity determination that
 can be applied to reasonably large samples ($\text{N}>100$) of low-redshift galaxies.
  The utility of the direct-method has been demonstrated by the observation that direct-method
 metallicities tightly correlate with metallicities obtained from oxygen recombination lines that
 more directly measure the oxygen abundance, where the relation has a slope of unity but
 an offset of $\sim0.2$~dex from a one-to-one relation \citep{bla15}.
  Metal recombination lines are $\sim10^4$ times weaker than strong lines and thus are not a practical
 metallicity indicator for any large sample.
  While the accuracy of the direct method is desirable, its use is limited because
 auroral lines are typically $\sim50-100$ times weaker than strong optical emission lines
 at low-metallicities (12+log(O/H)$\lesssim8.2$) and become weaker exponentially as metallicity increases,
 making it extremely difficult to detect these lines in individual metal-rich galaxies.
  For these reasons, samples of local galaxies with auroral line detections have sizes of
 only a few hundred and do not extend to 12+log(O/H)$\gtrsim8.4$ \citep{izo06, pil10}.

When auroral lines are not detected, galaxy oxygen abundances may be estimated from methods using
 only strong optical emission lines.  The ``strong-line method'' utilizes
 empirical or theoretical calibrations between strong optical emission line ratios and
 oxygen abundance.  Empirical calibrations are based on
 samples of individual \hii regions with direct-method metallicities \citep[e.g.,][]{pet04}.
  Theoretical calibrations instead make use of the predictions of photoionization models
 to determine the relations between line ratios and oxygen abundance \citep[e.g.,][]{kew02,kob04,tre04,dop16}.
  Because the strong-line method does not depend on the detection of any intrinsically weak
 emission lines, it can be applied to much larger samples of galaxies than the direct method.
  Strong-line metallicities have been estimated for sample sizes of $>10^4$ galaxies
 \citep[e.g., ][]{tre04} thanks
 to large spectroscopic surveys such as the Sloan Digital Sky Survey \citep[SDSS; ][]{yor00}.

Both the strong-line and direct methods share an inherent flaw when used to determine
 galaxy metallicities: they assume that the object of interest is a single \hii region.
  Empirical strong-line calibrations utilize \hii regions as the calibrating dataset, and
 will therefore not produce a reliable metallicity if the target does not follow the same
 relations between line ratios and oxygen abundance as \hii regions.  Theoretical strong-line
 calibrations are produced from photoionization models of single \hii regions (or in many cases
 a single slab of illuminated gas) and thus also assume that the target behaves similarly to an
 individual star-forming region.  When modeling galaxy emission spectra, it is common practice
 to illimunate the gas with the spectrum of a stellar population synthesis model
 \citep[e.g., Starburst99;][]{sb99} instead of a single stellar population (as in classical
 \hii regions).  However, this treatment fails to account for the variety of physical conditions
 of gas throughout the galaxy and the correlation of stellar properties with those variations.
  The direct method suffers from a similar problem, in that it
 assumes that the auroral and strong emission lines are produced in a single homogeneous \hii
 region ionized by a single star cluster.  

Galaxies are not single \hii regions, but are instead complex objects with a multiphase gaseous
 interstellar medium (ISM) and intricate substructure.
  The warm ($\sim10^4$~K) ionized phase includes \hii regions with a range of properties, as well
 as diffuse ionized gas (DIG) not contained in \hii regions.
  \hii regions are the line-emitting component associated with recent star formation,
 in which gas in close proximity to young, massive stars is ionized and
 emits both recombination and collisionally-excited lines.

While the light from \hii regions
 is of primary interest in determining gas-phase metallicity, other important sources of line emission
 exist in the ISM.  DIG contributes
 significantly to optical line emission in local galaxies.  Studies based on narrowband H$\alpha$
 imaging suggest that DIG emission contributes $30-60\%$ of the total H$\alpha$ flux in local
 spiral galaxies \citep{zur00,oey07}.  Additionally, DIG has different physical conditions and ionizing
 spectra from those of \hii regions, and therefore likely follows different line ratio excitation
 sequences \citep{zha17}.
  Hard ionizing radiation from accreting black holes incident on the ISM also produces
 line emission in galaxies harboring an active galactic nucleus (AGN), but in this study we
 ignore this source of line emission and focus only on galaxies dominated by star formation.
  Because of the diversity of ISM sub-components, applying the aforementioned methods to estimate
 galaxy metallicities while treating the galaxy as a single \hii region
 will inevitably result in some level of bias.

The observed global galaxy spectrum is a flux-weighted combination of light on a
 line-by-line basis from each of
 these line-emitting components falling in the spectroscopic slit or fiber.
  For typical spectroscopic apertures
 (i.e., SDSS fibers), this mixture includes multiple \hii and DIG regions
 with a spread in physical properties.
  A proper interpretation of the observed galaxy emission line spectrum cannot be obtained unless
 the mixture of these components is accounted for.

Robust galaxy gas-phase oxygen abundances are absolutely required when comparing
 observed metallicity scaling relations such as the MZR and FMR with simulations of galaxy chemical evolution.
  In this work, we reevaluate the reliability of oxygen abundances estimated from
 the emission lines of global galaxy spectra.
  For this analysis, we create
 simple models based on empirical auroral and strong emission line relations.
  These models include flux-weighting
 effects from the combination of emitting regions with a spread in physical properties, incorporating
 up-to-date line ratio and electron temperature relations for \hii regions.
  An important novel component of our models is the inclusion of emission from DIG regions based
 upon recent empirical results on DIG line
 ratios from the ongoing SDSS-IV MaNGA IFU survey \citep{zha17}.
  In Section~\ref{sec:models}, we motivate and describe the models and empirical relations upon which
 they are based.  We present results from the models and compare to both composites and individual
 local galaxies from SDSS in Section~\ref{sec:results}.  We characterize the biases in line ratios,
 electron temperature, and oxygen abundance measurements from global galaxy spectra in
 Section~\ref{sec:biases}, and discuss the effects on metallicity measurements for local galaxies.
  In Section~\ref{sec:application}, we apply corrections to the local mass-metallicity
 and fundamental metallicity relations and discuss DIG contamination in the context of other recent
 $z\sim0$ galaxy line-ratio studies.  In Section~\ref{sec:highz}, we discuss the implications
 for metallicity measurements from both the direct and strong-line methods for high-redshift galaxies.
  Finally, we summarize and make concluding remarks in Section~\ref{sec:summary}.
  Those readers who wish to skip over the details of the model framework may refer to Section~\ref{sec:biases}
 for the presentation of the biases in properties derived from global galaxy spectra, and subsequent
 sections for applications of the results.

Throughout this paper, we adopt shorthand abbreviations to refer to emission line ratios
 and present them here for the convenience of the reader.
  We normalize strong emission line fluxes to the H$\beta$ flux, following the practice
 of \hii region studies.  We use the following abbreviations for strong-line ratios throughout this work:
\begin{equation}
\text{O}3 = \log([\mbox{O}~\textsc{iii}]\lambda\lambda4959,5007/\text{H}\beta) ,
\end{equation}
\begin{equation}
\text{O}2 = \log([\mbox{O}~\textsc{ii}]\lambda\lambda3726,3729/\text{H}\beta) ,
\end{equation}
\begin{equation}
\text{N}2 = \log([\mbox{N}~\textsc{ii}]\lambda\lambda6548,6584/\text{H}\beta) ,
\end{equation}
\begin{equation}
\text{S}2 = \log([\mbox{S}~\textsc{ii}]\lambda\lambda6716,6731/\text{H}\beta) ,
\end{equation}
\begin{equation}
\text{O3N}2 = \text{O3} - \text{N2} .
\end{equation}
  These strong-line ratios are always reddening-corrected unless otherwise noted.
  The strong-line ratios that utilize a single doublet component and/or the Balmer line with the closest
 proximity in wavelength to the forbidden line, more common in galaxy studies, can be found from these ratios:
 log([O\iii]$\lambda$5007/H$\beta)=\text{O3} - 0.125$;
 log([N\ii]$\lambda$6584/H$\alpha)=\text{N2} - 0.581$;
 log([S\ii]$\lambda\lambda$6716,6731/H$\alpha)=\text{S2} - 0.456$;
 and log[([O\iii]$\lambda$5007/H$\beta$)/([N\ii]$\lambda$6584/H$\alpha)]=\text{O3N2} + 0.456$.
We also adopt abbreviations for the strong-to-auroral line ratios from which electron temperatures
 are estimated:
\begin{equation}
\text{Q}3 = [\mbox{O}~\textsc{iii}]\lambda\lambda4959,5007/\lambda4363 ,
\end{equation}
\begin{equation}
\text{Q}2 = [\mbox{O}~\textsc{ii}]\lambda\lambda3726,3729/\lambda\lambda7320,7330 ,
\end{equation}
\begin{equation}
\text{Q2N} = [\mbox{N}~\textsc{ii}]\lambda\lambda6548,6584/\lambda5755 .
\end{equation}
Whenever it occurs, the term ``metallicity" is used synonymously with gas-phase oxygen abundance (O/H) unless otherwise mentioned.
  We assume a $\Lambda$CDM cosmology with $H_0=70$~km~s$^{-1}$~Mpc$^{-1}$, $\Omega_m=0.3$, and $\Omega_{\Lambda}=0.7$.

\section{Modeling galaxies as ensembles of line-emitting regions}\label{sec:models}

There is clear evidence that global galaxy spectra cannot be described by \hii region
 photoionization models or \hii region empirical datasets alone.
  Local star-forming galaxies follow distinct excitation
 sequences from those of \hii regions in the [O\iii]$\lambda$5007/H$\beta$ vs.
 [S\ii]$\lambda\lambda$6716,6731/H$\alpha$,
[O\iii]$\lambda$5007/H$\beta$ vs. [O\ii]$\lambda\lambda$3726,3729/H$\beta$, and
 [O\iii]$\lambda$5007/H$\beta$ vs. [O~\textsc{i}]$\lambda$6300/H$\alpha$ diagrams \citep{cro15}.
  Such differences, alongside other pieces of evidence from past studies presented below,
 motivate a modeling approach that treats galaxies as collections of multiple emitting regions
 spanning a range of excitation levels and oxygen abundances.
  
In this section, we discuss past work modeling global galaxy spectra as ensembles of emitting regions,
 describe the \hii region and DIG datasets, present the line ratio relations and other
 inputs to the models, and outline the method by which the mock galaxy spectra are created.  We evaluate
 the performance of these models in Section~\ref{sec:results} by comparing to observations of galaxies
 from SDSS.

\subsection{Previous investigations of global galaxy biases}\label{sec:previous}

\citet{kob99} investigated the question of whether chemical abundances could be reliably estimated
 from global galaxy spectra.  Using a sample of six dwarf galaxies with both
 individual \hii region and global galaxy spectra, the authors found that the global spectra
 systematically overestimated electron temperatures by $\sim1,000-3,000$~K while underestimating
 direct-method 12+log(O/H) by $\sim0.05-0.2$ dex compared to the mean properties of the individual
 \hii regions.  These offsets were attributed to flux-weighting
 effects when combining light from multiple \hii regions with different levels of
 excitation.  \citeauthor{kob99} also investigated the same question for local spiral galaxies by comparing
 measurements from individual \hii regions to mock global spectra constructed using a weighted sum of
 the \hii region spectra in radial bins.  This analysis suggested that strong-line methods reproduced the mean
 metallicity of the individual \hii regions without significant systematic effects despite the range
 of abundances in the individal \hii region distributions.  However, their spiral galaxy models did
 not incorporate dust reddening and, critically, contributions from DIG emission, which were poorly
 constrained at the time.  Additionally, their sample of spiral and dwarf galaxies
 with direct-method measurements was very small (N=6) and only contained metal-poor
 (12+log(O/H)$\leq8.15$), low-mass objects.  An analysis utilizing a more representative
 sample spanning a wide dynamic range in mass and metallicity is needed to test metallicity estimates from modern
 spectroscopic surveys.

\citet{pil10} found that
 SDSS galaxies with auroral temperature measurements do not follow the \hii region relationship
 between electron temperature as measured from O$^+$ (T$_2$) and O$^{++}$ (T$_3$),
 known as the T$_2-$T$_3$ relation, but instead have lower \ttwo\ at fixed \te.
  A similar galaxy ionic temperature offset of $\sim1,000-1,500$~K lower \ttwo\ at fixed \te\ compared
 to the \hii region T$_2-$T$_3$ relation has
 been observed when \ttwo\ and \te\ are inferred from composite spectra constructed from local
 SDSS star-forming galaxies \citep{and13,cur17}.  Such composite spectra leverage the large-number
 statistics of SDSS to measure auroral-line ratios over an unprecedentedly wide range of galaxy
 properies.
  \citet{pil10} were able to
 roughly reproduce this offset by combining the spectra of 2-3 \hii regions falling on the \ttwo-\te\
 relation but spanning a wide range of temperatures, suggesting that such an offset could be the
 result of combining light from multiple line-emitting regions with different physical properties.
  However, the models of \citet{pil10} did not include any DIG component and thus were not
 representative of typical $z\sim0$ star-forming galaxies.
  \citet{pil12no} expanded on these results by simulating composite nebular spectra using a set of
 high-quality, self-consistent \hii region observations as input components.  These authors combined
 emission on a line-by-line basis from multiple \hii regions with abundances within a certain range
 of a central metallicity value, and found that the combination of multiple \hii regions can explain
 observed auroral-line properties of SDSS galaxies.  Furthermore, the bias in inferred nebular abundances
 relative to the central metallicity value increases with increasing width of the metallicity range.
  We note that \citet{pil12no} did not include any DIG emission regions in their composite spectra models.
  Collectively, these results imply that galaxy auroral and strong-line ratios do not behave in the
 same manner as those of individual \hii regions.  We build upon these previous studies of global galaxy spectra
 by creating models that treat galaxies ensembles of line-emitting regions with varying physical
 conditions, and crucially include a prescription for DIG emission.

\subsection{An empirical approach to modeling galaxy spectra}\label{sec:philosophy}

In order to characterize the biases in measurements of electron temperature, oxygen abundance,
 and strong-line ratios from global galaxy spectra, we constructed a set of models that
 are based on observed line-ratio relations of \hii and DIG regions.  We treat a galaxy as
 a collection of \hii and DIG regions with a distribution of physical properties, and create mock global
 galaxy spectra by summing the line fluxes from each individual component.
  These models are simple
 in nature, and minimize the number of free parameters that can be tuned
 to match observations of real galaxies.  In the description that follows, we will attempt
 to make it clear when we were forced to make
 assumptions due to a lack of constraining observations.

We chose to base our models on empirical relations rather
 than photoionization models for two reasons.  First, photoionization models have a large number of
 free parameters that can be fine-tuned to match a set of observations, often allowing for multiple
 degenerate solutions.  The interpretation of emission lines through photoionization models depends
 on the various required assumptions
 such as the shape of the relation between N/O and O/H, the method of accounting for the depletion of gas-phase
 elements onto dust grains, and the properties of the ionizing stellar population.
  In contrast, we prioritize simplicity over flexibility, minimizing the number of free parameters.

Second, DIG emission cannot be properly represented in photoionization models because the
 relative contributions of various ionization sources for DIG are still not agreed upon.
  The DIG ionization
 mechanism appears to be photoionization from some combination of leaking Lyman-continuum
 from O and B stars in \hii regions \citep{vog06,haf09,rey12} and
 evolved intermediate-mass post-AGB stars \citep{flo11,zha17}.  Although O and B stars appear to provide
 most of the DIG ionization energy, there is an ongoing discussion about the importance of evolved stars.
  It has been suggested that additional nonionizing heating sources such as shocks are required to 
explain DIG observations \citep{rey99,seo11}.
  Some emission attributed to DIG may also originate from dust-scattering of emission line photons
 produced in \hii regions \citep{bar15,asc16}.
  These effects are difficult to include in photoionization models, and introduce significant
 uncertainties.

By utilizing observed line ratio relations for both DIG and \hii regions, we minimize the number of
 free parameters and only sample regions of parameter space where real objects are found.  Thus,
 the main uncertainty concerning the applicability of these models stems from how representative the
 input datasets are of the full range of such emitting regions.

\subsection{Abundances and line emission of \hii regions}\label{sec:hiiregions}

We obtain line ratio relations for \hii regions from the sample of \citet{pil16}, which includes
 965 observations of \hii regions with auroral line measurements and direct-method abundances.
  We supplement this sample with recent observations of extragalactic \hii regions from
 \citet{cro16} and \citet{tor16}, bringing the total sample size to 1052.
  While all of the \hii regions in this sample have measurements of at least one auroral line,
 some of these auroral-line measurements have low S/N or are otherwise unreliable. 
  In order to construct a representative sample of \hii regions with a range of metallicities and
 ionization parameters,  we select a reference sample of high-quality, self-consistent observations
 from this parent sample using the counterpart method following \citet{pil12}.

The counterpart method is a technique for estimating metallicity that is based on the assumption
 that a set of \hii regions with the same physical properties such as density, electron temperature, and
 chemical abundance will have identical strong-line ratios.
  A high-quality reference sample of \hii regions with reliable auroral measurements can be selected
 by requiring the auroral-line ratios of an \hii region to closely match those of \hii regions with
 similar strong-line ratios, automatically excluding low-S/N measurements and strong outliers.
  Here, we only use the counterpart method to cull the \hii region parent sample of low-quality measurements.
  All electron temperatures and metallicities are determined using the direct method in our analysis.
  For the selection of the reference sample, we require the difference between the direct method
 and counterpart method oxygen and nitrogen abundances to be less than 0.1 dex, and we interpolate over a
 metallicity interval of 0.2 dex around the metallicity of the closest counterpart
 when determining the counterpart-method O/H and N/H.
  After iterating over the parent sample several times, the selection converges, yielding a reference sample
 of 475 objects that we refer to as the ``reference \hii region sample.''
  Objects in this sample have detections of [O\ii]$\lambda\lambda$3726,3729, H$\beta$,
 [O\iii]$\lambda$5007, H$\alpha$, [N\ii]$\lambda$6584, [S\ii]$\lambda\lambda$6716,6731, and
 at least one of the auroral lines [O\iii]$\lambda$4363 and [N\ii]$\lambda$5755.  All line fluxes
 have been corrected for dust reddening.
 
Electron temperatures are calculated using a five-level atom approximation and up-to-date atomic data
 \citep{san16a}.
  For the transition probabilities, we use values from the NIST MCHF database \citep{nistmchf} for all
 ions.  We obtain the collision strengths from \citet{sto14} for O\iii, \citet{hud05} for N\ii, and
 \citet{tay07} for O\ii.  
  The O\ii\ ion is only relevant for the models and galaxy comparison samples since auroral
 [O\ii]$\lambda\lambda$7320,7330 is not tabulated for the reference \hii region sample.
The vast majority of the \hii
 region sample has electron densities of n$_{\text{e}}<200$~cm$^{-3}$ and thus falls in the low density
 regime where electron temperature calculations are insensitive to the density.  We assume an
 electron density of n$_{\text{e}}=100$~cm$^{-3}$ for all temperature calculations.
  Because the five-level atom code is not optimized for speed and we need to calculate electron
 temperatures many times for each mock galaxy spectrum, we fit analytic formulae to the
 strong-to-auroral line ratio as a function of temperature obtained from the five-level atom code.
  Electron temperatures are calculated for a range of line ratios and
 we fit a function of the form $R=a e^{b/\text{T}_{\text{e}}}$, where $R$ is the strong-to-auroral line
 ratio and T$_{\text{e}}$ is the electron temperature for each ionic species.  We obtain
 the following best-fit equations, which are accurate to $<1.5\%$ between 5,000~K and 30,000~K:
\begin{equation}\label{eq:to3}
\text{Q3}=7.892 \text{exp}(3.278\times10^4~\text{K}/\text{T}_3),
\end{equation}
\begin{equation}\label{eq:to2}
\text{Q2}=7.519 \text{exp}(1.928\times10^4~\text{K}/\text{T}_2).
\end{equation}
\begin{equation}\label{eq:tn2}
\text{Q2N}=7.789 \text{exp}(2.493\times10^4~\text{K}/\text{T}_2),
\end{equation}
To infer the electron temperature from an observed line ratio, we invert these expressions.

For objects that have measurements of both [O\iii]$\lambda$4363 and [N\ii]$\lambda$5755, \te\ and \ttwo\
 are calculated using equations~\ref{eq:to3} and~\ref{eq:tn2}.  For objects with a measurement of
 only one of these auroral lines, we calculate the corresponding ionic temperature and infer the
 temperature of the other ionic zone assuming the T$_2-$T$_3$ relation of \citet{cam86}:
\begin{equation}\label{eq:t2t3}
\text{T}_2 = 0.7 \text{T}_3 + 3,000~\text{K}.
\end{equation}
We note that calculating the oxygen abundance requires \ttwo([O\ii]), while \ttwo([N\ii]) is measured
 for the reference \hii region sample because [O\ii]$\lambda\lambda$7320,7330 fluxes were not tabulated.
  We make the assumption commonly adopted that O\ii\ and N\ii\ predominantly trace the same ionic
 zone in the nebula such that \ttwo([O\ii])=\ttwo([N\ii]) as expected from photoionization models.
  We note for completeness that recent
 observations of \hii regions with measurements of both
 ionic temperatures have shown that this relation has a large dispersion and called
 the one-to-one correspondence into question \citep{ber15,cro15,cro16}.

Ionic oxygen abundances are calculated using the formulae from \citet{izo06}
 in the low-density limit:
\begin{equation}\label{eq:oplush}
\begin{multlined}
12+\log(\text{O}^+/\text{H}^+) = 
 \text{O2} + 5.961 \\+ \frac{1.676}{\text{T}_2} - 0.040 \log(\text{T}_2) - 0.034 \text{T}_2,
\end{multlined}
\end{equation}
\begin{equation}\label{eq:o2plush}
\begin{multlined}
12+\log(\text{O}^{++}/\text{H}^+) = 
 \text{O3} + 6.200 \\+ \frac{1.251}{\text{T}_3} - 0.55 \log(\text{T}_3) - 0.014 \text{T}_3.
\end{multlined}
\end{equation}
Total oxygen abundance is calculated assuming the fraction of oxygen in higher ionic and neutral states
 is negligible \citep{izo06,pil12}:
\begin{equation}
\frac{\mbox{O}}{\mbox{H}}\approx\frac{\mbox{O}^+}{\mbox{H}^+}+\frac{\mbox{O}^{++}}{\mbox{H}^+}.
\end{equation}

\begin{figure}
 \includegraphics[width=\columnwidth]{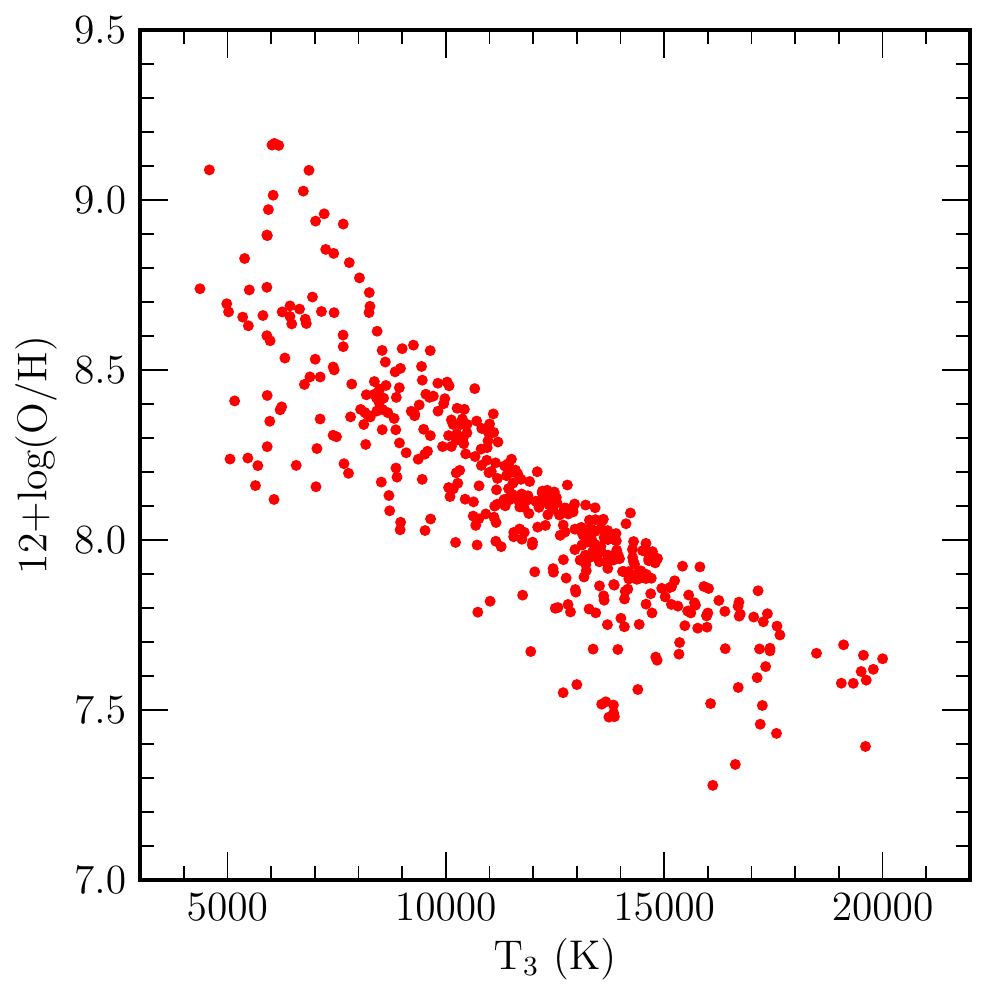}
 \caption{Direct-method 12+log(O/H) vs. T3 for the 475 individual \hii regions in the
 reference \hii region sample.
  Metallicities are calculated using up-to-date atomic data.
  12+log(O/H) is nearly linearly dependent on \te, which provides a good
 proxy for the direct-method metallicity.
}\label{fig:oht3}
\end{figure}

\begin{figure*}
 \centering
 \includegraphics[width=\textwidth]{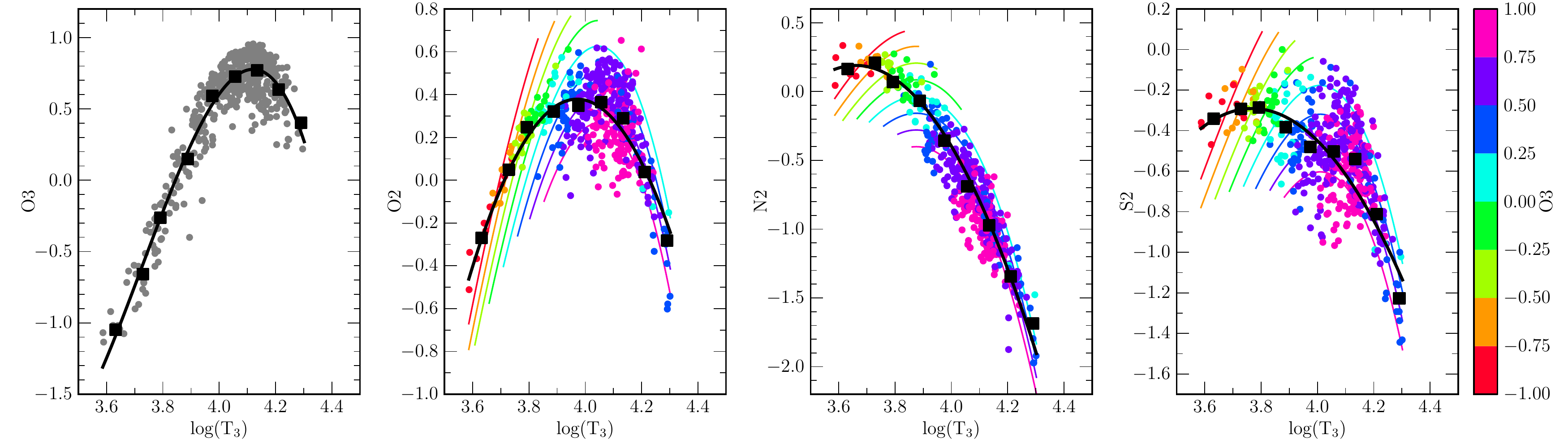}
 \caption{Strong-line ratios O3, O2, N2, and S2 as a function of \te\ for the reference \hii region sample.
  In each panel, black squares display the median strong-line ratio in bins of \te.  The black
 lines show best-fit polynomials to the median points presented in equations~\ref{eq:o3fit}-\ref{eq:s2fit}.
  In the right three panels, the points are color-coded according to O3.
  The anticorrelation between O3 and O2, N2, and S2 at fixed \te\ reflects ionization parameter dependence
 in the strong-line ratio vs. \te\ relations.  The solid colored lines display contours of constant O3
 according to the analytic parameterization of ionization parameter presented in
 equations~\ref{eq:delo2}-\ref{eq:dels2}.    
  Mock \hii region distributions are drawn from the analytic functional fits presented here, including
 the secondary ionization parameter dependence.
}\label{fig:hiiratios}
\end{figure*}

In order to create a distribution of realistic \hii regions with known metallicities
 in a mock galaxy, we first need to parameterize the strong line ratios of the observed \hii
 region sample because calculating direct-method metallicities requires the strong-line ratios O2
 and O3.
  Since we are interested in biases in abundance measurements, an obvious choice of parameter is the
 oxygen abundance.  However, because the direct-method oxygen abundance calculation depends on the
 strong line ratios O3 and O2, parameterizing by O/H will introduce covariances that are not
 observed in real \hii region samples.
  We instead parameterize the strong line ratios as a function of \te, which has no dependence
 on the strong-line ratios and is a good proxy
 for the oxygen abundance since the relationship between \te\ and 12+log(O/H) is nearly
 linear over the range of metallicities of interest here, as shown in Figure~\ref{fig:oht3}.
  Figure~\ref{fig:hiiratios} presents the strong line ratios O3, O2, N2, and S2
 as a function of \te\ for the reference \hii region sample.  For each line ratio, we find
 the median relation in bins of \te\ and fit polynomials to the median points to obtain
 functional forms of these relations:
\begin{equation}\label{eq:o3fit}
\text{O3} = -16.6 \log(\text{T}_3)^3 + 189 \log(\text{T}_3)^2 - 711 \log(\text{T}_3) + 884 ,
\end{equation}
\begin{equation}\label{eq:o2fit}
\text{O2} = -5.89 \log(\text{T}_3)^2 + 46.8 \log(\text{T}_3) - 92.8 ,
\end{equation}
\begin{equation}\label{eq:n2fit}
\text{N2} = -5.48 \log(\text{T}_3)^2 + 40.3 \log(\text{T}_3) - 73.9 ,
\end{equation}
\begin{equation}\label{eq:s2fit}
\text{S2} = -2.67 \log(\text{T}_3)^2 + 20.5 \log(\text{T}_3) - 38.3 .
\end{equation}

In Figure~\ref{fig:hiiratios}, the points in the right three panels are color coded by
 O3.  It is apparent that at fixed \te, each of these low-ionization line
 ratios is anticorrelated with O3.  This anticorrelation encodes the
 range of ionization parameters at fixed \te: higher ionization parameter \hii regions
 have higher O3 and lower low-ionization line ratios (N2, O2, and S2) at fixed \te.
  We include variations in ionization parameter in our models by encoding ionization
 parameter changes using $\Delta$X, the difference in line ratio X between the data point
 and the best-fit polynomial at fixed \te.  We subtract the best-fit polynomials from the data shown
 in the right three panels of Figure~\ref{fig:hiiratios} and
 fit linear functions with y-intercept of zero to
 $\Delta$O2, $\Delta$N2, and $\Delta$S2, all as a function of $\Delta$O3.
  In this way, we obtain fits with the following values:
\begin{equation}\label{eq:delo2}
\Delta\text{O2} = -0.47 \Delta\text{O3}
\end{equation}
\begin{equation}\label{eq:deln2}
\Delta\text{N2} = -0.42 \Delta\text{O3}
\end{equation}
\begin{equation}\label{eq:dels2}
\Delta\text{S2} = -0.56 \Delta\text{O3}
\end{equation}
Thus, the residuals of the O3 fit, $\Delta$O3, are used as input for the low-ionization line ratios.
  The standard deviation of the O3 residuals is 0.14 dex and is independent of \te.
  Adding the ionization parameter (i.e., $\Delta$O3) terms to the best-fit polynomials yields good fits
 to the data, with residuals having standard deviations of $\lesssim0.15$ dex for O2, N2, and S2.
  We note that the measurement uncertainty in \te\ is the main source of uncertainty in
 Figure~\ref{fig:hiiratios}, and accounts for some of the scatter about these fits.
  The solid lines in Figure~\ref{fig:hiiratios} show contours of constant O3 based on the combination of
 the polynomial fits and ionization parameter terms
 presented above.  These analytic functions represent the relationships between strong-line ratios
 and \te\ well for observed \hii regions.

These formulae allow us to obtain realistic strong-line ratios for \hii regions
 using input \te\ distributions.  The strong-line emission from mock \hii regions is
 combined with strong-line emission from DIG in order to produce global galaxy strong-line ratios
 that are a necessary component for the calculation of
 direct-method metallicity inferred from global galaxy spectra.
  We also investigate the impact of DIG emission and flux-weighted combination
 effects on diagnostic strong-line ratio diagrams and strong-line metallicity indicators.
  Additionally, we test whether our models simultaneously match the position of real galaxies in
 multiple strong-line ratio diagrams, a requirement for any realistic model of galaxy line emission.

\subsection{Line emission from diffuse ionized gas}\label{sec:digregions}

The models presented herein include emission from DIG, in addition to line emission from
 multiple \hii regions with varying abundances, for the first time.
  It is of critical importance to account for DIG emission when studying emission-line spectra of
 local star-forming galaxies since $\sim30-60$\% of H$\alpha$ emission
 in local spiral galaxies can be attributed to DIG \citep{zur00,oey07}.
  Because of the diffuse
 nature of DIG, its line emission has a low surface brightness compared to that of \hii regions
 and is thus difficult to observe.
  DIG was first identified with the discovery of a layer of warm ionized hydrogen permeating the
 Milky Way ISM \citep[i.e., the Reynolds Layer;][]{rey73}.
  Initial observations of DIG line ratios in other galaxies have come from longslit spectroscopy of extra-planar
 emission around edge-on galaxies \citep[e.g., ][]{ott01,ott02} or of low-surface-brightness
 emission in face-on galaxies probing only a small number of DIG regions \citep[e.g., ][]{hoo03}.
  These observations showed that low-ionization line ratios (N2, S2) of DIG
 are enhanced relative to those typical of \hii regions, but sample sizes were too small to establish
 DIG line ratio trends over a range of physical conditions.

\subsubsection{MaNGA observations of DIG line ratios}

In order to estimate DIG contribution to line emission in local star-forming galaxy spectra
 with a wide range of stellar masses and metallicities,
 we need to characterize DIG line ratios over a wide range of excitation levels throughout
 star-forming disks and tie
 DIG line ratios in a galaxy to the \hii region abundances in the same galaxy in some realistic way.
  To achieve this task, we characterize the DIG emission line ratios using data
 from the SDSS-IV Mapping Nearby Galaxies
 at APO \citep[MaNGA; ][]{bun15,yan16,law16} integral field spectroscopic (IFS) survey.  
  The MaNGA IFS dataset provides spatially-resolved spectroscopic observations of a large number
 of local star-forming galaxies for which such an analysis of DIG emission is possible \citep{zha17}.

\citet{zha17} recently showed how optical strong emission-line ratios in local galaxies vary with
 H$\alpha$ surface brightness (\sigha), with the strength of low ionization lines
 ([N\ii], [S\ii], [O\ii], and [O~\textsc{i}]) relative to Balmer lines
 increasing with decreasing \sigha\ at fixed radius.  However, \citeauthor{zha17}
 also found that O3 did not increase or
 decrease with \sigha\ on average.  Under the assumption that high-\sigha\ regions are dominated by
 \hii region emission and DIG emission becomes increasingly important as \sigha\ decreases, this
 result indicates that the O3 ratios of DIG and \hii regions are the same on average
 within a single galaxy.  Thus, we can match a sample of model \hii regions with DIG emitting regions in
 a way that mimics the ISM of real galaxies by matching in O3.  We note that \citet{zha17}
 found some stellar mass dependence for $\Delta$O3 vs.\ $\Delta$\sigha, such that
 DIG O3 is higher than that of \hii regions in the most massive third of their
 sample.  We do not include this stellar mass effect in our models because we have no direct way of assigning
 stellar mass to a mock galaxy, but this effect could be included in future models to increase the
 accuracy of the DIG representation.

In order to realistically match model distributions of DIG and \hii regions, we characterize the DIG
 excitation sequences of [N\ii]/H$\alpha$, [S\ii]/H$\alpha$, [O\ii]/H$\beta$, and [O\iii]/H$\beta$
 as a function of O3N2.  We utilize the sample of galaxies from SDSS Data Release 13 \citep{alb16} presented
 in \citet{zha17}.  From their sample of 365 blue, low-inclination galaxies, we selected a sample
 of 266 star-forming galaxies by requiring that the central region does not host an AGN according
 to the demarcation of \citet{kau03} in the [O\iii]/H$\beta$ vs.\ [N\ii]/H$\alpha$ diagram.  In
 order to determine the line ratios of the central region of a galaxy, we construct a 3$\arcsec$
 pseudo-fiber by summing the line fluxes of all spaxels within a 1.5$\arcsec$ radius of the galaxy
 center.  This pseudo-fiber mimics the aperture of an SDSS fiber, matching the observations upon which the
 \citet{kau03} demarcation are based.

For each galaxy in the DIG galaxy sample, we select all spaxels that have a signal-to-noise ratio S/N$\geq$3
 for [O\ii]$\lambda\lambda$3726,3729, H$\beta$, [O\iii]$\lambda$5007, H$\alpha$, [N\ii]$\lambda$6584,
 and [S\ii]$\lambda\lambda$6716,6731.  We assume that the highest \sigha\ spaxels are dominated by
 \hii region emission, while the lowest \sigha\ spaxels are dominated by DIG emission.  We identify
 spaxels with \sigha\ below the 10th percentile of the \sigha\ distribution in each galaxy as DIG-dominated.
  The threshold of 10\% was selected in order to provide the purest probe of DIG emission while
 still yielding a large sample of spaxels.
  Results do not change significantly when varying this threshold from 5\% to 15\%.
  The DIG-dominated spaxel sample can be thought of as the diffuse analog of the \hii region sample.
  Before calculating [O\ii]/H$\beta$, the [O\ii] and H$\beta$ fluxes are first corrected
 for reddening on a spaxel-by-spaxel basis assuming the attenuation law of
 \citet{car89} and an intrinsic ratio of H$\alpha$/H$\beta=2.86$.
  The line ratios [O\iii]/H$\beta$, [N\ii]/H$\alpha$, and [S\ii]/H$\alpha$
 are calculated without correcting for dust
 reddening given the close proximity in wavelength of the relevant emission lines.

\begin{figure*}
 \centering
 \includegraphics[width=\textwidth]{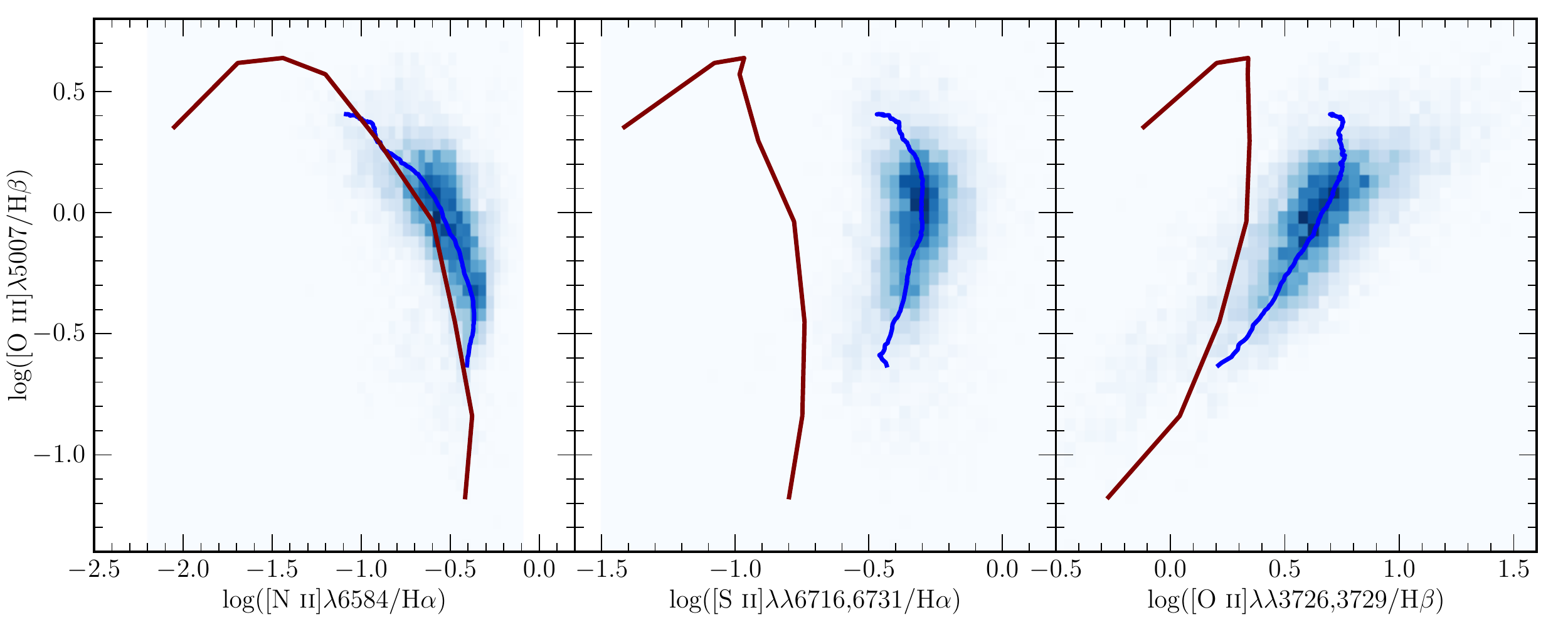}
 \caption{The O3N2 (left), O3S2 (middle), and O3O2 (right)
 strong-line ratio diagrams for DIG-dominated spaxels from MaNGA
 galaxies \citep{zha17}.  The blue two-dimensional histogram shows the distribution of strong-line ratios
 for the 10\% lowest surface-brightness spaxels for each galaxy in the DIG galaxy sample.  These spaxels
 are dominated by emission from DIG rather than \hii regions.  The running median of the DIG spaxel
 distribution in bins of O3N2 is displayed as a solid blue line.
  The maroon line shows the median strong-line excitation sequences of the reference \hii region sample
 in bins of \te.
}\label{fig:digratios}
\end{figure*}

We construct the DIG strong-line excitation sequences by taking the median line ratios of the DIG spaxel
 sample in bins of O3N2.
  We chose to bin in O3N2 rather than O3 because O3N2 monotonically increases
 with \te\ in a nearly linear fashion for the reference \hii region sample,
 while O3 is double-valued as a function of \te.
  We assume that O3 of DIG regions is also double-valued as a function of \te, in which case
 the median relation in bins of O3 would not be a good representation of the actual
 excitation sequence in the regime where the temperature-dependence of O3 weakens.
  While there are no constraints on the electron temperatures of DIG, we work under
 the assumption that DIG electron temperature decreases with increasing metallicity as for
 \hii regions.  Binning excitation sequences and matching \hii and DIG regions in O3N2 instead of O3 alone is also
 motivated by the fact that the sequences of \hii regions, $z\sim0$ star-forming galaxies, and
 DIG regions are nearly identical in the [O\iii]/H$\beta$ vs. [N\ii]/H$\alpha$ diagram as shown below
 in Section~\ref{sec:digextrap}, in agreement
 with the results of \citet{zha17} that O3 and N2 are minimally affected by DIG compared to other
 strong-line ratios.
  The DIG line ratio distributions and median excitation sequences in the
 [O\iii]/H$\beta$ vs. [N\ii]/H$\alpha$, [O\iii]/H$\beta$ vs. [S\ii]/H$\alpha$, and
 [O\iii]/H$\beta$ vs. [O\ii]/H$\beta$ diagrams
 are presented in Figure~\ref{fig:digratios}, which we refer to as the O3N2, O3S2, and O3O2 diagrams,
 respectively.  For comparison,
 the median line ratios of the reference \hii region sample in bins of \te\ are shown.  As implied by
 the results of \citet{zha17}, we find that at fixed O3, DIG regions display
 larger low-ionization line ratios than those of \hii regions.

\subsubsection{Extrapolations of DIG excitation sequences}\label{sec:digextrap}

We note that the reference \hii region sample spans a wide range of excitation levels with
 $-1.0\lesssim\text{O3}\lesssim1.0$, while the DIG spaxel sample only has sufficient statistics
 over a smaller range ($-0.6\lesssim\text{O3}\lesssim0.4$).  This limitation in the range of DIG
 excitation levels stems from the nature of the MaNGA sample used here, which mostly
 comprises moderate-metallicity galaxies leaving the low- and high-metallicity tails of the distribution
 poorly sampled.
  The dearth of low-O3 DIG measurements is also present due to the trend observed by \citet{zha17} in which
 DIG O3 is higher than \hii region O3 on average in massive (log(M$_*$/M$_{\odot})>10.08$) star-forming galaxies.
  Once the survey is completed, the full MaNGA sample will contain $\sim10$ times more
 galaxies than were available at the time of this study, which should allow for direct constraints of the
 metal-rich and metal-poor tails of the DIG excitation sequences.  To allow our models to be applicable
 for galaxies over a wide range of metallicities, we extend the DIG excitation sequences by making
 assumptions about the behavior of DIG line ratios in the low- and high-metallicity tails based on
 the position of \hii regions and $z\sim0$ SDSS galaxies in strong-line excitation diagrams.

In order to compare the positions of galaxies and \hii regions in line ratio diagrams,
 we select a comparison sample of $z\sim0$ star-forming galaxies from SDSS
 Data Release 7 \citep[DR7;][]{aba09} for which strong-line measurements are
 available.  We take global galaxy properties and emission-line measurements from the MPA-JHU
 SDSS DR7 catalogs.\footnote[3]{Available online at \texttt{http://www.mpa-garching.mpg.de/SDSS/DR7}}
  We use the same selection criterion employed by \citet{and13} , and later compare our models to their
 stacks of SDSS galaxies constructed from this sample of individual star-forming galaxies.
  We require SDSS galaxies to have $0.027\leq z\leq0.25$ and S/N$\geq5$ for each of the lines
 H$\beta$, H$\alpha$, and [N\ii]$\lambda$6584.  AGN are rejected by requiring
 log([N\ii]$\lambda$6584/H$\alpha)<-0.4$ as well as a location below the star-forming/AGN demarcation of
 \citet{kau03} in the [O\iii]/H$\beta$ vs. [N\ii]/H$\alpha$ diagram when S/N$\geq3$ for
 [O\iii]$\lambda$5007.
  This selection yields a sample of 209,513 local star-forming galaxies with a median redshift of
 $z_{\text{med}}=0.08$, which we refer to as the
 ``SDSS strong-line comparison sample.''\footnote[4]{\citeauthor{and13} additionally
 rejected galaxies for which the SDSS photometric flags
 indicated that the spectroscopic fiber targeted the outskirts of a large galaxy instead of
 a galaxy center, and removed low-mass targets for which the stellar mass was obviously
 incorrect through visual inspection.  We do not apply these additional criteria since such
 issues affect less than 0.5\%\ of the sample.}
  Before calculation of the line ratios, the emission-line fluxes were
 corrected for reddening using the attenuation law of \citet{car89}, assuming an intrinsic Balmer
 decrement of H$\alpha$/H$\beta=2.86$.
  In line-ratio diagrams involving [O\ii]$\lambda\lambda$3726,3729, [O\iii]$\lambda$5007, or
 [S\ii]$\lambda\lambda$6716,6731, only the subset of galaxies with $\text{S/N}\geq3$ in the relevant
 emission lines are plotted.

\begin{figure*}
 \centering
 \includegraphics[width=\textwidth]{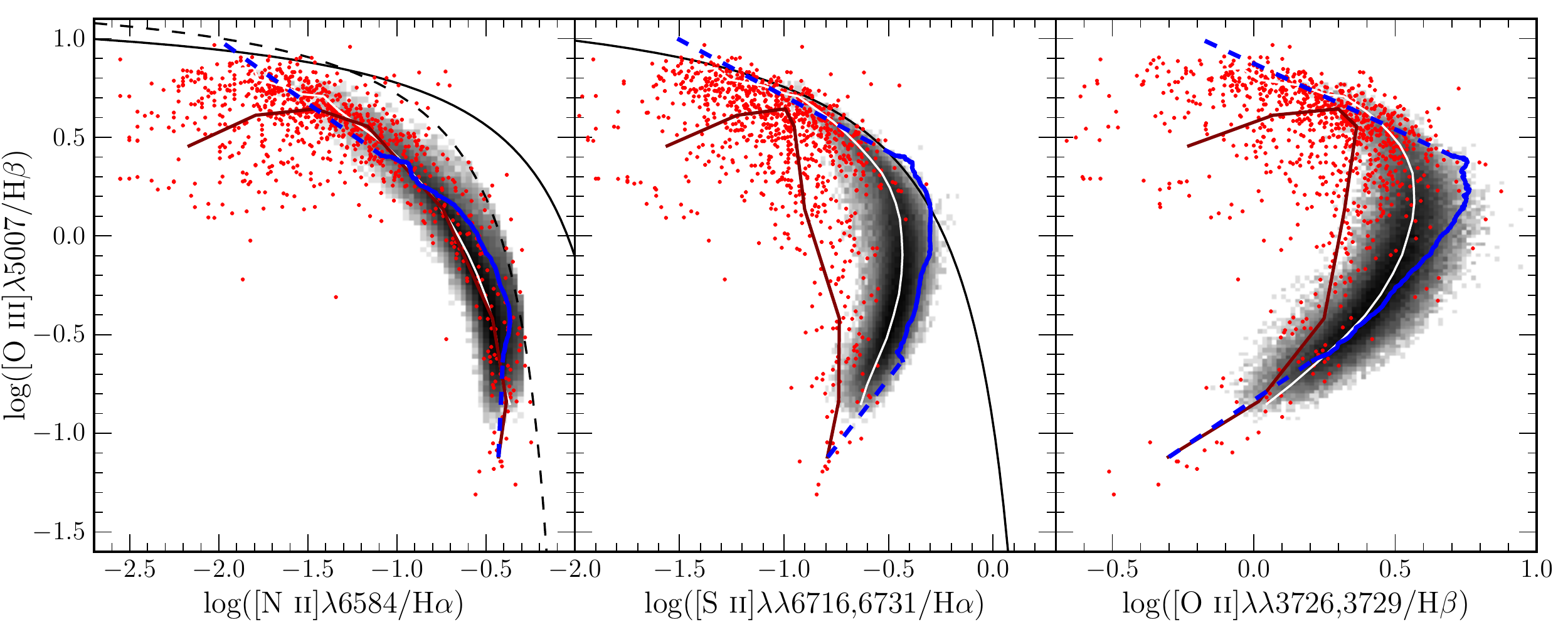}
 \caption{Excitation sequences of \hii regions, SDSS star-forming galaxies, and DIG regions in
 the O3N2 (left), O3S2 (middle), and O3O2 (right) diagrams.
  The gray two-dimensional histogram shows the distribution
 of $z\sim0$ star-forming galaxies from the strong-line comparison sample, where the white line
 represents the running median of the galaxies in bins of O3N2.
  Individual \hii regions in the reference \hii region sample are represented by red points, where the
 maroon line gives the running median of the \hii regions.
  The solid blue line shows the running median of the DIG-dominated spaxels from the MaNGA
 DIG galaxy sample in bins of O3N2, while the dashed blue lines denote the linear extrapolations
 that we assume for the DIG excitation sequences.
  In the left and middle panels, the solid black line shows the ``maximum-starburst'' line of
 \citet{kew01}, while the empirical demarcation between AGN and star-forming galaxies of
 \citet{kau03} is displayed as the black dashed line in the left panel.
}\label{fig:allratios}
\end{figure*}

The excitation sequences of \hii regions and $z\sim0$ SDSS star-forming galaxies in the
 O3N2, O3S2, and O3O2 diagrams
 are presented in Figure~\ref{fig:allratios}.
  These plots demonstrate the necessity of including
 DIG emission in order to properly interpret SDSS star-forming galaxy line ratios.  In the O3N2 diagram,
 \hii regions and SDSS galaxies follow nearly identical sequences, suggesting that the DIG O3N2 sequence
 is similar to that of \hii regions, as observed in the DIG line ratios from MaNGA data \citep{zha17}.
  In the O3S2 and O3O2 diagrams,
 SDSS galaxies are offset towards significantly
 higher [S\ii]/H$\alpha$ and [O\ii]/H$\beta$ at fixed [O\iii]/H$\beta$ compared to \hii regions,
 suggesting that the galaxy spectra contain a significant DIG contribution based on the observed
 DIG line ratio relations in Figure~\ref{fig:digratios}.
  The interpretation that DIG is largely responsible for the offset between \hii regions and SDSS galaxies
 in these diagrams is supported by the observation of \citet{mas16} that SDSS galaxies
 display a dependence on H$\alpha$ surface-brightness (\sigha) perpendicular to these excitation sequences,
 such that those galaxies with the lowest \sigha\ are offset farthest from the \hii region sequences.
  Under the assumption that DIG accounts for a larger fraction of line emission in galaxies with lower \sigha\
 \citep{oey07}, the results of \citeauthor{mas16} imply that DIG emission is most important in
 those galaxies farthest offset from the \hii region sequences, while galaxies with large \sigha\ and
 highly-concentrated star formation appear more similar to \hii regions in these line ratio spaces.

In the O3S2 diagram, \hii regions and SDSS galaxies show the largest offset at moderate excitation
 ([O\iii]/H$\beta\sim0$) where the DIG [S\ii]/H$\alpha$ reaches a maximum.  Above and below this point, the \hii
 region and SDSS sequences appear to converge suggesting that the DIG line ratios become similar to
 \hii region line ratios in the low- and high-excitation tails.  A similar behavior is observed in the
 O3O2 diagram, where the \hii regions and SDSS galaxies converge at the lowest and highest metallicities.
  We therefore adopt simple linear extrapolations of the DIG excitation sequences that approach the
 point of convergence of \hii regions and SDSS galaxies
 at both high and low metallicities.  This behavior is consistent with the turnover in DIG
 [N\ii]/H$\alpha$ and [S\ii]/H$\alpha$ at moderate [O\iii]/H$\beta$ observed in the data.
  The adopted extrapolations are shown as blue dashed lines in Figure~\ref{fig:allratios}.
  It is possible that these linear extrapolations are not accurate representations of the DIG excitation
 sequences, however the relative locations of the \hii region and SDSS sequences
 suggest these extrapolations
 provide a good approximation.  The region of largest uncertainty is the extreme metal-poor regime in which
 the \hii region sequences turn over in [O\iii]/H$\beta$  while the DIG sequence extrapolations continue increasing
 in [O\iii]/H$\beta$.
  \citet{zha17} demonstrated that [O\iii]/H$\beta$ of DIG and \hii regions is the same on average at fixed galactocentric
 radius for the least-massive third of their sample (log(M$_*$/M$_{\odot})<9.43$), suggesting an agreement between
 DIG and \hii region [O\iii]/H$\beta$ in low-metallicity environments.  However, the MaNGA survey only targeted galaxies
 with log(M$_*$/M$_{\odot})\gtrsim9.0$ that do not have low enough stellar masses to populate the extreme
 low-metallicity tail.  The nature of DIG line ratios in this regime therefore cannot be directly constrained.
  It is possible that the DIG sequences also turn over like the \hii region sequences, but this uncertainty
 only affects a regime where a small fraction of SDSS galaxies lie, and will therefore minimally impact
 our results.

The low- and high-metallicity convergence of the SDSS and \hii region sequences could also arise
 from a changing DIG contribution with metallicity, such that emission line contribution from DIG
 is largest at moderate
 metallicity and is small at low and high metallicities.  In this case, the DIG line ratios need not
 converge with the galaxy and \hii region line ratios in either extreme regime.  However, individual
 SDSS galaxies do not
 show any evidence for a strong dependence of the fractional DIG contribution on O3 or \mstar\
 (see Section~\ref{sec:stackmodels} below), disfavoring an explanation based on a dynamic level
 of DIG contribution.

\subsection{Model framework}\label{sec:framework}

We create the individual line-emitting components of mock galaxies using the line-ratio
 relations of \hii and DIG regions described above, and construct fake global galaxy spectra
 by combining light from the individual components in a manner that mimics the ISM structure of
 real galaxies.
  Below we describe the methodology used to create one mock galaxy spectrum, which is repeated
 many times using a range of input parameters to build up a statistical sample of mock galaxies.

\begin{figure}
 \includegraphics[width=\columnwidth]{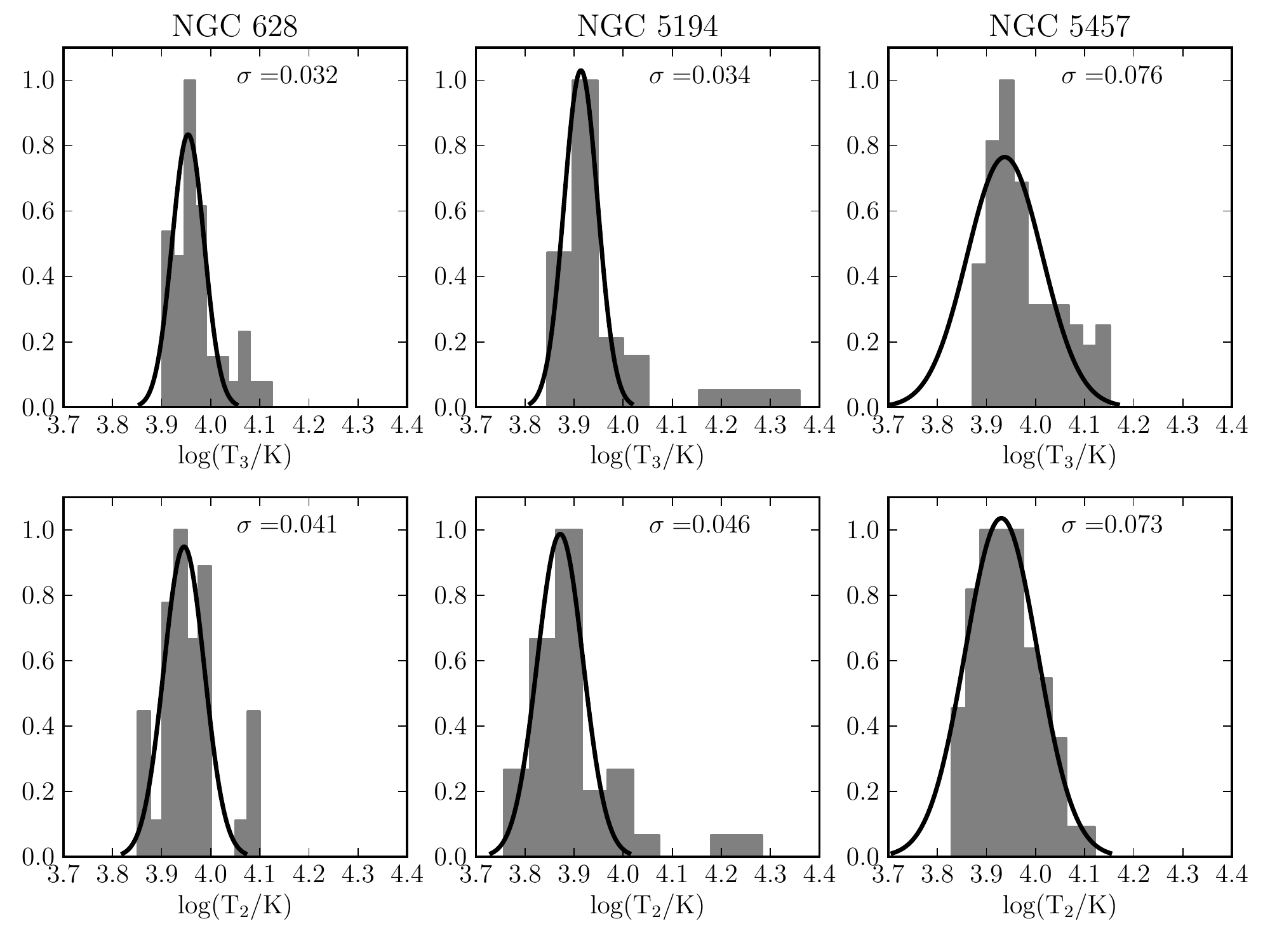}
 \caption{\te\ and \ttwo\ distributions of individual \hii regions within the local spiral galaxies
 NGC~628 \citep{ber15}, NGC~5194 \citep{cro15}, and NGC~5457 \citep{cro16}.
  Black lines show log-normal fits to the electron temperature distribution in each panel,
 where the best-fit width of the log-normal is given in the upper right corner.
}\label{fig:chaoste}
\end{figure}

First, we begin with a population of \hii regions.  As described in Section~\ref{sec:hiiregions},
 the strong line ratios are parameterized by the electron temperature \te, and thus the oxygen
 abundance is primarily a function of \te\ in the models.  We produce a population of \hii
 regions by randomly selecting N$_{\text{HII}}$ samples from an input distribution of \te.
  We adopt a log-normal shape for this \te\ distribution,
 in which the free parameters are the central temperature T$_{\text{cent}}$ and the logarithmic width
 $\sigma_{\text{T}}$.  A log-normal distribution is observationally motivated by the distributions of
 \te\ and \ttwo\ 
 of individual \hii regions in local star-forming spirals.  These distributions are shown in
 Figure~\ref{fig:chaoste}
 using data from three galaxies in the CHAOS survey \citep{ber15,cro15,cro16}.
  The central temperature T$_{\text{cent}}$
 is representative of the characteristic metallicity of the galaxy star-forming regions, while the
 width of the distribution $\sigma_{\text{T}}$ corresponds to the range of metallicities spanned
 by individual \hii regions.  The CHAOS galaxies are characterized by $\sigma_{\text{T}}=0.03-0.08$~dex.
  We note that a log-normal distribution has symmetric wings in log(T), but the CHAOS galaxies display
 high-temperature wings, with no corresponding low-temperature wings.
  This absence is likely an observational bias because of the
 exponential decline in auroral line strength with decreasing temperature, supported by the fact that
 the lowest-temperature measurements in each CHAOS galaxy tend to fall only just
 above the S/N$\geq3$ cut on auroral line strength.

We obtain the strong-line ratios for the \hii regions using the \te\ distribution as input
 to equations~\ref{eq:o3fit}-\ref{eq:s2fit}.
  We add an ionization parameter term to these median relations by assigning
 $\Delta$O3 to each mock \hii region by randomly drawing from a normal distribution with a mean of zero
 and a standard deviation of 0.14~dex, equal to the observed scatter about the median in the
 O3 vs. \te\ diagram for the reference \hii region sample.
  $\Delta$O2, $\Delta$N2, and $\Delta$S2 are then calculated for
 each mock \hii region using equations~\ref{eq:delo2}-\ref{eq:dels2}.  The final strong-line ratios
 for each mock \hii region are obtained by adding $\Delta$O3, $\Delta$O2, $\Delta$N2, and $\Delta$S2
 to the O3, O2, N2, and S2 values obtained from the polynomial fits of equations~\ref{eq:o3fit}-\ref{eq:s2fit}.
  Each of these strong-line ratios
 has H$\beta$ as the denominator, so we assume an \hii region H$\beta$ flux distribution in order to obtain
 the strong-line fluxes.  For simplicity, we assign the same H$\beta$ flux to each \hii region.
  This assumption does not affect our results because H$\beta$ flux does not
 show any dependence on either \te\ or 12+log(O/H) in the CHAOS \hii regions on a galaxy-by-galaxy basis,
 suggesting that the brightness of an \hii region does not depend on its abundance properties
 \citep{ber15,cro15,cro16}.
  Thus, using a distribution of H$\beta$ fluxes is simply a source of scatter but has no
 systematic effect on any of our results.
  The combination of strong-line ratios and H$\beta$ fluxes yields the intrinsic fluxes of
 [O\ii]$\lambda\lambda$3726,3729, H$\beta$, [O\iii]$\lambda\lambda$4959,5007,
 [N\ii]$\lambda\lambda$6548,6584, and [S\ii]$\lambda\lambda$6716,6731 for each \hii region.
  The H$\alpha$ flux is obtained assuming an intrinsic ratio of H$\alpha$/H$\beta=2.86$
  We then combine the [O\iii]$\lambda\lambda$4959,5007 and [O\ii]$\lambda\lambda$3726,3729 fluxes
 with the \te\ values to produce the intrinsic fluxes of the auroral lines [O\iii]$\lambda$4363 and
 [O\ii]$\lambda\lambda$7320,7330 using equations~\ref{eq:to3} and~\ref{eq:to2}.

\begin{figure}
 \includegraphics[width=\columnwidth]{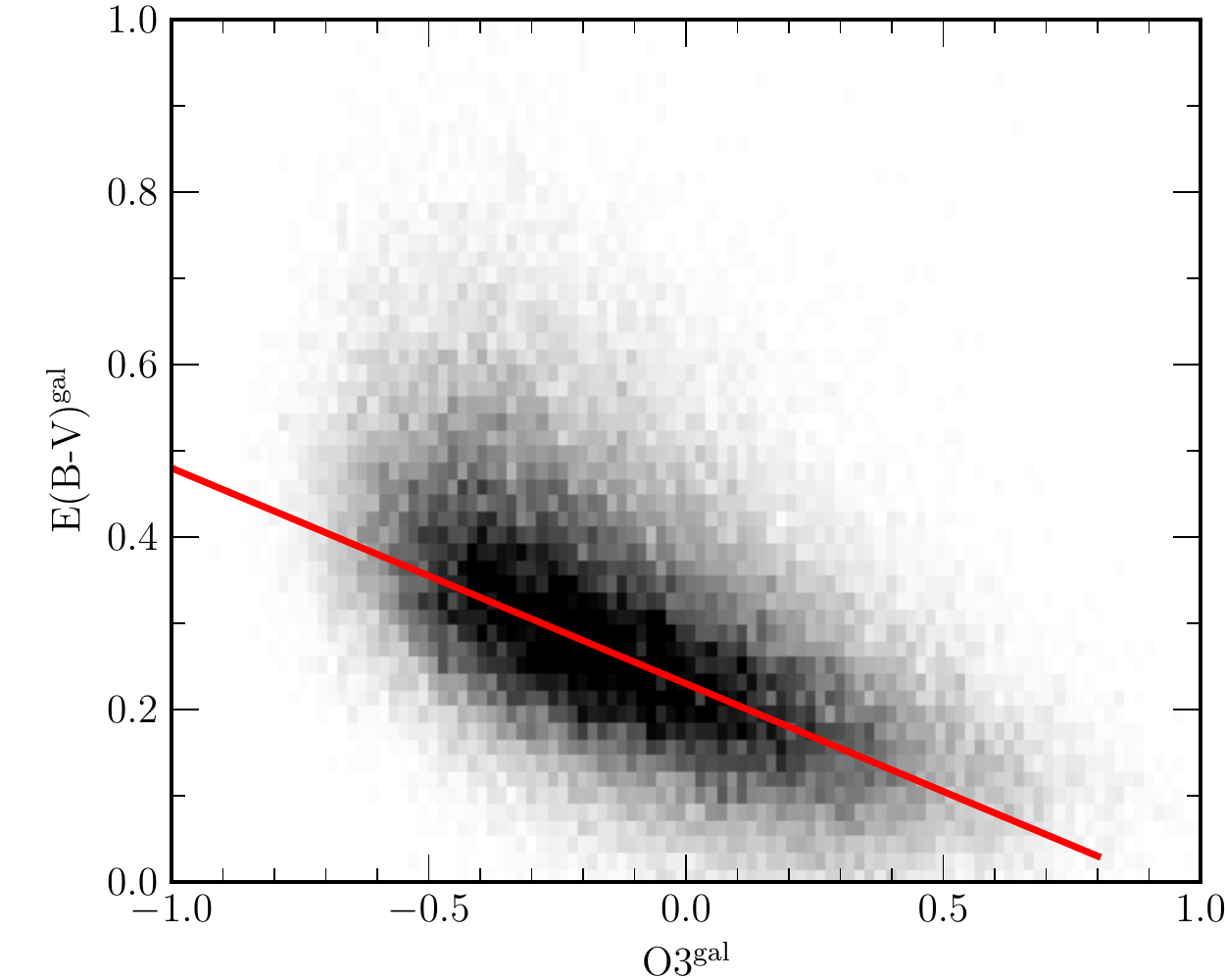}
 \caption{E(B-V)\gal\ as a function of O3\gal\ for the SDSS strong-line comparison sample.
  The amount of reddening increases with increasing metallicity and decreasing excitation,
 as represented by decreasing O3\gal.
  The red line shows the linear representation of this relationship, given by 
 $\text{E(B-V)}^{\text{gal}}=-0.25\times\text{O}3^{\text{gal}} + 0.23$.
}\label{fig:ebvsdss}
\end{figure}

The emission lines from each individual \hii region are then reddened.
  The E(B-V) values of individual \hii regions in the CHAOS sample do not correlate with
 \te\ or 12+log(O/H) on a galaxy-by-galaxy basis, suggesting that using a random E(B-V)
 assuming some distribution shape is appropriate.
  For each \hii region, we draw an E(B-V) value from a normal distribution with a width of 0.15 magnitudes, with
 negative E(B-V) values set to zero.
  The E(B-V) distributions of the CHAOS \hii regions suggest that this shape and width
 is appropriate for local star-forming spirals.
  While individual \hii region E(B-V) shows no correlation with metallicity,
 E(B-V)\gal\ inferred from the global galaxy spectrum does correlate with global galaxy properties
 such as O3\gal, the O3 ratio inferred from global galaxy spectra,
 as shown in Figure~\ref{fig:ebvsdss} for the strong-line comparison sample.
  The anticorrelation between E(B-V)\gal\ and O3\gal\ suggests that the center of the E(B-V)
 distribution for the \hii regions depends on the characteristic metallicity of the galaxy,
 reflecting the relationship between galaxy reddening and chemical enrichment \citep{hec98}.
  We adopt a linear representation of the data in Figure~\ref{fig:ebvsdss}, and use the median O3 of the
 simulated \hii regions to set the center of the E(B-V) distribution using this linear relation.
  We note that the relationship shown in Figure~\ref{fig:ebvsdss} is derived from the global galaxy
 ratio O3\gal, which we later conclude is biased with respect to the median \hii region ratio O3\shii.
  Iteratively including this bias in the E(B-V) relation changes the central E(B-V) values
 by $<0.04$~magnitudes, which has no impact on our results.
  We redden the strong and auroral line fluxes from each modeled \hii region individually using its assigned
 E(B-V), assuming the attenuation law of \citet{car89}.

We produce a number of DIG regions equal to the number of \hii regions, and each DIG region
 is associated with an \hii region.  DIG emission is observed to be spatially correlated with
 \hii regions \citep{zur00} and the diffuse gas immediately surrounding an \hii region likely has
 a similar metallicity to that of the gas in that \hii region.
  Each DIG region is assigned the same O3N2 value as its associated \hii region.  While
 the results of \citet{zha17} suggest that DIG and \hii region O3 is the same on average,
 O3 is double-valued as a function of electron temperature and metallicity and thus does not
 provide a good parameterization of these properties, as described in Section~\ref{sec:digregions} above.
  If we matched DIG and \hii regions in O3, the double-valued nature of the line ratio would make it unclear
 how to match DIG and \hii regions in the regime where O3 vs. \te\ is flat.
  We instead match in O3N2, which increases monotonically with increasing \te\ for \hii regions.
  Matching in O3N2 instead of O3 alone should still provide a realistic way of associating DIG and \hii regions
 based on the close agreement of the DIG and \hii region sequences in the O3N2 diagram and on the
 result of \citet{zha17} that the systematic difference between \hii and DIG region N2 is small.
  Given O3N2, the strong-line ratios O3, N2, S2, and O2 of each DIG region
 are assigned using the DIG excitation sequences shown in Figure~\ref{fig:allratios}.
  We use the
 running median of the DIG sequences if $-0.6<$log([O\iii]$\lambda$5007/H$\beta)<0.4$
 and use the linear extrapolations of the DIG sequences otherwise.

The DIG fraction, \fdig, is defined as the fraction of the total intrinsic Balmer
 line flux of the galaxy that originates from DIG.
  The H$\alpha$ and H$\beta$ line fluxes of each DIG region are assigned such that the intrinsic DIG
 Balmer line fluxes account for the fraction \fdig\ of the combined intrinsic Balmer line flux
 of the \hii and DIG region.
  The same \fdig\ is used for each \hii-DIG region pair in a single mock galaxy.
  The [O\iii]$\lambda\lambda$4959,5007, [N\ii]$\lambda\lambda$6548,6584, [S\ii]$\lambda\lambda$6716,6731, and
 [O\ii]$\lambda\lambda$3726,3729 fluxes are then calculated using the
 strong-line ratios and Balmer line fluxes.
  In order to calculate the auroral line fluxes of the DIG regions, we need the DIG electron
 temperatures \te\ and \ttwo.  However, there are no observational constraints on the electron
 temperature of DIG because of its low surface brightness and the intrinsic weakness of the auroral lines.
  We initially assume that the electron temperatures \te\ and \ttwo\  of each DIG region
 are equal to the electron
 temperatures of the associated \hii region, but we reevaluate this assumption
 in Section~\ref{sec:stackmodels} below.
  With the assumed DIG \te\ and \ttwo, the intrinsic DIG auroral-line fluxes [O\iii]$\lambda$4363 and
 [O\ii]$\lambda\lambda$7320,7330 are calculated using equations~\ref{eq:to3} and~\ref{eq:to2}.
  The line fluxes of each DIG region are then reddened assuming the same extinction as the
 associated \hii region.

The global galaxy spectrum is produced by summing the reddened flux from each \hii and DIG region on a
 line-by-line basis for [O\ii]$\lambda\lambda$3726,3729, [O\iii]$\lambda$4363, H$\beta$,
 [O\iii]$\lambda\lambda$4959,5007, [N\ii]$\lambda$5755, H$\alpha$, [N\ii]$\lambda\lambda$6548,6584,
 [S\ii]$\lambda\lambda$6716,6731, and [O\ii]$\lambda\lambda$7320,7330.
  We then analyze the global galaxy spectrum as if it were real global galaxy spectroscopic data.
  The global line fluxes
 are corrected for reddening using the summed H$\alpha$ and H$\beta$ fluxes assuming
 an intrinsic ratio H$\alpha$/H$\beta=2.86$ and the attenuation law of \citet{car89}.  The
 strongline ratios are calculated using the dereddened global line fluxes.

The global electron temperatures are calculated from the global dereddened strong-to-auroral line
 ratios using equations~\ref{eq:to3} and~\ref{eq:to2}, and global direct-method oxygen abundances are calculated
 using equations~\ref{eq:oplush} and~\ref{eq:o2plush}.
  We calculate global 12+log(O/H) under three assumptions: (1) both
 [O\iii]$\lambda$4363 and [O\ii]$\lambda\lambda$7320,7330 are detected, (2) only [O\iii]$\lambda$4363
 is detected, and (3) only [O\ii]$\lambda\lambda$7320,7330 is detected.  We refer to the three
 metallicities as 12+log(O/H)$_{\text{T2,T3}}$, 12+log(O/H)$_{\text{T3}}$, and 12+log(O/H)$_{\text{T2}}$,
 respectively.
  In cases 2 and 3, the unknown electron temperature is estimated from the known electron temperature
 using equation~\ref{eq:t2t3}.
  These three oxygen abundance values will be useful for comparing with different real datasets since
 it is not uncommon for only one of the auroral oxygen lines to be detected in galaxy spectra, even in stacks.

The process described above is repeated many times while varying T$_{\text{cent}}$ in order to
 build up a statistical sample of mock galaxy spectra, allowing us to average over sources of scatter to
 find median trends.  In order to quantify the bias between the distribution of \hii region properties
 and the global properties as inferred from the galaxy spectrum, for each line ratio or physical property
 we save both the global value inferred from the galaxy spectrum and
 the median value of the distribution of individual \hii regions for every mock galaxy.
  Properties derived from the global galaxy spectra will be indicated with the superscript ``gal,"
 while median properties of the \hii region distribution will be denoted by the superscript ``HII."

There are only four free parameters in these models.
  These free parameters are the number of \hii regions per galaxy, N$_{\text{HII}}$,
 the central temperature of the \hii region \te\ distribution, T$_{\text{cent}}$,
 the width of the \hii region \te\ distribution, $\sigma_{\text{T}}$,
 and the fraction of intrinsic Balmer flux originating from DIG emission, \fdig.
  In practice, $\sigma_{\text{T}}$ and \fdig\ are set to observationally-motivated
 values appropriate for the real dataset being modeled, while T$_{\text{cent}}$ is freely
 varied to produce galaxies with a range of metallicities.
  The value of N$_{\text{HII}}$ determines how well the \te\ distribution is sampled, and thus
 simply corresponds to a source of scatter if N$_{\text{HII}}$ is small, but does not
 change any trends.

\section{Comparison datasets and fiducial models}\label{sec:results}

We compare our models to observations of local star-forming galaxies in order to verify that the mock galaxy
 spectra produced following the methodology of Section~\ref{sec:framework}
 resemble spectra of real galaxies.
  Auroral emission lines are detected for a few hundred individual SDSS galaxies
 \citep{izo06,pil10}, and we assemble a sample of such galaxies for comparison
 in Section~\ref{sec:auroralmodels}.
  However, samples of individual SDSS galaxies for which auroral lines are detected are not
 representative of typical star-forming galaxies from which the $z\sim0$ MZR is
 constructed, generally having much higher SFR at fixed \mstar\ than average and sampling only
 the low-mass, low-metallicity tail of the local population.  Auroral line measurements across
 a wide dynamic range of galaxy properties have been obtained by stacking SDSS spectra
 \citep{and13,bro16,cur17}, providing a comparison sample that is more representative than
 samples of individual galaxies with auroral line detections.
  We therefore focus primarily on constructing models
 representing typical $z\sim0$ star-forming galaxies.
  We compare these models to the SDSS stacks from
 \citet{and13}, \citet{bro16}, and \citet{cur17} (hereafter AM13, B16, and C17, respectively),
 and quantify biases in metallicity measurements made from global galaxy spectra.
  The electron temperatures and oxygen abundances for all comparison samples
 are calculated using the same methods as for the mock galaxy spectra, outlined in Section~\ref{sec:models}.

\subsection{SDSS stacks with auroral line detections}\label{sec:stackmodels}

Auroral line measurements have been obtained across a wide range of stellar masses, SFRs, and
 excitation levels by stacking the spectra of $z\sim0$ star-forming galaxies from SDSS.  Creating composite
 spectra in bins of global galaxy properties allows for the detection of the weak auroral lines
 [O\iii]$\lambda$4363 and [O\ii]$\lambda\lambda$7320,7330 by leveraging the statistical power of
 hundreds or thousands of galaxies per bin to increase sensitivity.  This method has progressed
 metallicity studies of local galaxies by reducing the reliance on strong-line indicators.
  AM13 utilized measurements of composite spectra binned in \mstar\ alone, as well as \mstar\ and SFR,
 to investigate the
 MZR and FMR using direct-method metallicities.  B16 constructed composite spectra for
 SDSS galaxies in bins of \mstar\ and position above or below the mean $z\sim0$ relation between \mstar
 and specific star-formation rate (SSFR; SFR/\mstar),
 demonstrating a systematic dependence of strong-line indicators on position relative to the
 \mstar-SSFR relation.  C17 binned galaxies in O3 and O2,
 and utilized auroral line measurements from stacked spectra to provide fully empirical strong-line
 metallicity calibrations based on global galaxy spectra rather than \hii regions for the first time.
  Strong and auroral line ratio measurements of the stacked spectra from these studies provide a
 comparison sample that both spans a wide range of metallicities and is representative of the
 $z\sim0$ star-forming population.

Our goal is to quantify the mean bias in metallicity measurements inferred from global galaxy
 spectra relative to the characteristic metallicity of the \hii regions within a galaxy as a function
 of global galaxy properties.
  Thus, it is imperative that the samples that are used to test the performance of the models
 are themselves representative of the normal star-forming population of galaxies.
  While the stacked
 spectra from AM13, B16, and C17 are constructed from samples that largely overlap
 (AM13 and B16 use identical sample selection, while the selection criteria of C17 only slightly differ),
 we simultaneously compare to stacks from all three works in order to average over differences in
 selection, binning, and stacking methods.  Because of the binning methods of each work, some bins will
 contain galaxies that are wholly unrepresentative of the typical local population (e.g., \mstar-SFR bins
 that fall far from the mean $z\sim0$ \mstar-SFR relation).
  We only compare to
 those stacks from each work that closely follow the mean galaxy property relations of the local population.
  We use the stacks binned in \mstar\ only from AM13.
  From B16, we use those stacks
 that fall within $\pm0.5$~dex in SSFR of the mean $z\sim0$ \mstar-SSFR relation.  We select the C17 stacks
 for which the central O3 and O2 of the bin fall within $\pm0.1$~dex
 of the median relation of the strong-line comparison sample of individual SDSS galaxies.
  As described below, we choose observationally-motivated values of the DIG fraction \fdig,
 the number of \hii regions per mock galaxy  N$_{\text{HII}}$, and
 the \te\ distribution width $\sigma_{\text{T}}$
 appropriate for the sample of galaxies from which the stacked spectra were created.

\begin{figure}
 \includegraphics[width=\columnwidth]{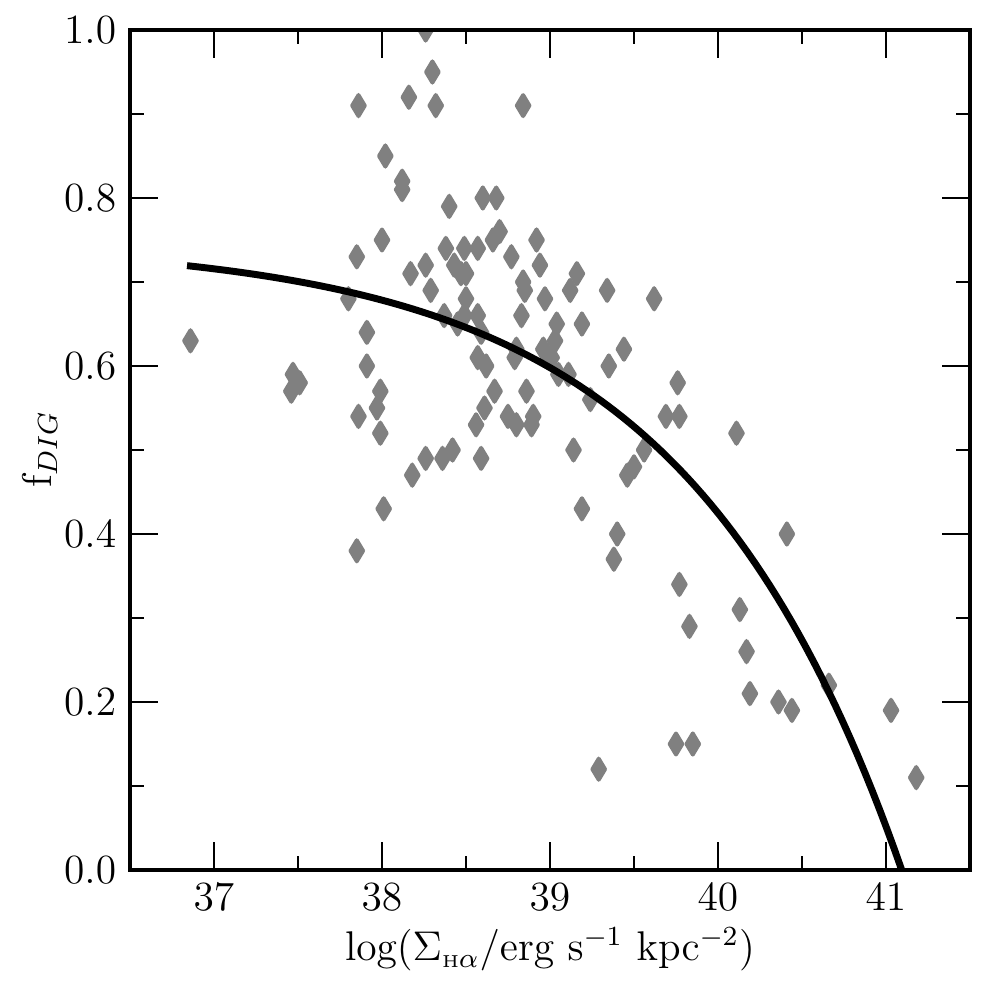}
 \caption{The fraction of Balmer emission originating from DIG, \fdig, vs. \sigha\ for galaxies from the
 SINGG H$\alpha$ survey \citep{oey07}, displayed as gray diamonds.  The best-fit function of the form
 \fdig$\sim\Sigma_{\text{H}\alpha}^{1/3}$, as suggested by \citeauthor{oey07} on theoretical grounds,
 is displayed as a black line and presented in equation~\ref{eq:fdig}. 
}\label{fig:oeyfit}
\end{figure}

We place constraints on \fdig\ using the H$\alpha$ surface brightness, \sigha, given by 
\begin{equation}\label{eq:sigha}
\Sigma_{\text{H}\alpha}=\frac{L_{\text{H}\alpha}^{\text{tot}}}{2\pi R_{\text{half,H}\alpha}^2},
\end{equation}
where $L_{\text{H}\alpha}^{\text{tot}}$ is the total H$\alpha$ luminosity, and
 $R_{\text{half,H}\alpha}$ is the half-light radius of the galaxy H$\alpha$ emission.
  \citet{oey07} demonstrated that \fdig\ decreases with increasing \sigha.
  The authors argued that a scenario in which \hii regions occupy a larger fraction of the ionized ISM
 volume as star formation becomes more concentrated predicts a dependence of
 \fdig$\sim\Sigma_{\text{H}\alpha}^{1/3}$, which agreed well with the data.
  Using the dataset from \citet{oey07}, we fit \fdig\ as a function of \sigha\ 
 assuming this theoretically-predicted functional form, and obtain
\begin{equation}\label{eq:fdig}
f_{\text{DIG}} = -1.50\times10^{-14}\times\Sigma_{\text{H}\alpha}^{1/3} + 0.748,
\end{equation}
where \sigha\ is given in units of erg~s$^{-1}$~kpc$^{-2}$.  The data and best-fit function
 are shown in Figure~\ref{fig:oeyfit}.

We constrain \fdig\ for the SDSS stack samples using \sigha\ of the individual SDSS galaxies
 in the strong-line comparison sample, which is nearly identical to the sample from which
 the AM13 and B16 stacks were constructed (see Section~\ref{sec:digregions}).
  We note that results do not change if the strong-line comparison sample is instead selected
 using the criteria of C17.
  To determine \sigha\ for the strong-line comparison sample, we first aperture-correct the
 intrinsic fiber H$\alpha$ luminosities.  We obtain aperture corrections by dividing the
 total SFR by the fiber SFR, and apply these correction factors to the fiber H$\alpha$ luminosities
 to obtain total H$\alpha$ luminosities \citep{bri04}.
  While measurements of the H$\alpha$ half-light radii are not available, we instead use the optical sizes.
  R-band sizes of local star-forming galaxies have been shown to be similar to H$\alpha$ sizes
 \citep{jam09}.  We use the elliptical Petrosian R-band half-light radii
 from the NASA-Sloan Atlas v1.0.1\footnote[5]{\texttt{http://www.nsatlas.org}}.
  Galaxy sizes are not available for all galaxies in the strong-line comparison sample.
  The NASA-Sloan Atlas contains
 size measurements for 79\% of the full sample ($\sim165,000$ galaxies) for which we compute
 the dust-corrected \sigha\ according to
 equation~\ref{eq:sigha}.
  The DIG fraction for each galaxy is then estimated using equation~\ref{eq:fdig}.

\begin{figure}
 \includegraphics[width=\columnwidth]{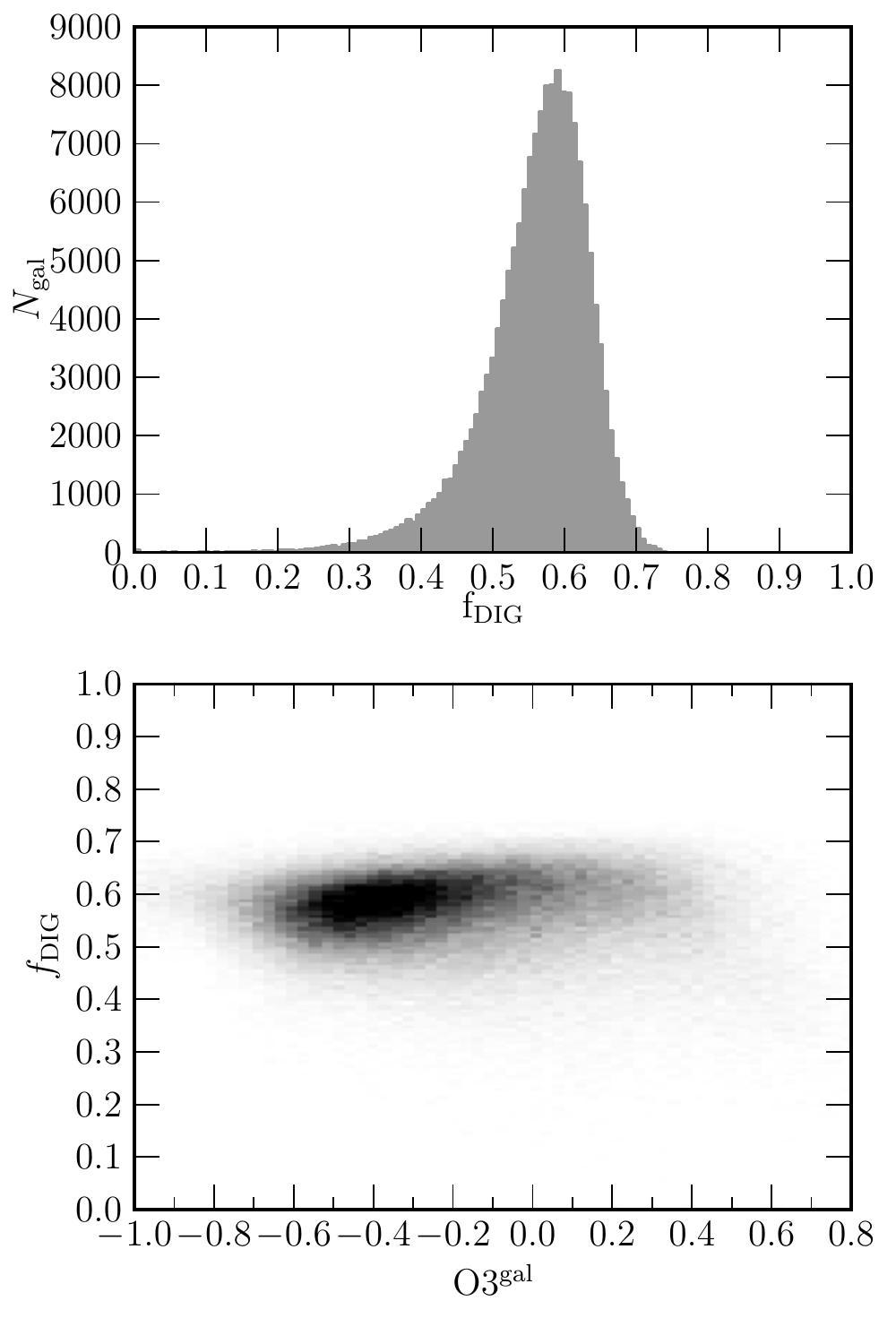}
 \caption{The distribution of \fdig\ (top) and \fdig\ as a function of O3 (bottom)
 for SDSS galaxies in the strong-line comparison sample.  The \fdig\ distribution has a median
 value of 0.55 and a standard deviation of 0.08, and is nearly Gaussian in shape.
  The DIG fraction \fdig\ does not show a strong dependence on excitation, as encoded by O3.
}\label{fig:sdssfdig}
\end{figure}

The distribution of \fdig\ values and \fdig\ vs. O3\gal\ for the strong-line comparison
 sample are shown in Figure~\ref{fig:sdssfdig}.
  The strong-line comparison sample has a median \fdig\ of 0.55 with a standard deviation of 0.08.
  The distribution shape is nearly Gaussian, with a more significant tail towards low \fdig.
  The DIG fraction shows no significant dependence on excitation across a wide dynamic range, with
 the median \fdig\ changing by $<5$\% as a function of O3\gal.
  Additionally, the scatter in \fdig\ also shows no strong dependence on O3\gal.  We therefore assign \fdig\ 
 to each mock galaxy by randomly drawing values from a normal distribution with a mean value of 0.55 and a
 standard deviation of 0.08.  It is important to note that \fdig\ is not dependent on any line ratios, and is thus
 independent of all of the line-ratio diagrams that we use to test the models.

As noted above, the number of \hii regions per galaxy, N$_{\text{HII}}$, does not affect any
 trends but instead corresponds to a source of uncorrelated scatter, which decreases as
 N$_{\text{HII}}$ increases and the \te\ distribution is better sampled.
  The chosen value of N$_{\text{HII}}$ has no systematic effects on our results.
  Nevertheless, we choose a value of N$_{\text{HII}}$ that is appropriate for SDSS fiber
 observations of local star-forming galaxies.  At $z_{\text{med}}=0.08$, the median redshift
 of the strong-line comparison sample, the 3\arcsec\ diameter of an SDSS fiber corresponds to a physical
 diameter of 4.5~kpc.  Based on \hii region identification in narrowband H$\alpha$ surveys \citep[e.g.,][]{zur00},
  a $\sim4.5$~kpc diameter aperture will typically contain tens of \hii regions, but may contain as
 few as a handful of \hii regions depending on fiber placement, the ISM structure, and level of
 star-formation activity in the galaxy.  We choose a value of N$_{\text{HII}}=25$ to roughly match the number
 of \hii regions expected to fall within an SDSS fiber aperture.

We set $\sigma_{\text{T}}=0.07$~dex, a value that appears to be reasonable for $z\sim0$ star-forming
 galaxies as described below.
  This value is within the range of observed $\sigma_{\text{T}}$ for local spirals in the CHAOS
 galaxy survey \citep[see Fig.~\ref{fig:chaoste},][]{ber15,cro15,cro16}.
  We note that the \te\ distribution width, $\sigma_{\text{T}}$, encodes both stochastic variations
 in metallicity due to inhomogeneities in the ISM and systematic
 variations from radial metallicity gradients, if the observed aperture covers a large area of the disk
 compared to the steepness of the gradient.  Radial oxygen abundance gradients of local star-forming galaxies
 tend to have slopes ranging from $\sim-0.01$ to $\sim-0.1$ dex kpc$^{-1}$ for galaxies with
 $8.5<\log(\text{M}_*/\text{M}_{\odot})<11.$, with less massive galaxies displaying steeper gradients on
 average \citep{san14,ho15}.  At the median redshift of the strong-line comparison sample,
 $z_{\text{med}}=0.08$, 1\arcsec\ corresponds to a physical length of 1.5~kpc.
  With the assumption that the 3\arcsec\ SDSS fiber is placed on the center of
 each galaxy, the light falling in the fiber probes the inner $\sim2$~kpc radially.
  Thus, the additional temperature
 variations due to metallicity gradients are likely only significant for the least-massive galaxies
 in SDSS.
  After measurement uncertainty is accounted for, the intrinsic scatter of \hii regions about the
 metallicity gradients in local spirals is $\sim0.05-0.1$ dex \citep{ken03,ros08,bre11,ber13,cro15,cro16},
 corresponding to $\sim0.02-0.07$ dex in \te\ indicitave of the minimum $\sigma_{\text{T}}$ in the
 absence of metallicity gradients for local star-forming galaxies.  The shallow gradients of the CHAOS
 galaxies (0.02-0.04~dex kpc$^{-1}$) suggest that stochastic variations in metallicity account for
 the majority of the width of the electron temeperature distributions in Figure~\ref{fig:chaoste}.

In summary, the model matched to typical SDSS star-forming galaxies assumes a Gaussian \fdig\ 
 distribution characterized by a mean and standard deviation of 0.55 and 0.08, respectively,
 N$_{\text{HII}}=25$, and $\sigma_{\text{T}}=0.07$~dex.
  We create 2500 mock galaxy spectra following the method described in Section~\ref{sec:framework},
 where T$_{\text{cent}}$ is drawn from a logarithmic uniform distribution from
 log(T$_{\text{cent}}/$K$)=3.7$~to~4.3 (5,000 to 20,000~K), the \te\ range of the reference \hii region sample.
  We infer median line ratio and electron temperature relations from these 2500 mock galaxy spectra.
  We create an additional model for comparison with the same parameters except \fdig$=0$ such
 that the mock galaxies are constructed from \hii regions only and include no DIG emission.
  We refer to this model with no DIG emission as the \hiimod model.

\begin{figure*}
 \centering
 \includegraphics[width=\textwidth]{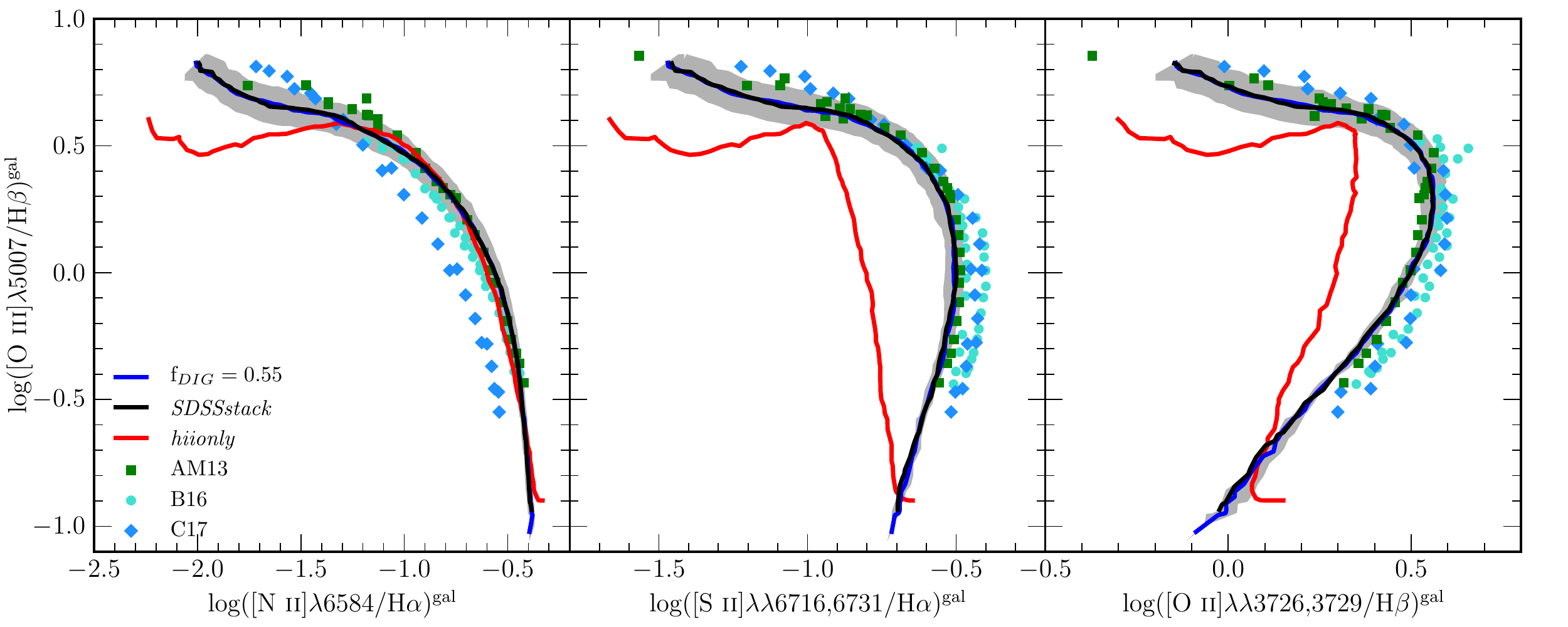}
 \caption{The O3N2 (left), O3S2 (middle), and O3O2 (right) strong-line ratio
 diagrams for stacks of SDSS galaxies and models under different sets
 of assumptions and input parameters.  The stacks of $z\sim0$ SDSS star-forming galaxies from AM13, B16, and
 C17 are shown as green squares, light blue circles, and dark blue diamonds, respectively.
  The blue line shows the running median of mock galaxies in bins of O3N2\gal\ for the model with \fdig=0.55
 and equal \hii and DIG \ttwo\ at fixed metallicity.  The shaded gray region represents the 68th-percentile width
 of the distribution of mock galaxies around this running median.
  The black line displays the running median for the \stackmod model with \fdig=0.55 under the assumption that
 DIG \ttwo\ is 15\% lower than \hii region \ttwo\ at fixed metallicity, and is identical to the blue line
 in these strong-line ratio diagrams.
 The red line shows the \hiimod model that does not include DIG emission (\fdig=0.0).
}\label{fig:bptdigfrac055}
\end{figure*}

\begin{figure}
 \includegraphics[width=\columnwidth]{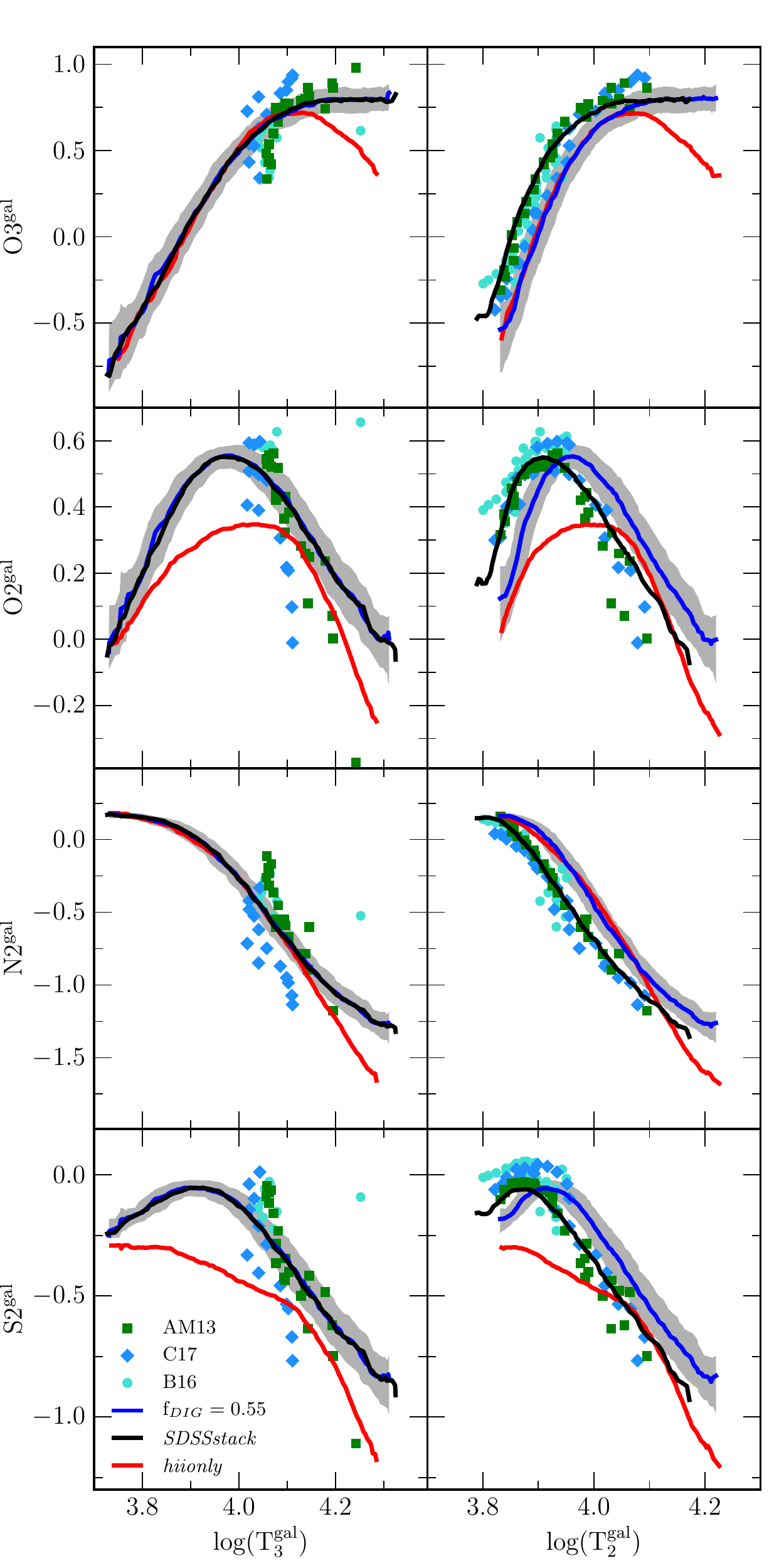}
 \caption{Global galaxy strong-line ratios O3\gal, O2\gal, N2\gal, and S2\gal\ as a function of
 \tegal\ (left column) and \ttwogal\ (right column).  Lines and points are the same as in
 Figure~\ref{fig:bptdigfrac055}.
  The \fdig=0.55 model is systemtacially offset towards higher \ttwo\ at fixed line ratio compared to
 the SDSS stacks.  This offset is not present in the \stackmod model, in which we assume that DIG \ttwo\
 is lower than \hii region \ttwo\ by 15\% at fixed metallicity.
}\label{fig:vstdigfrac055}
\end{figure}

\begin{figure}
 \includegraphics[width=\columnwidth]{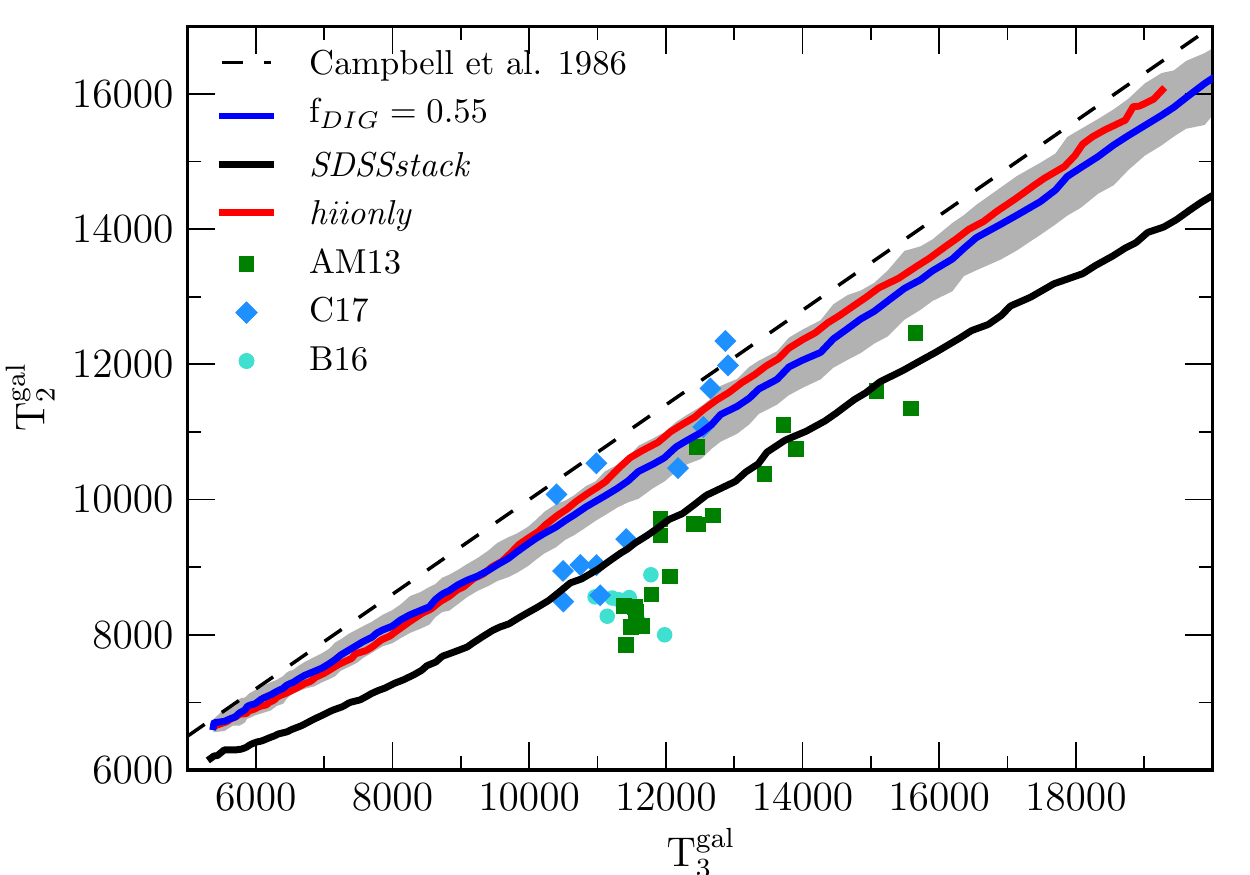}
 \caption{The global galaxy ionic temperature diagram of \ttwogal\ vs. \tegal.
  The dashed black line shows the \hii region \ttwo-\te\ relation of \citet{cam86} given in
 equation~\ref{eq:t2t3}.  All other lines and points are the same as in Figure~\ref{fig:bptdigfrac055}.
}\label{fig:t2t3digfrac055}
\end{figure}

We compare the model with \fdig=0.55 matched to SDSS stacks to AM13, B16, and C17 stacks in
 Figures~\ref{fig:bptdigfrac055}-\ref{fig:t2t3digfrac055}, and include the \hiimod model for
 comparison.
  The O3N2, O3S2, and O3O2 strong-line ratio diagrams are shown in Figure~\ref{fig:bptdigfrac055}.
  The \fdig=0.55 model shows excellent agreement with excitation sequences followed by the SDSS stacks.
  The \hiimod model fails to reproduce the O3S2 and O3O2 sequences, displaying lower S2 and O2 at
 fixed O3 than the AM13, B16, and C17 stacks at nearly all values of O3.  The largest disagreement
  occurs in the moderate metallicity regime where O$3\sim0.0$.
 This failure of the \hiimod model confirms that combinations
 of \hii regions alone cannot simultaneously reproduce line ratio sequences in all line-ratio spaces.
  DIG emission properties are distinct from those of \hii regions in S2 and O2,
 which strongly affects global galaxy line ratios and must be taken into consideration.
  The close agreement of the \fdig=0.55 model to the observations in the O3S2 and O3O2 diagrams
 suggests that both the DIG excitation sequences in Figure~\ref{fig:allratios} and the \fdig\ relation in
 equation~\ref{eq:fdig} are reasonable.

The strong-line ratios O3\gal, O2\gal, N2\gal, and S2\gal\
 are shown as a function of \tegal\ and \ttwogal\ in Figure~\ref{fig:vstdigfrac055}.  \tegal\ measurements
 are only available for the SDSS stacks  with $\text{T}_3^{\text{gal}}>10^4$~K.
  The reason for this limited range is twofold.
  First, the strong-to-auroral line ratio [O\iii]$\lambda\lambda$5007,4959/$\lambda$4363 becomes
 exponentially weaker at lower \te\ while O3 also drops off significantly due to a smaller
 fraction of oxygen in the O$^{++}$ state, leading to an extremely weak [O\iii]$\lambda$4363 at low \tegal\ that
 may not even be detected in stacks.  Second, [O\iii]$\lambda$4363 is blended with [Fe\ii]$\lambda$4360,
 which significantly contaminates [O\iii]$\lambda$4363 measurements for high-metallicity, low-\tegal\ galaxies
 in which [Fe\ii]$\lambda$4360 is stronger
 \citep{and13,cur17}.  \tegal\ measurements are not shown for AM13, B16, and C17 stacks
with [Fe\ii]$\lambda$4360/[O\iii]$\lambda4363>0.5$ for which
 [O\iii]$\lambda$4363 measurements were deemed unreliable.
  Due to the combined effect of these two limitations, \tegal\ measurements are only shown for stacks of galaxies
 with log(M$_*$/M$_{\odot})\lesssim9.4$.
  The \fdig=0.55 model matches the observed SDSS stacks well in the \tegal\ diagrams within the amount of
 scatter displayed by the SDSS stacks.
  The \hiimod
 model fails to produce high enough O2\gal\ and S2\gal\ values at moderate \tegal\ to match the observations.
  Due to the limited dynamic range of the \tegal\ measurements for the SDSS stacks, the turnover points
 of the models (log$(\text{T}_3^{\text{gal}})\approx3.95$ for O2\gal\ and S2\gal)
 that would provide an excellent test for agreement are not sampled by the SDSS stacks.

In the right column of Figure~\ref{fig:vstdigfrac055},
 we again compare the predicted model line ratios with those observed in the SDSS composites,
 this time as a function of \ttwogal.
  The auroral line [O\ii]$\lambda\lambda$7320,7330 does not suffer
 from contamination or severe dropoff in brightness at high metallicities, and is thus robustly
 measured across a much wider range of temperatures than [O\iii]$\lambda$4363.
  The \hiimod model shows large discrepancies in O2\gal\ and S2\gal\, again demonstrating the importance
 of accounting for DIG emission in global galaxy spectra.
  The shapes of the \fdig=0.55 model sequences match those of the observed sequences
 well, but with a systematic offset towards higher \ttwogal\ at fixed line ratio
 that is seen in all four strong-line ratios.

Since strong-line ratios of \hii regions as a function of \te\ and \ttwo\ are directly constrained by
 observations, this offset in \ttwogal\ must either  originate from incorrect assumptions about
 DIG region line ratios or
 electron temperatures, or from a systematic effect in the binning and stacking process that is not
 captured in our models that only produce individual galaxy spectra.
  It is improbable that incorrect strong-line DIG excitation sequences (see Figure~\ref{fig:allratios}) are the
 cause of this offset, because shifting the strong-line ratios of DIG regions alone
 would lead to a mismatch in the strong-line vs. strong-line sequences shown in Figure~\ref{fig:bptdigfrac055},
 which agree well under the current set of assumptions.  Additionally, no systematic offset
 is observed the strong-line vs. \tegal\ plots in Figure~\ref{fig:vstdigfrac055},
 and changing the DIG strong-line excitation sequences would introduce a disagreement in these diagrams as well.
  The \ttwogal\ offset also cannot be resolved by decreasing \fdig\ as \ttwogal\ increases.
  This adjustment would introduce disagreement in Figure~\ref{fig:bptdigfrac055}
 and the left column of Figure~\ref{fig:vstdigfrac055} while still failing to match the
 high-\ttwogal\ tail of the observations where
 even the \hiimod model overestimates O2\gal, N2\gal, and S2\gal\ and underestimates O3\gal\ at fixed \ttwogal.

The \ttwo-\te\ diagram is shown in Figure~\ref{fig:t2t3digfrac055}.
  In this diagram, the SDSS stacks display lower \ttwogal\ at fixed \tegal\ than the \hii
 region relation of \citet{cam86} (equation~\ref{eq:t2t3}).  Both the \hiimod\ and  \fdig=0.55 models fall
 below the \hii region \ttwo-\te\ relation, but neither show as large of an offset as the SDSS stacks.
  The small difference between the \hiimod\ (\fdig=0.0) and \fdig=0.55 models is predominantly due to the
 different relation between O2 and \ttwo\ for DIG regions compared to that of \hii regions, which changes
 the relative weight of regions of different \ttwo\ to the \ttwogal\ estimate from the global galaxy spectrum.
 In order to match the observations, the \fdig=0.55 model must have higher \tegal\ at fixed \ttwogal or
 lower \ttwogal\ at fixed \tegal.  No systematic
 offset in \tegal\ is observed in Figure~\ref{fig:vstdigfrac055}, suggesting that the mismatch
 between model and observations in the \ttwo-\te\ diagram is caused by a mismatch in \ttwogal\ alone.
  The \ttwogal\ offsets in Figure~\ref{fig:vstdigfrac055} appear to be roughly equivalent for each
 strong-line ratio, with the model being $\sim0.05$ dex higher in \ttwogal\ at fixed line ratio
 than the observations, corresponding to an offset of $\sim1,000-1,500$~K at \ttwo$=8,000-12,000$~K
 that closely matches the \ttwogal\ discrepancy between the \fdig=0.55 model and observations at fixed \tegal.
  We conclude that the discrepancy between the \fdig=0.55 model and SDSS composites
 originates from \ttwogal\ alone.

The \ttwogal\ discrepancy between model and observations could arise from either an incorrect
 assumption about DIG electron temperatures in the models or some aspect of the stacking process that is not
 captured by our individual galaxy models, since we are comparing to data from stacked spectra.  We do not have
 a way of testing the latter hypothesis without matching the scatter in SDSS excitation sequences in
 detail and obtaining actual line profiles from the models in addition to line fluxes,
 which our models do not do since they are only designed to investigate mean properties of the
 local galaxy population.  However, both AM13 and C17 performed their stacking procedures on individual
 SDSS galaxies with auroral-line detections and found that the inferred electron temperatures and
 metallicities of the stacks were not systematically offset from the mean values of the individual galaxies.
  These tests suggest that the offset in \ttwogal\ does not originate from a systematic effect of the binning
 and stacking procedures.

The \ttwogal\ discrepancy then most likely arises from an incorrect assumption about \ttwo\
 of DIG regions.
  Up to this point, we have assumed that \ttwo\ and \te\ of a DIG region are the same as those of the
 associated \hii region.  The observed \ttwogal\ offset instead suggests that the DIG \ttwo\ is lower than
 that of \hii regions at fixed metallicity, while \te\ remains roughly equivalent.  We find that
 the model can be brought into excellent agreement with the SDSS composities if the \ttwo\ of each
 DIG region is assumed to be 15\% (0.06 dex) lower than that of the associated \hii region.
  Models incorporating the revised DIG \ttwo\ assumption are shown in
 Figures~\ref{fig:bptdigfrac055}-\ref{fig:t2t3digfrac055} as black lines, and we refer to this
 model as the \stackmod model.
  Results are unchanged for diagrams that do not involve \ttwogal, while the shift towards
 lower \ttwogal\ brings the adjusted model into excellent agreement with the SDSS stacks in
 Figures~\ref{fig:vstdigfrac055} and ~\ref{fig:t2t3digfrac055}.
  We adopt the \stackmod model as the fiducial description
 for typical $z\sim0$ star-forming galaxies for the remainder of this work.

Our assumption that DIG \ttwo\ is lower than \ttwo\ of \hii regions on average is in conflict
 with past studies of Milky Way and extragalactic DIG, which suggest that DIG electron temperature
 is higher on average than that of \hii regions.
  Studies of DIG strong-line ratios in the Milky Way \citep{haf99} and other galaxies \citep{ott01,ott02,hoo03}
 have suggested that DIG is hotter than \hii regions on average based on the larger N2, S2, and O2 ratios
 observed for DIG.  Additionally, \citet{rey01} detected the auroral line [N\ii]$\lambda$5755 for DIG along
 one line-of-sight in the Milky Way, and found that DIG along this line-of-sight has a higher temperature
 than bright Galactic \hii regions. 
  However, the Milky Way results only probe a number of distinct sight lines and
 most extragalactic studies of DIG line-ratios observe extra-planar DIG in edge-on galaxies.
  It is not clear how representative such observations are of the DIG regions observed in face-on disk
 galaxies in the MaNGA DIG galaxy sample, and if past comparisons between DIG and \hii region temperature
 have been made at fixed nebular abundance.
  Additional observations of temperature-sensitive auroral lines for DIG regions are required to
 unequivocally settle the question of DIG electron temperature.
  Even if DIG \ttwo\ is not lower than that of \hii regions in reality, adopting this
 assumption in our model framework captures an important systematic effect that is present in
 global galaxy spectra and will contribute to the bias in direct-method metallicity.

One concern is whether the disagreement in the \ttwo-\te\ diagram and strong-line ratio
 vs. \ttwo\ diagrams can be resolved by adjusting other model parameters instead of making
 an assumption that DIG \ttwo\ is 15\% lower than \hii region \ttwo\ at fixed metallicity.
  In particular, the width of the \te\ distribution, $\sigma_{\text{T}}$, can change the
 magnitude of the offset in the \te-\ttwo\ diagram.  This effect has been previously shown by
 \citet{pil12no}, who demonstrated that the \ttwo-\te\ offset increases as the range of
 metallicities of combined \hii regions increases (equivalent to increasing $\sigma_{\text{T}}$
 in our framework).  We have set $\sigma_{\text{T}}=0.07$~dex based on empirical observations
 of \hii regions in individual galaxies \citep{ber15,cro15,cro16}, but it is worthwhile to
 investigate whether different values of $\sigma_{\text{T}}$ may resolve the \ttwo\
 discrepancies.  Such an investigation is presented in Appendix~\ref{appendix2}.
  To briefly summarize, while increasing $\sigma_{\text{T}}$ can reproduce the SDSS stack
 \ttwo-\te\ offset without any different assumptions regarding DIG \ttwo, the required values
 of $\sigma_{\text{T}}$ lead to significant changes in the predicted global galaxy strong-line
 ratios that do not match the observations.  Adjusting $\sigma_{\text{T}}$ is thus not
 a viable option for resolving the \ttwo\ discrepancies, and we continue under the assumption
 that DIG \ttwo\ is 15\% lower than \hii region \ttwo\ at fixed metallicity.

\subsection{The SDSS auroral-line comparison sample}\label{sec:auroralmodels}

We also test our model framework against observations of individual SDSS galaxies
 with electron temperature measurements.
  We use the sample of 181 SDSS galaxies from \citet{pil10} for which
 both [O\iii]$\lambda$4363 and [O\ii]$\lambda\lambda$7320,7330 have been detected.
  We expand this
 sample by adding 271 galaxies from \citet{izo06}.  Izotov et al. identified 309 SDSS galaxies for
 which [O\iii]$\lambda$4363 was measured, and we add those galaxies that were not already included
 as part of the \citet{pil10} sample.
  We note that while all of these additional galaxies have measurements of [O\iii]$\lambda$4363 and
 [O\ii]$\lambda\lambda$7320,7330, only 86 have measurements of
 [O\ii]$\lambda\lambda$3726,3729 due to their redshifts, and thus not all of them have estimates
 of \ttwogal.
  Those objects lacking [O\ii]$\lambda\lambda$3726,3729 observations
 are not plotted in diagrams involving O2\gal\ or \ttwogal.
  We refer to this combined sample as the ``auroral-line comparison sample.''

We determine the DIG fraction for the auroral-line comparison sample following the same method used for the
 larger SDSS sample.  We apply aperture correction factors to the reddening-corrected H$\alpha$ luminosities, and
 determine \sigha\ using the R-band half-light radius for each galaxy in the auroral-line comparison sample.
  The DIG fractions are found using equation~\ref{eq:fdig}.  The auroral-line comparison sample
 has an \fdig\ distribution that is nearly Gaussian, with a mean and standard deviation of 0.4 and 0.13,
 respectively, and no strong dependence on the level of excitation in the global galaxy spectrum.
  The lower average \fdig\ for this sample compared to that of
 the full SDSS sample reflects the extreme star-forming
 nature of galaxies in the auroral-line comparison sample.
  Model \fdig\ values are drawn randomly from this normal distribution.
  We adopt the same value as in the \stackmod model for the number of \hii regions per fiber,
 N$_{\text{HII}}=25$, but find that a smaller width of the \te\ distribution better fits the auroral-line sample,
 instead using $\sigma_{\text{T}}=0.02$~dex.
  A smaller value of $\sigma_{\text{T}}$ is likely more appropriate for the low-mass, high-sSFR galaxies
 in the auroral-line comparison sample.  Metals can be distributed more
 homogeneously throughout the ISM in such galaxies than
 in massive or low-sSFR galaxies because of an increase in feedback
 efficiency, as suggested by flatter metallicity gradients with decreasing \mstar\ and
 increasing sSFR \citep{ho15,ma17}.
  We continue to assume that DIG \ttwo\ is 15\% lower than \hii region \ttwo\ at fixed metallicity,
 as described in Section~\ref{sec:stackmodels}.  Despite the comparatively narrow \tegal\ range of the
 auroral-line comparison sample, we again randomly draw 2500 samples of T$_{\text{cent}}$ from a
 logarithmic uniform
 distribution over log(T$_{\text{cent}}/$K$)=3.7$~to~4.3, and create mock galaxy spectra
 following the method described in Section~\ref{sec:framework}.  We refer to this model as
 the \aurmod model.

\begin{figure*}
 \centering
 \includegraphics[width=\textwidth]{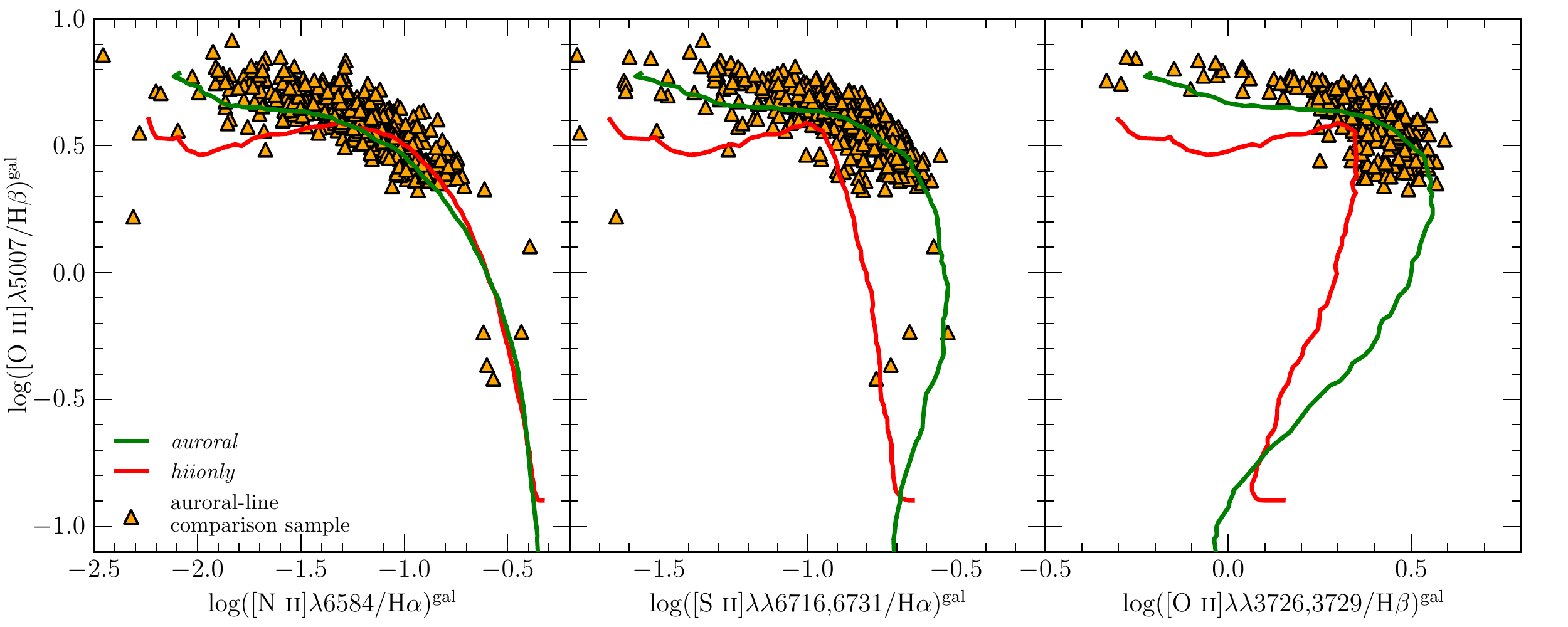}
 \caption{The O3N2 (left), O3S2 (middle), and O3O2 (right) strong-line ratio
 diagrams for individual SDSS galaxies with auroral line detections and models under different sets
 of assumptions and input parameters.  Orange triangles show SDSS galaxies in the auroral-line comparison
 sample from \citet{pil10} and \citet{izo06}.
  The green line displays the running median of mock galaxies in bins of O3N2\gal\ for the \aurmod model
 with \fdig=0.40 and $\sigma_{\text{T}}=0.02$~dex,
 under the assumption that DIG \ttwo\ is 15\% lower than \hii region \ttwo\ at fixed metallicity.
  The red line shows the \hiimod model that includes no DIG emission (\fdig=0.0) and follows the same
 DIG \ttwo\ assumption as the \aurmod model.
}\label{fig:bptdigfrac040}
\end{figure*}

\begin{figure}
 \includegraphics[width=\columnwidth]{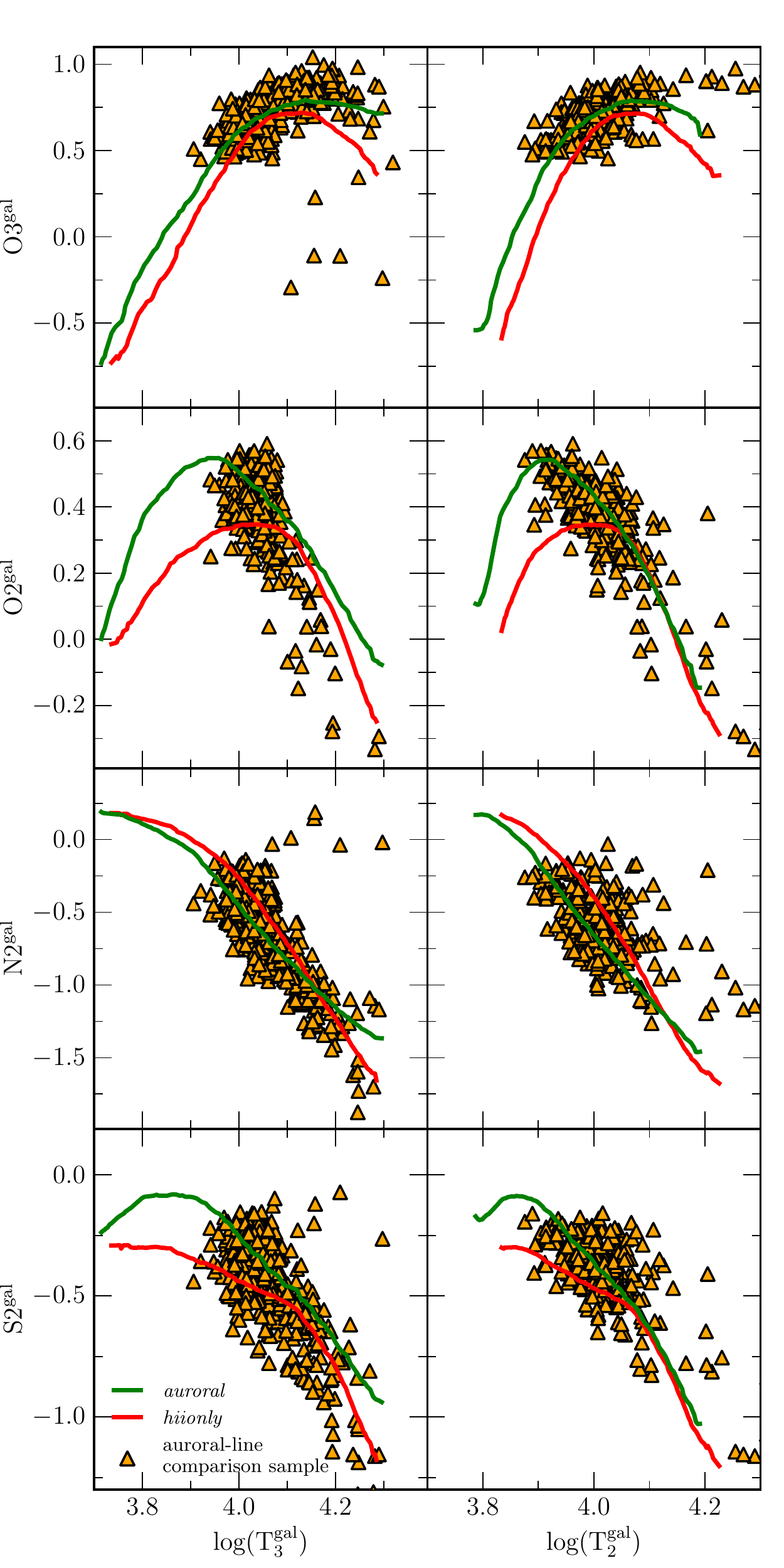}
 \caption{Global galaxy strong-line ratios O3\gal, O2\gal, N2\gal, and S2\gal\ as a function of
 \tegal\ (left column) and \ttwogal\ (right column).  Lines and points are the same as in
 Figure~\ref{fig:bptdigfrac040}.
}\label{fig:vstdigfrac040}
\end{figure}

\begin{figure}
 \includegraphics[width=\columnwidth]{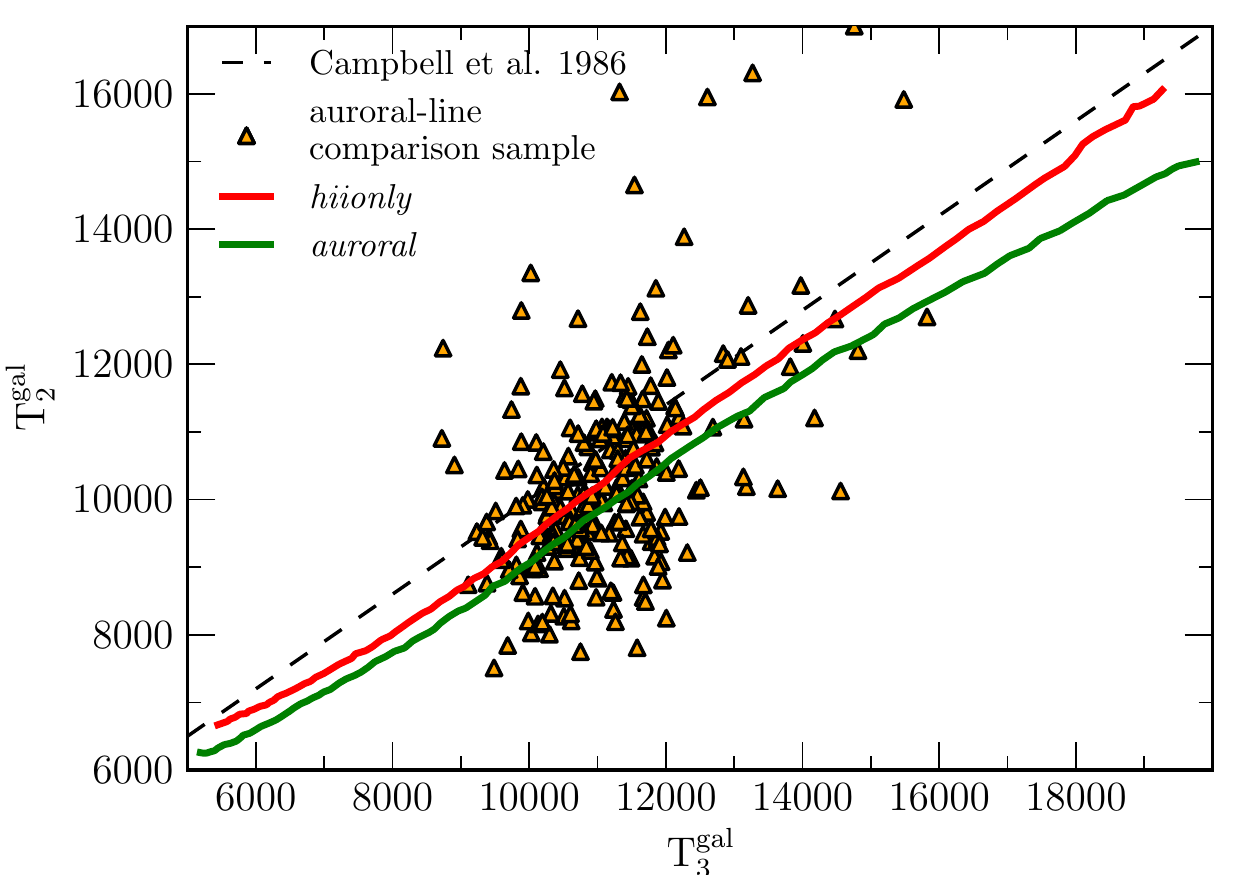}
 \caption{The global galaxy ionic temperature diagram of \ttwogal\ vs. \tegal.
  The dashed black line shows the \hii region \ttwo-\te\ relation of \citet{cam86} given in
 equation~\ref{eq:t2t3}.  All other lines and points are the same as in Figure~\ref{fig:bptdigfrac040}.
}\label{fig:t2t3digfrac040}
\end{figure}

We compare the \aurmod model to individual galaxies in the auroral-line comparison sample in
 Figures~\ref{fig:bptdigfrac040}-\ref{fig:t2t3digfrac040}.
  We find that the strong-line ratio diagrams (Fig.~\ref{fig:bptdigfrac040}), line ratios
 as a function of \ttwogal\ (Fig.~\ref{fig:vstdigfrac040}, right column), and the \ttwo-\te\ diagram
 (Fig.~\ref{fig:t2t3digfrac040}) show excellent agreement between the \aurmod model
 and observations.  However, as a function of \tegal\ (Fig.~\ref{fig:vstdigfrac040}, left column) the model overestimates
 N2\gal, S2\gal, and O2\gal\ at fixed \tegal\ in the high-temperature regime (log(\tegal$)>4.1$).
  The discrepancies are largest in the high-temperature, low-metallicity regime for which we had to
 extrapolate the DIG excitation sequences, and thus may suggest that the DIG excitation sequence
 extrapolations are not completely accurate in this regime, especially for O2\gal.
  We caution that results for metal-poor galaxies with observed
 O3N2\gal$\gtrsim2.0$ and log(\te)$\gtrsim4.15$ rely heavily on the extrapolation of the DIG sequences
 and should therefore be treated with caution.
  It is also possible that the extreme star-forming and metal-poor nature of some of the galaxies
 in the auroral-line comparison sample require some physics that is not captured in the framework
 of our simple models.  
  However, the model framework appears to perform well overall even for a sample of extreme
 star-forming galaxies that are unrepresentative of the local star-forming galaxy population.
  It is also of note that the assumption that DIG \ttwo\ is lower than \hii region \ttwo\ at
 fixed metallicity is required to match observations of \textit{individual} galaxies with auroral-line
 measurements in addition to stacks of SDSS galaxies, suggesting that the \ttwo\ offset
 is not a result of some systematic effect introduced by the stacking process.

\section{Results}\label{sec:biases}

Using these models, we characterize the biases in strong-line ratios,
 electron temperatures, and direct-method metallicity measurements from global galaxy spectra.
  These biases arise as a consequence of DIG contamination of emission lines and
 flux-weighting effects when combining emission from multiple \hii regions with a range of excitation
 levels.  The bias of a particular property is quantified by taking the difference between
 the value measured from a mock global galaxy spectrum and the median value of the same property
 for the individual \hii regions in that mock galaxy.  We represent the bias in property X
 with the symbol $\Delta$X.  For each property, the superscript ``gal'' (X\gal) indicates that the property
 is derived from the observed global galaxy spectrum, while the superscript ``HII'' (X\shii) is used to
 indicate the median value of the property for the distribution of \hii regions within each galaxy.
  Obtaining measurements that are representative of the
 \hii region distribution in a galaxy is desirable because both strong-line and direct-method
 metallicity estimates are based on \hii regions (real or simulated) rather than ensembles of \hii regions
 surrounded by DIG.
  Furthermore, the \hii regions trace only the most recent generation of star formation.
  Therefore, they provide a metallicity that is ideal for comparing to
 cosmological hydrodynamical simulations, which trace the metallicity of star-forming particles.

The bias determinations presented below can be used to correct properties measured from global
 galaxy spectra in order to obtain values representative of the median \hii region distributions.
  The \stackmod model is designed to reproduce the mean properties and trends of $z\sim0$ star-forming galaxies,
 but does not accurately represent deviations of individual galaxies from mean relations.

\subsection{Biases in the strong-line ratios}\label{sec:strongbias}

We quantify the typical global galaxy bias in the strong-line ratios O3, O2, N2, and S2
 for $z\sim0$ star-forming galaxies from SDSS.  These strong-line ratios can be combined to construct
 strong-line metallicity indicators used in calibrations that are widely applied to estimate galaxy
 metallicities.  Additionally, the strong-line ratios O3 and O2 are used in
 the calculation of direct-method metallicity (equations~\ref{eq:oplush} and~\ref{eq:o2plush}).
  It is thus of great importance to eliminate biases in galaxy strong-line ratios before using
 either the direct method or strong-line calibrations to determine galaxy metallicities.

\begin{figure}
 \includegraphics[width=\columnwidth]{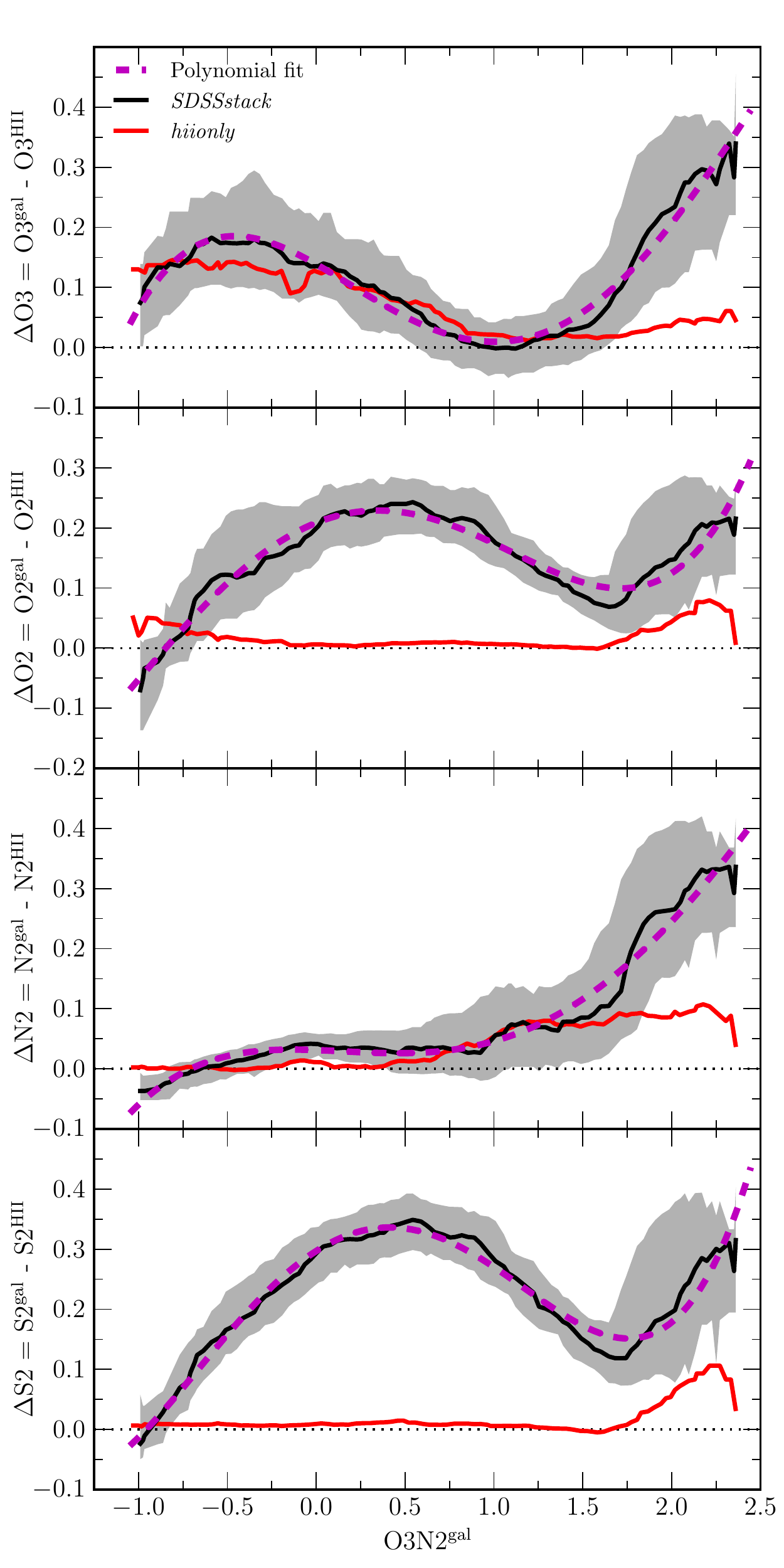}
 \caption{The difference between the global galaxy line ratio and median line ratio of the \hii region
 distribution, $\Delta$X, as a function of O3N2\gal\ for the strong-line ratios X=O3, O2, N2, and S2.
  The red line shows the running median of the 2500 mock galaxy realizations in bins of O3N2\gal\
 for the \hiimod model with \fdig=0.0.
  The running median of the \stackmod model with \fdig=0.55 is displayed as a black line, where the gray
 shaded region corresponds to the 68th-percentile width of the distribution of mock galaxies around the
 running median.
  In each panel, the dashed magenta line shows the best-fit fourth-order polynomial to the bias in the
 global galaxy line ratio, $\Delta$X, for the \stackmod model.  The best-fit coefficients are presented
 in Table~\ref{tab:coeffs}.
}\label{fig:strongbias}
\end{figure}

In Figure~\ref{fig:strongbias}, we present
 the global galaxy biases in O3, O2, N2, and S2 as a function of O3N2\gal, the O3N2 ratio as observed
 in global galaxy spectra.
  To determine the typical biases, we take the running median of individual mock galaxy spectra
 from each model in bins of O3N2\gal.
  The biases are quantified as a
 function of O3N2\gal\ instead of each individual line ratio (i.e., we show $\Delta$O3 vs. O3N2\gal\
 as opposed to $\Delta$O3 vs.~O3\gal)
 because O3N2 increases monotonically with metallicity and \te, and does not saturate over the range of
 metallicities of interest here.
  O3 and O2 are double-valued such that it would be necessary to determine on which branch a galaxy
 lies in order to correct the line ratio,
 and S2 and N2 saturate at high metallicities, limiting
 the utility of bias estimates as a function of these line ratios.
  Parameterizing by O3N2 instead should not severely limit the number of galaxies to which these
 corrections may be applied since O3N2 only involves strong lines that are easily detected in low-redshift
 star-forming galaxies down to low metallicities and stellar masses.
  We include the \hiimod model for comparison in order to understand how much of the bias
 arises from DIG contamination.  Biases in the \hiimod model arise purely from flux-weighting
 effects due to combining light on a line-by-line basis from multiple \hii regions with different metallicities.
  Any additional bias in models including DIG is driven by the inclusion of DIG emission in the
 global spectrum.

The top panel of Figure~\ref{fig:strongbias} shows the bias in O3.
  O3\gal\ can be biased high by as much as $+0.3$~dex in typical local star-forming galaxies, with the maximum
 bias occuring at O3N2\gal$\gtrsim2.0$.
  The \stackmod and \hiimod models display a similar level of O3\gal\ bias at O3N2\gal$<1.5$.  In this regime,
 DIG emission has little effect on the global O3 bias because DIG and \hii regions are matched in
 O3N2, and SDSS galaxies, DIG, and \hii regions follow similar excitation sequences
 in the O3N2 diagram.  The positive bias at O3N2\gal$<1.5$,
 reaching $+0.2$~dex at O3N2$^{\text{gal}}\sim-0.5$, is predominantly driven by flux-weighting effects
 when combining light from \hii regions with a range of metallicities due to the shape of the
 O3 vs.~\te\ relation of \hii regions (Figure~\ref{fig:hiiratios}).
  Since H$\beta$ luminosity does not correlate with
 electron temperature for \hii regions within a single galaxy \citep{ber15,cro15,cro16},
 when the slope of the O3 vs.~\te\
 relation is steep the global [O\iii]$\lambda\lambda$4959,5007 flux is dominated by high-O3 \hii regions
 while \hii regions of all metallicities contribute equally to the global H$\beta$ flux on average.
  Thus, a steep slope in the O3 vs.~\te\ relation leads to an O3\gal\ value that is higher than O3\shii,
 the median of the individual \hii regions. The positive bias is largest where the slope is steepest.
  In contrast, when the O3 vs.~\te\ relation is flat near the turnover point (O3N$2^{\text{HII}}\sim1.9$,
 log(\te$)\sim4.1$)
 the bias from flux-weighting effects becomes small, as seen in the \hiimod model at high O3N2\gal.
  At O3N2\gal$>1.5$, the \stackmod model diverges from the \hiimod model, displaying a significant positive
 $\Delta$O3.  In this low-metallicity regime, the DIG and \hii region sequences diverge in the
 O3N2 diagram such that DIG regions have higher O3 and N2 than \hii regions at fixed O3N2.  This divergence
 leads to a positive O3\gal\ bias that increases with increasing O3N2\gal.
  We note that mock galaxy spectra with O3N2\gal$\gtrsim2.0$ rely almost entirely on the
 linear extrapolation of the DIG excitation sequence, and predictions in this regime should therefore
 be used with caution.

A similar relative behavior between the \stackmod and \hiimod models is observed in the N2\gal\ bias,
 shown in the second panel from the bottom in Figure~\ref{fig:strongbias}.
  At O3N2\gal$<1.5$, the two models closely follow one another, while they diverge at O3N2\gal$>1.5$
 where DIG emission plays a role in the \stackmod model.
  The explanation is the same as for the O3\gal\ bias, except that in this case the N2 vs.~\te\
 relation is flat at low \te\ (low O3N2) and becomes steeper with increasing \te.  Flux-weighting
 effects therefore lead to no bias at low O3N2\gal\ and a slight increase in the bias at higher O3N2\gal\
 reaching $+0.1$~dex due to \hii regions alone.
  The additional bias in the \stackmod model at O3N2\gal$>1.5$ is again due to the divergence of DIG
 and \hii regions in the O3N2 diagram, such that DIG regions have higher N2 than \hii regions at fixed
 O3N2.

DIG emission plays a much more important role in the O2\gal\ bias, displayed in the second panel from
 the top in Figure~\ref{fig:strongbias}.  The \hiimod model
 shows a negligible O2\gal\ bias of $<0.1$~dex at all O3N2\gal\ values.  The lack of a significant
 bias in the \hiimod model results from the shape of the O2 vs.~\te\ relation, which peaks at
 log(T$_3^{\text{HII}})\sim4.0$ (O3N2$^{\text{HII}}\sim1.4$)
 and does not have a severely steep
 slope in either extreme.  The \stackmod model displays a larger $\Delta$O2 of $>0.1$~dex over
 most of the O3N2\gal\ range, peaking at $+0.25$~dex,
 and is primarily caused by DIG contamination in global galaxy spectra.
  DIG displays higher O2 at fixed O3N2 than \hii regions, as can be seen in the O3N2 and O3O2
 diagrams (Figure~\ref{fig:digratios}), leading to an overestimate of O2 relative to the median O2
 of the the \hii region distribution.
  The behavior of the \stackmod O2\gal\ bias can
 be understood through the divergence of the DIG excitation sequence from that of \hii regions in the
 O3O2 diagram at both low metallicities (O3N2\gal$\gtrsim1.75$) and moderate metallicities (O3N2\gal$\sim0.5$).

The S2\gal\ bias, shown in the bottom panel of Figure~\ref{fig:strongbias},
 behaves similarly to that of O2\gal.  A flux-weighted combination of \hii regions
 alone only leads to a small positive $\Delta$S2 at O3N2\gal$\gtrsim1.75$.
  Elevated S2 in DIG regions leads to a bias in S2\gal\ values as high as $+0.35$~dex at O3N2\gal$\sim0.5$.
  S2\gal\ displays a larger bias than O2\gal\ because the DIG and \hii region excitation sequences
 have a larger separation in the O3S2 diagram than in the O3O2 diagram (Figure~\ref{fig:digratios}).
  
In order to correct for these strong-line ratio biases in observed galaxy samples, we fit each
 bias as a function of observed O3N2\gal\ with a fourth-order polynomial of the form
\begin{equation}\label{eq:ratpoly}
\Delta R = c_0 + c_1 x + c_2 x^2 + c_3 x^3 + c_4 x^4 ,
\end{equation}
where $x=\text{O3N2}^{\text{gal}}$ and $R$ is the strong-line ratio O3, O2, N2 or S2.
  The best-fit polynomials are shown in Figure~\ref{fig:strongbias} and the coefficients are
 given in Table~\ref{tab:coeffs}.  These bias functions may be subtracted from observed
 galaxy strong-line ratios to obtain the median strong-line ratios of the \hii region distributions,
 correcting for DIG contribution and flux-weighting effects.
  We note that the corrections presented above are only appropriate for a sample of galaxies representative
 of typical $z\sim0$ star-forming galaxies with \fdig=0.55, and should not be applied to unrepresentative
 samples of galaxies.  See Appendix~\ref{appendix} for bias characterizations over a range of \fdig.

\begin{table}
 \centering
 \caption{Global galaxy bias coefficients}\label{tab:coeffs}
 \renewcommand{\arraystretch}{1.2}
 \begin{tabular}{ c | l l l l l }
   \hline\hline
   \multicolumn{6}{c}{Strong-line ratios\tablenotemark{a}} \\
   $\Delta R$ & $c_0$ & $c_1$ & $c_2$ & $c_3$ & $c_4$ \\[0pt]
   \hline
   O3  &  0.138 & -0.168 & -0.0749 & 0.140 & -0.0262 \\
   O2  &  0.208 & 0.118 & -0.173 & -0.00540 & 0.0260 \\
   N2  &  0.0312 & -0.0111 & -0.0277 & 0.0640 & -0.0103 \\
   S2  &  0.296 & 0.188 & -0.214 & -0.0463 & 0.0457 \\
   \hline\hline
   \multicolumn{6}{c}{Electron temperatures\tablenotemark{b}} \\
   $\Delta T_{\text{e}}$ & $c_0$ & $c_1$ & $c_2$ & $c_3$ & $c_4$ \\[0pt]
   \hline
   T$_3$  &  -2,171 &  6,813 & -2,537, & -2,278 & 1,109 \\
   T$_2$  &  18,280 &  -75,610 &  114,500 & -78,200 & 19,690 \\
   \hline\hline
   \multicolumn{6}{c}{Direct-method oxygen abundances\tablenotemark{c}} \\
    $\Delta$log(O/H)  & $c_0$ & $c_1$ & $c_2$ & $c_3$ & $c_4$ \\[0pt]
   \hline
   T$_3$ and T$_2$\tablenotemark{d} &  0.121 & -0.337 &  0.629 & -0.267 &  0.0333 \\
   T$_3$ only\tablenotemark{e}  &  -0.0266 & -0.591 &  0.530 &  0.311 & -0.362 \\
   T$_2$ only\tablenotemark{f}  & 0.340 & -0.459 &  0.420 & -0.0143 & -0.00841 \\
   \hline
 \end{tabular}
 \tablenotetext{1}{Coefficients for equation~\ref{eq:ratpoly}.}
 \tablenotetext{2}{Coefficients for equation~\ref{eq:temppoly}.}
 \tablenotetext{3}{Coefficients for equation~\ref{eq:ohpoly}.}
 \tablenotetext{4}{The direct-method 12+log(O/H) case where both \te\ and \ttwo\ are directly determined from the galaxy spectrum.}
 \tablenotetext{5}{The case where only \te\ is estimated directly, while \ttwo\ is inferred using equation~\ref{eq:t2t3}.}
 \tablenotetext{6}{The case where only \ttwo\ is estimated directly, while \te\ is inferred using equation~\ref{eq:t2t3}.}
\end{table}

\subsection{Biases in the electron temperatures}\label{sec:tempbias}

\begin{figure}
 \includegraphics[width=\columnwidth]{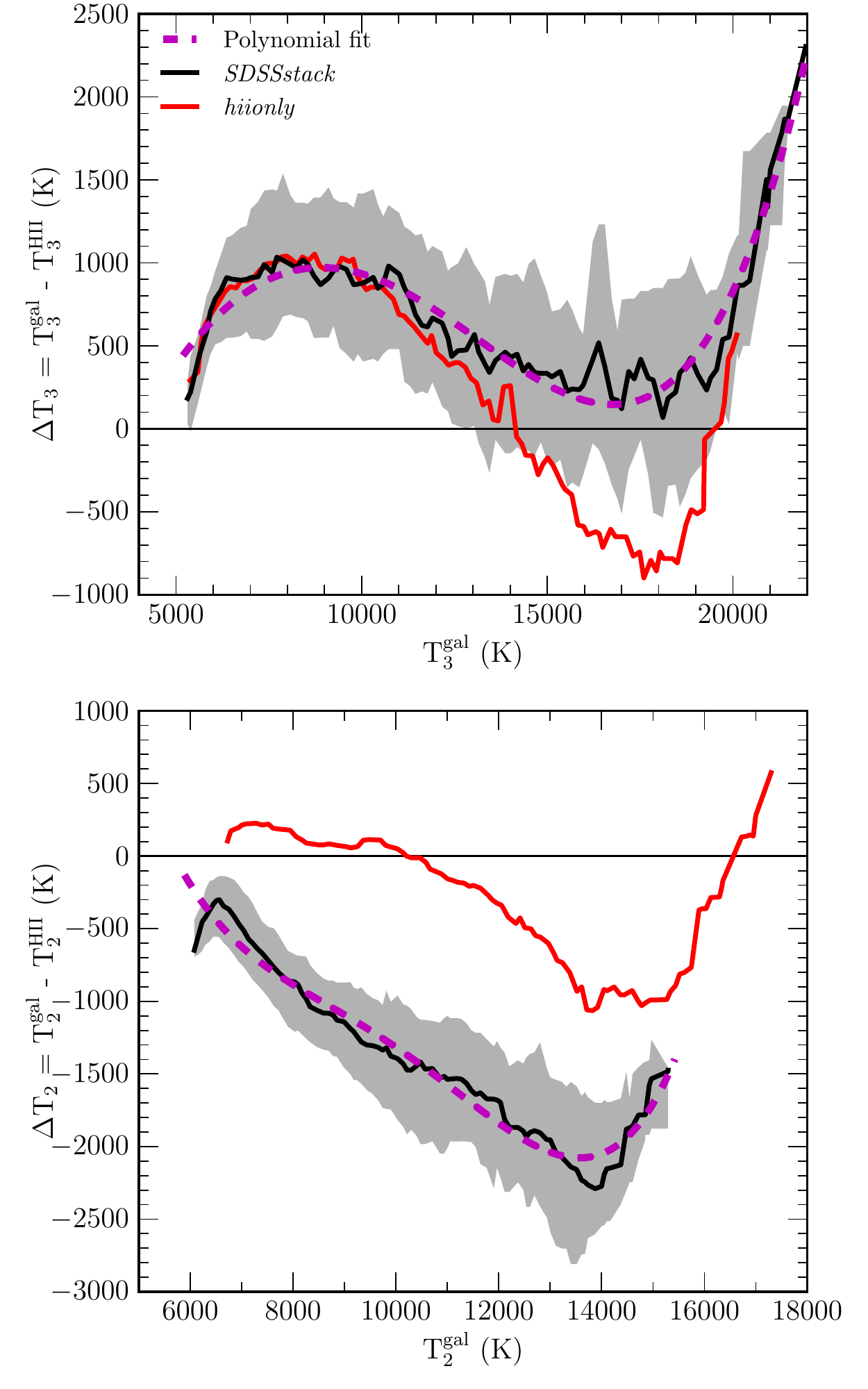}
 \caption{
The difference between the global galaxy electron temperature, inferred from the observed galaxy spectrum,
 and the median electron temperature of
 the \hii region distribution as a function of electron temperature.  Results for \te\ are shown
 in the top panel, while the bias in \ttwo\ is presented in the bottom panel.
  Lines and shading are the same as in Figure~\ref{fig:strongbias}.
  The best-fit coefficients are given in Table~\ref{tab:coeffs}.
}\label{fig:tempbias}
\end{figure}

Electron temperatures as inferred from global galaxy spectra also display biases with respect
 to the median electron temperature of the \hii regions.  We quantify the bias
 in \tegal\ (\ttwogal) by taking the running median of $\Delta$\te\ ($\Delta$\ttwo)
 of the individual mock galaxy spectra in bins of \tegal\ (\ttwogal).
  The typical biases in \te\ and \ttwo, as inferred from global galaxy spectra,
 are shown in Figure~\ref{fig:tempbias}.

In the \hiimod model, \tegal\ is biased by as much as $\pm1,000$~K with respect to \te\shii, the median value
 of the \hii region distribution.  Below \tegal$=14,000$~K, the \hiimod \tegal\ is biased high, while
 $\Delta$\te\ is negative at \tegal$=14,000-19,000$~K.  Since the \hiimod model does not include any DIG
 emission, these biases are a result of flux-weighting effects when combining light from multiple \hii
 regions of different electron temperatures.  \te\ sets the strong-to-auroral line ratio
 $\text{Q3}=[$O\iii]$\lambda\lambda$4959,5007/$\lambda$4363.  As \te\ increases, Q3 decreases exponentially
 (equation~\ref{eq:to3}) such that hotter \hii regions contribute
 more strongly to the global strong-to-auroral line ratio.
   In isolation, this trend would lead to $\Delta$\te\ that is always
 positive and increases significantly at high \tegal.  However, since H$\beta$ luminosity does not
 correlate with \te\ for individual \hii regions in single galaxies, on average [O\iii]$\lambda$5007 luminosity
 depends on \te\ according to the O3 vs.~\te\ relation shown in Figure~\ref{fig:hiiratios}.
  Thus, hotter \hii regions have more luminous [O\iii]$\lambda$5007 on average
 below T$_3<14,000$~K (log(\te$/K)<4.15$).  On the other hand, at T$_3>14,000$~K,
 hotter \hii regions typically have \textit{lower} [O\iii]$\lambda$5007 luminosity.  Thus, in the \hiimod model,
 there is a positive bias in \tegal\ at \tegal$<14,000$~K due to the increased weight of hotter \hii regions
 that have both higher O3 and lower Q3.
  At \tegal$=14,000-19,000$~K, cooler \hii regions with higher O3 and higher Q3 contribute more to \tegal\
 because the steepness of the dropoff in the O3 vs.~\te\ relation dominates over the decrease in
 Q3 with increasing \te\ in this regime.
  However, the exponential nature of the Q3 vs.~\te\ relation eventually
 dominates over the falling O3 vs.~\te\ relation, leading to a sharp increase in the \tegal\ bias above
 T$_3=19,000$~K as hotter \hii regions again receive more weight in the global \tegal\ calculation.

In the \stackmod model, the \tegal\ bias mimics that of the \hiimod model at \tegal$<12,000$~K, the
 regime  where \hii regions and DIG follow similar excitation sequences in the O3N2 diagram.
  At \tegal$>12,000$~K, the DIG and \hii region sequences diverge in the O3N2 diagram, such that DIG
 O3 continues increasing with \te\ whereas \hii region O3 turns over.  This difference in the DIG
 O3 vs.~\te\ behavior leads to $\Delta$\te\ that is always positive in the \stackmod model that
 includes DIG emission.  Similar to the \hiimod model, the exponential dependence of Q3 on \te\
 begins to dominate at \tegal$>19,000$~K due to flux-weighting effects,
 as evidenced by a sharp increase in $\Delta$\te.

The \ttwogal\ bias for the \hiimod model ranges from $-1,000$~K to $+500$~K.
  The bias can again be understood as a consequence of combining light
 from \hii regions with a range of temperatures.  The strong-to-auroral line ratio
 $\text{Q2}=[$O\ii]$\lambda\lambda$3726,3729/$\lambda\lambda$7320,7330 depends on \ttwo\ according
 to equation~\ref{eq:to2}.
  As before, the global bias is determined by the interplay of the auroral-line Q2 vs. \ttwo\
 and strong-line O2 vs. \ttwo\ relations.
  The O2 vs.~\te\ relation is shown in Figure~\ref{fig:hiiratios}.
  Since \ttwo\ is linearly dependent on \te, the O2 vs. \ttwo\ relation will have the same shape
 as the O2 vs.~\te\ relation modulo a linear transform to the temperature axis.
  At T$_2<10,000$~K, hotter \hii regions have both higher [O\ii]$\lambda\lambda$3726,3729
 luminosity and lower Q2, leading to a positive $\Delta$\ttwo, although this bias is fairly
 small since the O2 vs.~\ttwo\ relation slope is not extreme in this regime.
  At T$_2=10,000-15,000$~K, the O2 vs.~\ttwo\ relation drops off steeply such that cooler
 \hii regions have higher [O\ii]$\lambda\lambda$3726,3729 luminosity and dominate the
 \ttwogal\ measurement, leading to a negative $\Delta$\ttwo\ that reaches $-1,000$~K at \ttwogal$=14,000$~K.
  The exponential fall of Q2 with increasing \ttwo\ begins to dominate at \ttwogal$>15,000$~K, leading to
 a rapid increase in $\Delta$\ttwo.

The \ttwogal\ bias for the \stackmod model is always negative and can be large, underestimating
 \ttwohii\ by as much as 2,000~K.  There are two effects
 driving the difference between the \ttwogal\ bias of the \stackmod and \hiimod models.  First,
 inclusion of DIG emission significantly increases the O2\gal\ ratio.  However, $\Delta$O2
 is not a strong function of O3N2\gal\ (a good proxy for electron temperature), and will thus not have a large
 effect on the \ttwogal\ bias, which is sensitive to the slope of the $\Delta$O2 vs. O3N2\gal\ relation
 rather than the normalization.
  When the slope of the $\Delta$O2 vs. O3N2\gal\ relation is flat, the bias in O2\gal\ is not a function
 of electron temperature and thus does not strongly affect the globally-derived \ttwogal.
  The dominant factor separating the \stackmod and \hiimod models in
 $\Delta$\ttwo\ is our inference that DIG \ttwo\ is 15\% lower than \hii region \ttwo\ at fixed metallicity.
  This choice was motivated by differences between observations and a model in which DIG and \hii region \ttwo\
 was always equal.  An offset between model and stacks of SDSS galaxies was observed in all plots
 involving \ttwo\ but was not present in plots that only include \te, suggesting that DIG \ttwo\ is not equivalent
 to \hii region \ttwo\ at fixed metallicity.
  The difference in DIG and \hii region \ttwo\ effectively shifts the globally-derived \ttwogal\ lower
 and results in a large negative $\Delta$\ttwo\ when DIG emission is included, ultimately resulting in
 a significant underestimation of \ttwohii\ from global galaxy spectra at all metallicities.

We fit $\Delta$\te\ and $\Delta$\ttwo\ as a function of \tegal\ and \ttwogal, respectively, using a
 fourth-order polynomial:
\begin{equation}\label{eq:temppoly}
\Delta T_{\text{e}} = c_0 + c_1 y + c_2 y^2 + c_3 y^3 + c_4 y^4 ,
\end{equation}
where $T_{\text{e}}$ is \te\ or \ttwo, and $y=\text{T}_3^{\text{gal}}/10^4$~K or $\text{T}_2^{\text{gal}}/10^4$~K.
  The best-fit coefficients are presented in Table~\ref{tab:coeffs}.

\subsection{Biases in direct-method metallicity measurements}\label{sec:ohbias}

\begin{figure*}
 \includegraphics[width=\textwidth]{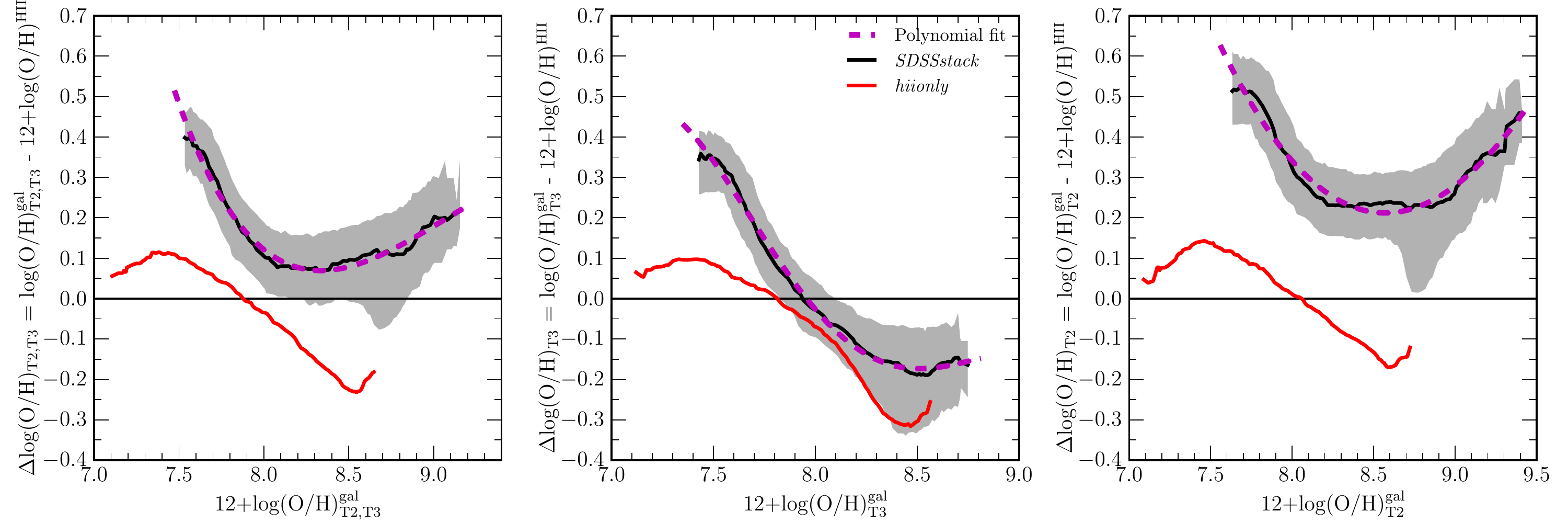}
 \caption{
The difference between the global galaxy direct-method metallicity, inferred from the observed galaxy spectrum,
 and the median metallicity of the \hii region distribution.
  We show the bias in global galaxy metallicity for three cases: both \te\ and \ttwo\ are measured from
 the galaxy spectrum (left panel); only \te\ is measured directly and \ttwo\ is estimated using the
 \ttwo-\te\ relation of equation~\ref{eq:t2t3} (middle panel); and only \ttwo\ is measured directly
 and \te\ is estimated using equation~\ref{eq:t2t3} (right panel).
  Lines and shading are the same as in Figure~\ref{fig:strongbias}.
  The best-fit fourth-order polynomial coefficients are presented in Table~\ref{tab:coeffs}.
}\label{fig:ohbias}
\end{figure*}

We use the same method employed above to characterize the bias in the direct-method oxygen
 abundance as inferred from global galaxy spectra.  Since it is common for only one auroral
 line to be measured in a galaxy spectrum, we evaluate the bias in metallicity for the cases where
 (1) both \tegal\ and \ttwogal\ are measured directly from the galaxy spectrum,
 (2) only \tegal\ is measured and \ttwogal\ is inferred from the \ttwo-\te\ relation of equation~\ref{eq:t2t3},
 and (3) only \ttwogal\ is measured and \tegal\ is inferred from the \ttwo-\te\ relation of equation~\ref{eq:t2t3}.
  The bias in global direct-method metallicity, $\Delta$log(O/H),
 as a function of direct-method metallicity inferred from global galaxy spectra
 is presented in Figure~\ref{fig:ohbias} for each of these three cases.  The bias is always calculated
 with respect to the median direct-method metallicity of the individual \hii region distribution
 in each mock galaxy.
  Having determined the biases in both strong-line ratios and electron temperatures,
 we can elucidate the origin of direct-method oxygen abundance biases.  Once again, we separately report
 the results from the \hiimod and \stackmod models
 to understand the additional effects that DIG contamination introduces.

\begin{figure}
 \includegraphics[width=\columnwidth]{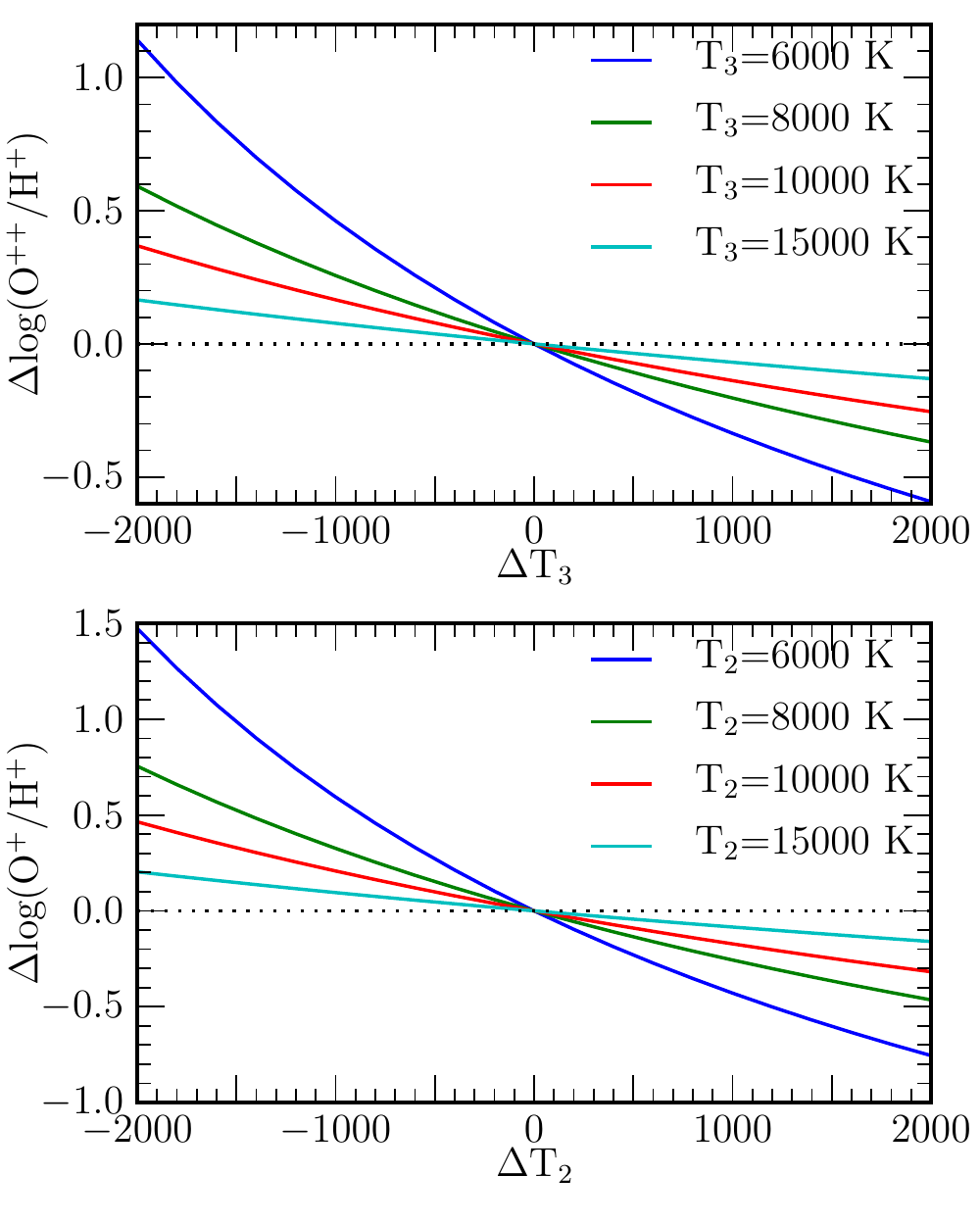}
 \caption{Temperature dependence of the ionic oxygen abundances at fixed strong-line ratio.
  The top panel displays the change in O$^{++}$/H$^+$ with changes in \te\
 according to equation~\ref{eq:o2plush} for a range of \te\, as indicated
 by the solid lines where color corresponds to the \te\ value.  The bottom panel shows the same relationship
 for O$^+$/H$^+$ and \ttwo\ according to equation~\ref{eq:oplush}.
  In each panel, the dotted line indicates zero change in the ionic abundance.
}\label{fig:deltaohvst}
\end{figure}

The formulae for the calculation of the ionic abundances O$^+$/H and O$^{++}$/H
 (equations~\ref{eq:oplush} and~\ref{eq:o2plush}) are functions of both the strong-line ratio
 of each ion (O2 or O3) and the corresponding ionic electron temperature (\ttwo\ or \te).
  O$^+$/H has a linear dependence
 on O2 and O$^{++}$/H has a linear dependence on O3 such that a bias in either of these strong-line
 ratios will result in an equivalent bias in the corresponding ionic abundance.  We plot the temperature
 dependence of the ionic abundance formulae at fixed strong-line ratio
 for a range of \te\ and \ttwo\ in Figure~\ref{fig:deltaohvst}.
  At high temperature (low metallicity) the temperature dependence is weak such that
 even a large bias in electron temperature does not significantly bias the ionic abundance.
  The temperature dependence is strong at low temperature (high metallicity) such that
 even a moderate bias of $\pm500$~K can change the ionic abundance by $\sim0.2$~dex.
  How much the bias in a particular ionic abundance affects the total oxygen abundance depends on the
 relative population of oxygen in O$^+$ and O$^{++}$, which is a function of metallicity.

We first focus on the case where both \tegal\ and \ttwogal\ are measured directly from the galaxy spectrum.
  In the low-metallicity limit (12+log(O/H$)^{\text{gal}}_{\text{T2,T3}}\lesssim8.0$),
 most of the oxygen is in O$^{++}$ such that changes in O$^+$/H will have a negligible effect on the
 total oxygen abundance.  The global metallicity bias is thus dominated by biases in O3\gal\ and \tegal\
 in this regime.  At low
 metallicity (high O3N2\gal), O3\gal\ is biased high by $<+0.05$~dex in the \hiimod model,
 contributing only a small
 amount to $\Delta$log(O/H$)_{\text{T2,T3}}$.
  \tegal\ is biased $\sim500$~K low at low metallicity in the \hiimod model.
  When combined, these two effects lead to a bias in the direct-method metallicity of
 $+0.1$~dex in the low-metallicity limit from combinations \hii regions alone.
  The \stackmod model has a positive \tegal\ bias that increases sharply at the low-metallicity
 extreme.  While higher \tegal\ will bias the global metallicity low, the temperature dependence of direct-method
 metallicity is weakest at high temperature.  The metallicity bias for the \stackmod model at low metallicity
 is instead dominated by the positive bias in O3\gal, which reaches +0.3 dex in the low-metallicity limit.
  The strong-line bias leads to an overestimate of galaxy direct-method metallicity that grows
 from $+0.1$~dex at 12+log(O/H$)^{\text{gal}}_{\text{T2,T3}}=8.0$ to $+0.4$~dex at
 12+log(O/H$)^{\text{gal}}_{\text{T2,T3}}=7.6$.

In the high-metallicity limit (12+log(O/H$)^{\text{gal}}_{\text{T2,T3}}\gtrsim8.5$),
 O$^+$ is the most numerous ionic form of oxygen and biases in
 O3\gal\ and \tegal\ will be subdominant drivers of $\Delta$log(O/H$)_{\text{T2,T3}}$.
  In the \hiimod model, O2\gal\ is relatively unbiased at all metallicities while \ttwogal\ is biased
 high by $\sim250$~K at \ttwogal$\sim7500$~K (high metallicity).
  While the \ttwogal\ bias is not large, the direct-method
 metallicity is highly sensitive to temperature changes at low temperature, such that a bias of
 only $\sim250$~K in \ttwo\ leads to a bias in the direct-method metallicity of $-0.2$~dex.
  When DIG emission is included in the \stackmod model, O2\gal\ is biased high
 and \ttwogal\ is biased low,
 $\Delta$log(O/H$)_{\text{T2,T3}}$ to $+0.2$~dex at 12+log(O/H$)^{\text{gal}}_{\text{T2,T3}}=9.0$.
  Including DIG emission leads to significantly
 different behavior of $\Delta$log(O/H) from the case where emission from \hii regions alone is
 considered.  Because DIG exhibits higher low-ionization line ratios and lower \ttwo\
 than \hii regions, DIG contamination in global galaxy spectra leads to an
 overestimate of 12+log(O/H)\shii, the median metallicity of the \hii region population.

The cases where only one ionic temperature is measured directly can be understood as
 modulations of the case where both \ttwogal\ and \tegal\ are known.
  Biases from strong-line ratios will remain the same, while the bias arising from the unknown
 temperature will differ.
  We note that the \hiimod bias shows little change when only one ionic temperature is
 known.  This consistency occurs because \hii regions closely follow the the \ttwo-\te\ relation of
 equation~\ref{eq:t2t3} from which the unknown temperature is inferred.

The case where only \tegal\ is measured from the galaxy spectrum is shown in the middle panel
 of Figure~\ref{fig:ohbias}.
  In this case, \ttwogal\ is estimated using the \ttwo-\te\ relation of equation~\ref{eq:t2t3}.
  While \hii regions follow this relation,
 ionic temperature measurements from global galaxy spectra show that galaxies do not,
 instead having lower \ttwo\ at fixed \te\ than \hii regions.  Assuming that
 galaxies follow the same ionic temperature relation as \hii regions is a common
 assumption in the literature \citep[e.g.,][]{izo06,ber12,jon15}.
  For the \stackmod model, this assumption leads to an overestimate of \ttwogal\
 by $1,000-1,500$~K when inferred using equation~\ref{eq:t2t3}.  This discrepancy has little
 effect at low metallicity where O$^+$ is negligible, but leads to a negative bias
 in oxygen abundance of $-0.15$~dex at 12+log(O/H$)^{\text{gal}}_{\text{T3}}=8.4-8.7$.

The right panel of Figure~\ref{fig:ohbias} shows the case where only \ttwogal\ is measured
 directly from the galaxy spectrum.
  While this case is less common in the literature for individual galaxies
 than the case where only [O\iii]$\lambda$4363 is detected,
 stacked spectra of high-metallicity galaxies often only yield [O\ii]$\lambda\lambda$7320,7330
 detections \citep{lia07,and13,bro16,cur17}.
  In this case, \tegal\ is underestimated by $\sim2,000$~K at all values of \ttwogal when \tegal\
 is inferred from the \ttwo-\te\ relation of equation~\ref{eq:t2t3}.
  This incorrect \tegal\ value leads to a large $\Delta$log(O/H$)_{\text{T2}}$ value of $+0.3$ to $+0.5$~dex
 at 12+log(O/H$)^{\text{gal}}_{\text{T2}}<8.0$ for the \stackmod model.
  Overesimating \tegal\ by $2,000$~K leads to an overestimate of O$^{++}$/H by $\sim+0.6$~dex
 because of the temperature dependence of O$^{++}$/H at low \te, as shown for \te=6,000~K
 in the top panel of Figure~\ref{fig:deltaohvst}.
  While O$^{++}$ is not the main form of oxygen in this high-metallicity, low-temperature regime,
  the $\sim+0.6$~dex overestimate of O$^{++}$/H causes O$^{++}$/H to contribute
 strongly enough to affect the total oxygen abundance.
  Thus, the global bias in direct-method metallicity is higher
 at all metallicities when \ttwogal\ is known and \tegal\ is inferred from an \hii region
 \ttwo-\te\ relation than when both \ttwogal\ and \tegal\ are measured directly.
  The \stackmod model predicts that the additional bias when only \ttwogal\ is known
 compared to the case when both \ttwogal\ and \tegal\ are measured
 is $\sim0.15$~dex and is nearly constant with metallicity.  This value is in excellent
 agreement with the observation of \citet{and13} that the metallicities of their
 \mstar-binned stacks were 0.18~dex higher on average if the metallicity was calculated
 using \ttwogal\ alone instead of both \ttwogal\ and \tegal (see their Figure~6).
  We note that if a \ttwo-\te\ relation fit to galaxy stacks instead of \hii regions was used to infer
 the unknown ionic temperature, then the bias when only one temperature is measured would
 closely match the bias shown in the left panel of Figure~\ref{fig:ohbias} where both ionic temperatures are known.
  The differences between the three panels arise solely because galaxies do not fall on the
 \hii region \ttwo-\te\ relation.

We fit the direct-method metallicity bias in each of the three ionic temperature cases
 with a fourth-order polynomial:
\begin{equation}\label{eq:ohpoly}
\Delta\text{log(O/H)} = c_0 + c_1 z + c_2 z^2 + c_3 z^3 + c_4 z^4 ,
\end{equation}
where $z=12$+log(O/H$)^{\mathrm{gal}}-8$ for the appropriate electron temperature case (\ttwo\ and \te, \ttwo\ only,
 or \te\ only).
  The best-fit coefficients are presented in Table~\ref{tab:coeffs}.
  These functions may be used to correct direct-method
 metallicities of galaxies with measured auroral lines in order to obtain a metallicity measurement
 that is characteristic of the \hii region population.

\subsection{Corrections for individual galaxies or unrepresentative samples}

We have presented best-fit polynomials that allow for the correction of strong-line ratios,
 electron temperatures, and direct-method oxygen abundances obtained from global galaxy spectra
 to values that are representative of the distribution of \hii regions in each galaxy.  These
 corrections are based on a model that is matched to the typical $z\sim0$ star-forming population
 as represented by stacks of SDSS galaxies from AM13, B16, and C17.
  The best-fit polynomials presented above thus provide robust corrections for samples of galaxies
 that are also representative of the typical local star-forming population, that is, having
 \fdig=0.55 on average.
  It may be of interest, however, to correct the line ratios, temperatures, and oxygen abundances
 of individual galaxies that do not fall on the mean relations, or to provide corrections for an unrepresentative
 sample of galaxies, as would be necessary for galaxies that do not follow the mean \mstar-SFR
 relation when investigating the SFR dependence of the MZR.  We provide a recipe for correcting
 individual galaxies or unrepresentative samples for which \fdig=0.55 is not appropriate
 in Appendix~\ref{appendix}.

\section{Application to the $z\sim0$ MZR and FMR}\label{sec:application}

In this section, we show examples of how the biases determined from our model framework
 can be used to correct
 local metallicity scaling relations, removing the effects of flux-weighting and DIG contamination.
  We apply corrections to the $z\sim0$ direct-method MZR and FMR, and investigate the effects of
 strong-line ratio biases on the MZR when using strong-line metallicity calibrations.
  We additionally demonstrate how the expected decrease in \fdig\ with increasing SFR can
 explain the observed trends in strong-line ratio at fixed direct-method metallicity from
 \citet{bro16} and \citet{cow16}.

\subsection{The $z\sim0$ direct-method MZR}\label{sec:mzrdirect}

We investigate the effects of the biases in direct-method galaxy metallicity presented in
 Section~\ref{sec:ohbias} on measurements of the local MZR.
  Using measurements from composite spectra of local
 star-forming galaxies in bins of \mstar, AM13 presented the local MZR over three orders of
 magnitude in \mstar\ and an order of magnitude in 12+log(O/H).  The increase in sensitivity from
 stacking enabled AM13 to probe an order of magnitude lower in \mstar\ (log(\mstar/$\msun)=7.5$)
 than most previous MZR studies based on strong-line metallicities \citep[e.g.,][]{tre04},
 and measure direct-method metallicities
 representative of galaxies with such a wide dynamic range in properties for the first time.

\begin{figure}
 \includegraphics[width=\columnwidth]{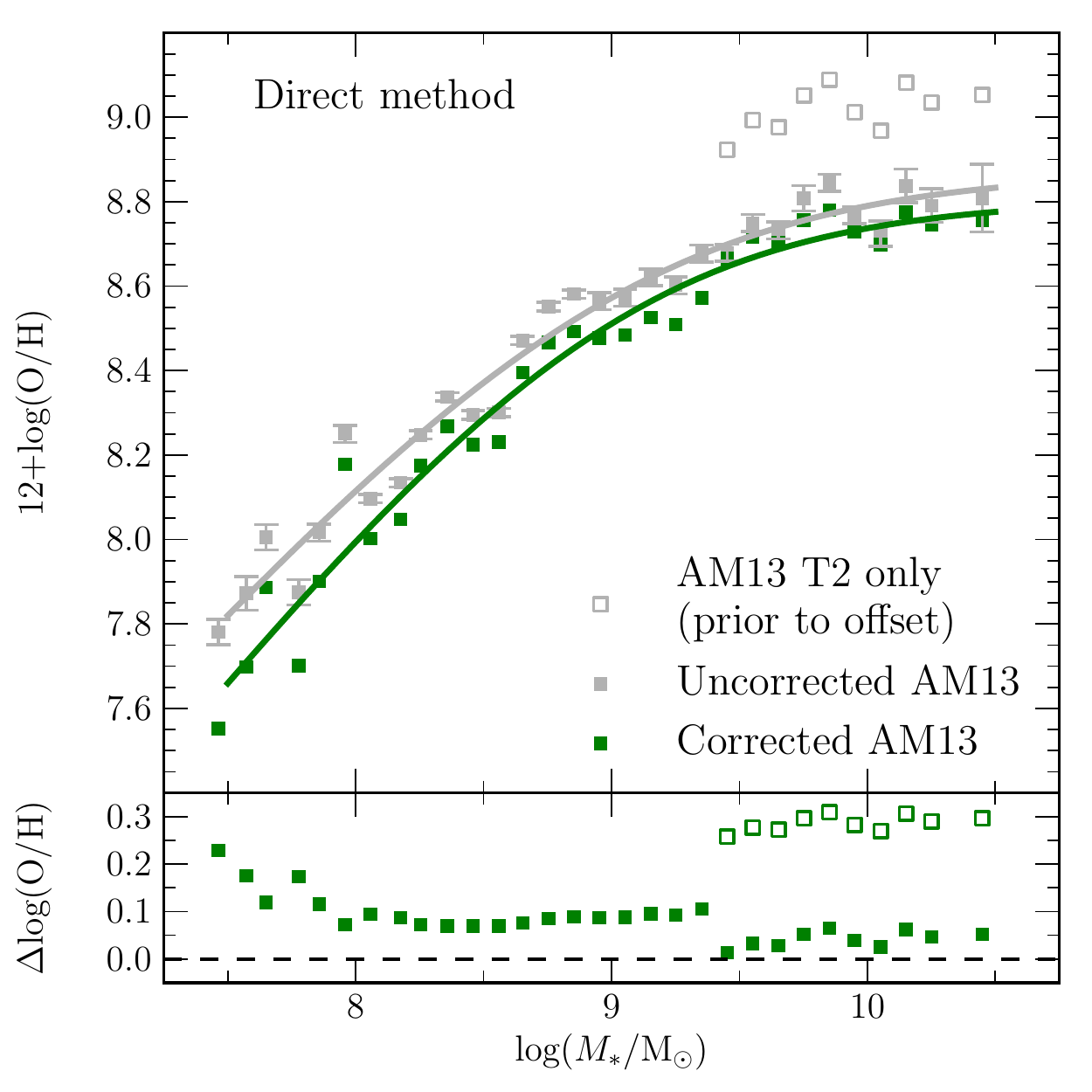}
 \caption{\textsc{Top:}
  The direct-method MZR for stacks of SDSS star-forming galaxies in bins of \mstar\ from AM13.
  Filled gray squares use uncorrected AM13 metallicities recalculated using updated atomic data,
 while filled green squares show the metallicities after correcting for the effects of flux-weighting
 and DIG emission.  Error bars on the uncorrected points only include measurement uncertainties, and would
 thus be identical for the corrected
 points.  Hollow gray squares present 12+log(O/H)$_{\text{T2}}$ for stacks with only \ttwo\
 estimates, prior to offsetting to account for the difference between 12+log(O/H)$_{\text{T2}}$ and
 12+log(O/H)$_{\text{T2,T3}}$.  The gray and green lines show the best-fit MZR function of \citet{mou11}
 (equation~\ref{eq:mzr}) using uncorrected and corrected metallicities, respectively.
  The corrected MZR displays a lower normalization and steeper low-mass slope than before correction.
  \textsc{Bottom:}  The difference between uncorrected and corrected metallicities as a function of \mstar.
  Hollow points present $\Delta$log(O/H) between uncorrected 12+log(O/H)$_{\text{T2}}$ prior to applying
 the offset and the corrected metallicity.  The corrections produced by our models naturally account
 for the offset between 12+log(O/H)$_{\text{T2}}$ and 12+log(O/H)$_{\text{T2,T3}}$.
 }\label{fig:mzrdirect}
\end{figure}

The direct-method MZR from AM13 stacks is presented in the top panel of Figure~\ref{fig:mzrdirect}.
  We show both the MZR using direct-method metallicities as inferred from the stacked spectra without correcting
 for any biases
 (gray points) and the galaxy metallicities after correcting for the biases presented
 in Section~\ref{sec:ohbias} (green points).
  The original AM13 metallicities were recalculated using our methodology,
 which includes updated atomic data.  Accordingly, our AM13 metallicities prior to correction
 are systematically shifted
 with respect to those reported in AM13, yielding $\sim0.1$~dex higher metallicities in the highest mass bins,
 and slightly lower metallicities in the lowest mass bins.
  We followed the methodology of AM13 to estimate 12+log(O/H) for
 those high-mass stacks that do not have clean detections of [O\iii]$\lambda$4363 by adjusting the metallicity
 as calculated using \ttwo\ only by an amount equal to the median difference between
 12+log(O/H)$_{\text{T2,T3}}$ and 12+log(O/H)$_{\text{T2}}$ (see their Section~3.2 and Figure~6).
  We find this median offset to be $-0.24$~dex, slightly larger in magnitude than the offset of
 $-0.18$~dex reported in AM13 owing to the different atomic data and
 ionic abundance determinations used here.
  The values of 12+log(O/H)$_{\text{T2}}$ prior this offset adjustment for stacks with \ttwo\ only
 are shown as unfilled gray squares.
  Correcting the stacks with no \te\ measurement yields an uncorrected MZR that shows no obvious break
 at the point where [O\iii]$\lambda$4363 is no longer cleanly detected.

Corrected direct-method metallicites are obtained by applying the best-fit corrections shown in
 Figure~\ref{fig:ohbias}.  Stacks with measurements of both [O\iii]$\lambda$4363 and
 [O\ii]$\lambda\lambda$7320,7330 are corrected using the 12+log(O/H)$_{\text{T2,T3}}$ fit, while
 stacks with only [O\ii]$\lambda\lambda$7320,7330 are corrected by applying the 12+log(O/H)$_{\text{T2}}$
 fit to the uncorrected 12+log(O/H)$_{\text{T2}}$ prior to the offset adjustment.
  Our models naturally account for the offset in metallicity when only \ttwo\ is measured without the
 need for an ad hoc adjustment to the normalization as in AM13.
  It is important to note that the range of uncorrected galaxy metallicities
 (12+log(O/H)$_{\text{T2,T3}}^{\text{gal}}=7.8-8.7$; 12+log(O/H)$_{\text{T2}}^{\text{gal}}=8.9-9.1$)
 fall within the range of the models and do not fall close to the lowest or highest model galaxy
 metallicities where extrapolations are heavily relied upon.  Thus, our choice of extrapolations does not
 strongly impact our results.  The bottom panel of Figure~\ref{fig:mzrdirect} shows the difference between
 the uncorrected and corrected AM13 metallicities, where the original uncorrected AM13 metallicites have
 been recalculated with our updated atomic data.

We fit the uncorrected and corrected direct-method MZRs with the asymptotic logarithmic formula of
 \citet{mou11}, also used by AM13:
\begin{equation}\label{eq:mzr}
\text{12+log(O/H)}=\text{12+log(O/H)}_{\text{asm}} - \text{log}\left[1+\left(\frac{M_{\text{TO}}}{\text{M}_*}\right)^{\gamma}\right] .
\end{equation}
This function is a power law of slope $\gamma$ at low stellar masses, and approaches the asymptotic
 metallicity $\text{12+log(O/H)}_{\text{asm}}$ at high stellar masses, where the turnover mass $M_{\text{TO}}$
 controls the transition point between the two behaviors.
  The best-fit values for the uncorrected AM13 direct-method MZR are
 $[\text{12+log(O/H)}_{\text{asm}}, M_{\text{TO}}, \gamma]=[8.87\pm0.03, 8.99\pm0.09, 0.67\pm0.02]$
 (compare to [8.80, 8.90, 0.64] in AM13),
 and the fit is shown as a gray line in Figure~\ref{fig:mzrdirect}.
  The best fit to the corrected AM13 direct-method MZR is shown as a green line, with best-fit
 parameters $[\text{12+log(O/H)}_{\text{asm}}, M_{\text{TO}}, \gamma]=[8.80\pm0.02, 8.98\pm0.08, 0.75\pm0.03]$.

We find lower corrected metallicities at all stellar masses compared to the uncorrected metallicites
 (reflected in the 0.07~dex lower $\text{12+log(O/H)}_{\text{asm}}$), with the lowest-mass bins displaying
 the largest shift.  This trend results in a steeper low-mass slope of 0.75 after correcting for
 the effects of flux-weighting and DIG contamination, compared to a slope of 0.67 for the uncorrected AM13 MZR.
  Accurately determining the low-mass slope of the MZR is of primary importance since it is set by
 the scaling of outflow efficiency with stellar mass, as parameterized by the mass loading
 factor defined as the ratio of outflow rate and SFR, which in turn reflects
 the nature of galactic winds \citep{fin08}.  In particular, energy-driven galactic winds predict a steeper
 low-mass slope than momentum-driven winds \citep{pee11}.
  The turnover mass identifies the stellar mass at which galactic winds become inefficient and unable
 to remove sufficient material in large-scale outflows, and is unaffected by our corrections.
  In summary, correcting for flux-weighting and DIG contamination results in a lower normalization
 and steeper low-mass slope of the $z\sim0$ direct-method MZR, and these changes have a significance
 of $\sim2\sigma$.

\subsection{Strong-line MZR at $z\sim0$}\label{sec:mzrstrong}

The direct-method MZR provides a robust measurement of the shape of the $z\sim0$ MZR since it is
 constructed using a reliable metallicity determination that can be applied to a large number of galaxies
 through the stacking process.
  Nevertheless, we investigate the effects of biases in global galaxy line ratios on the MZR as
 determined using strong-line metallicity calibrations.  Such calibrations have been widely applied
 in the local universe, and strong-line calibrations are currently the only method available to determine
 gas-phase metallicities of high-redshift galaxies due to the difficulty of detecting faint auroral lines
 at cosmological distances.  Whenever investigating redshift evolution of the MZR, it is crucial that all samples
 being compared at least have metallicities determined using the same calibration to eliminate known
 systematic differences between various strong-line calibrations \citep{kew08}.
  Correcting strong-line MZRs for biases can thus provide more robust determinations of the evolution of the MZR.
  However, potential evolution of physical conditions of star-forming regions with redshift may ultimately
 require a reevaluation of strong-line calibrations at high redshift \citep{ste14,san15,sha15}.
  Even so, eliminating observational biases from $z\sim0$ strong-line MZR measurements provides a more robust
 baseline relative to which metallicity evolution can be inferred.

\begin{figure*}
 \includegraphics[width=\textwidth]{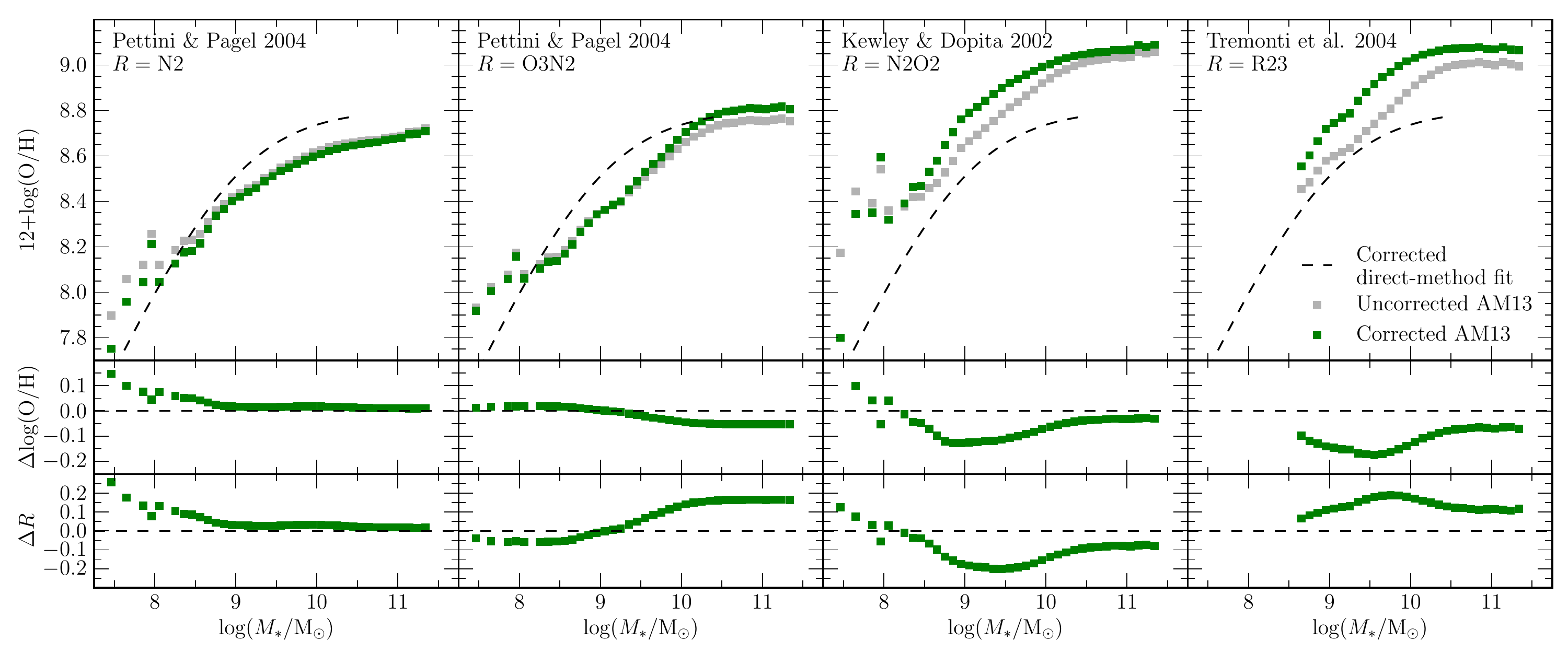}
 \caption{The MZR based on strong-line metallicities for the \mstar\ stacks of AM13 is displayed in the top row.
  We show results for four commonly-used strong-line indicators: the empirical
 N2 and O3N2 calibrations of \citet{pet04}, the theoretical N2O2 calibration of \citet{kew02},
 and the theoretical R23 calibration of \citet{tre04}.  In each panel, filled gray squares use the
 uncorrected strong-line ratio to infer the metallicity, while filled green squares utilize the strong-line
 ratio after correcting for flux-weighting effects and DIG emission.  For reference, the corrected direct-method
 MZR best-fit function from Section~\ref{sec:mzrdirect} is shown as a dashed black line.
  The middle row displays the difference $\Delta$log(O/H) between uncorrected and corrected metallicity,
 while the bottom row presents the difference $\Delta R$ between the uncorrected and corrected strong-line
 ratio.  The bias in strong-line metallicity, primarily driven by DIG contamination, can exceed 0.1~dex.
}\label{fig:mzrstrong}
\end{figure*}

We analyze the $z\sim0$ strong-line MZRs using measurements of strong-line ratios from the AM13 stacks.
  We investigate the effects of global galaxy strong-line ratio biases on the MZR
 using four widely-applied metallicity calibrations:
 two empirical calibrations (Pettini \& Pagel 2004 N2 and O3N2; PP04N2 and PP04O3N2, respectively)
 and two theoretical calibrations (Kewley \& Dopita 2002 N2O2 and Tremonti et al. 2004 R23;
 KD02N2O2 and T04R23, respectively).  The original AM13 strong-line metallicities are calculated using
 the dust-corrected line fluxes appropriate to each calibration as reported in AM13.
  In order to determine the corrected AM13 strong-line metallicities, we first apply corrections to each
 strong-line indicator based on the best-fit polynomials presented in Figure~\ref{fig:strongbias}
 and then estimate strong-line metallicities using each calibration.  The individual strong-line ratio
 biases in Figure~\ref{fig:strongbias} may be combined to provide a correction to any strong-line metallicity
 indicator.
  The uncorrected and corrected strong-line MZRs using each of the four calibrations are presented in
 the top row of Figure~\ref{fig:mzrstrong}, while the bias in log(O/H) is shown in the center row, and
 the bias in the strong-line ratio is presented in the bottom row.
  The best-fit corrected AM13 direct-method MZR is shown for comparison.

There is a large spread in the normalization of the MZR when using different strong-line calibrations,
 as first pointed out by \citet{kew08}.  Empirical calibrations based on \hii region samples
 with auroral-line measurements (PP04N2, PP04O3N2) yield metallicities that are $\sim0.3$~dex lower than those
 obtained from theoretical calibrations based on photoionization models (KD02N2O2, T04R23).
  It is unsurprising that the empirical calibrations produce metallicities that most closely
 match the direct-method AM13 MZR, since the calibration dataset is dominated by objects with direct-method
 metallicities.
  Correcting for flux-weighting and DIG effects does not reduce the offset between empirical and
 theoretical calibrations, but instead increases the magnitude of the disagreement by shifting
 the KD02N2O2 and T04R23 MZRs towards higher metallicity at fixed \mstar.  That the conflict between
 theoretical and empirical calibrations remains suggests that the
 disagreement between MZRs based on theoretical and empirical calibrations is not a result of observational biases
 in global galaxy spectra.  Instead, the problem appears to be a manifestation of a long-standing
 disagreement in normalization of the metallicity scale between direct-method and theoretical strong-line
 calibrations observed for extragalactic \hii regions \citep{ken03}.

It is unclear which method provides a metallicity scale closer to the truth since there are potential
 systematic issues on both sides.
  Empirical direct-method calibrations may be biased towards
 lower metallicities due to the presence of temperature gradients and inhomogeneities within the ionized gas
 \citep{sta05,bre07}, although this problem primarily affects high-metallicity, low-temperature \hii regions.
  Direct-method metallicities may indeed have a normalization bias, but have been shown to tightly correlate
 with metallicities determined from oxygen recombination lines with a slope of unity \citep{bla15}
 and an offset of $\sim-0.2$~dex.
  For photoionization models, it is difficult to determine the proper combination of input parameters
 and physical conditions that produce realistic \hii regions because of degeneracies among parameters.
  Additionally, observed nearby \hii regions often have filamentary gas structures and cluster stars distributed
 throughout the ionized gas \citep[e.g., 30 Dor;][]{pel11},
 a very different geometry from the ionizing point source and uniform-density
 sphere or slab of gas utilized in most photoionization codes \citep{kew02,gut16}.

All strong-line MZRs display a high-mass flattening, although this is more apparent with some calibrations
 than others.  In general, the turnover mass is higher than that measured with the direct-method.
  The PP04N2 and PP04O3N2 turnover masses do not change significantly once the galaxy metallicities are
 corrected for flux-weighting effects and DIG contamination.
  In contrast, the KD02N2O2 and T04R23 MZRs have turnover masses that are shifted lower
 when using corrected metallicities, bringing the turnover mass into better agreement with that of
 the direct-method MZR.

AM13 found that various strong-line calibrations produce MZRs that have low-mass slopes much shallower
 than that of the direct-method MZR.  We also find that all strong-line MZRs using uncorrected metallicities
 have low-mass slopes close to $\gamma\sim0.3-0.4$, significantly shallower than the slope of 0.75
 for the direct-method MZR.  For all strong-line MZRs except PP04O3N2, correcting for flux-weighting
 effects and DIG emission yields steeper low-mass slopes.  The KD02N2O2 and T04R23 slopes appear to be
 close to that of the direct-method MZR after correction, relieving some tension between the
 theoretical strong-line and direct-method MZR shapes.

In summary, after correction for flux-weighting effects and DIG contamination in global galaxy line ratios,
 theoretical strong-line calibrations appear to match the direct-method MZR low-mass slope and turnover mass,
 but retain a large offset in normalization.  Empirical strong-line calibrations provide a much closer match
 in normalization, but display higher turnover mass and shallower low-mass slope than those measured
 with the direct-method.  Tensions between empirical and theoeretical strong-line metallicities remain
 even after correcting for contamination from DIG emission.

\subsection{The direct-method $z\sim0$ FMR}\label{fmrdirect}

\begin{figure*}
 \includegraphics[width=\textwidth]{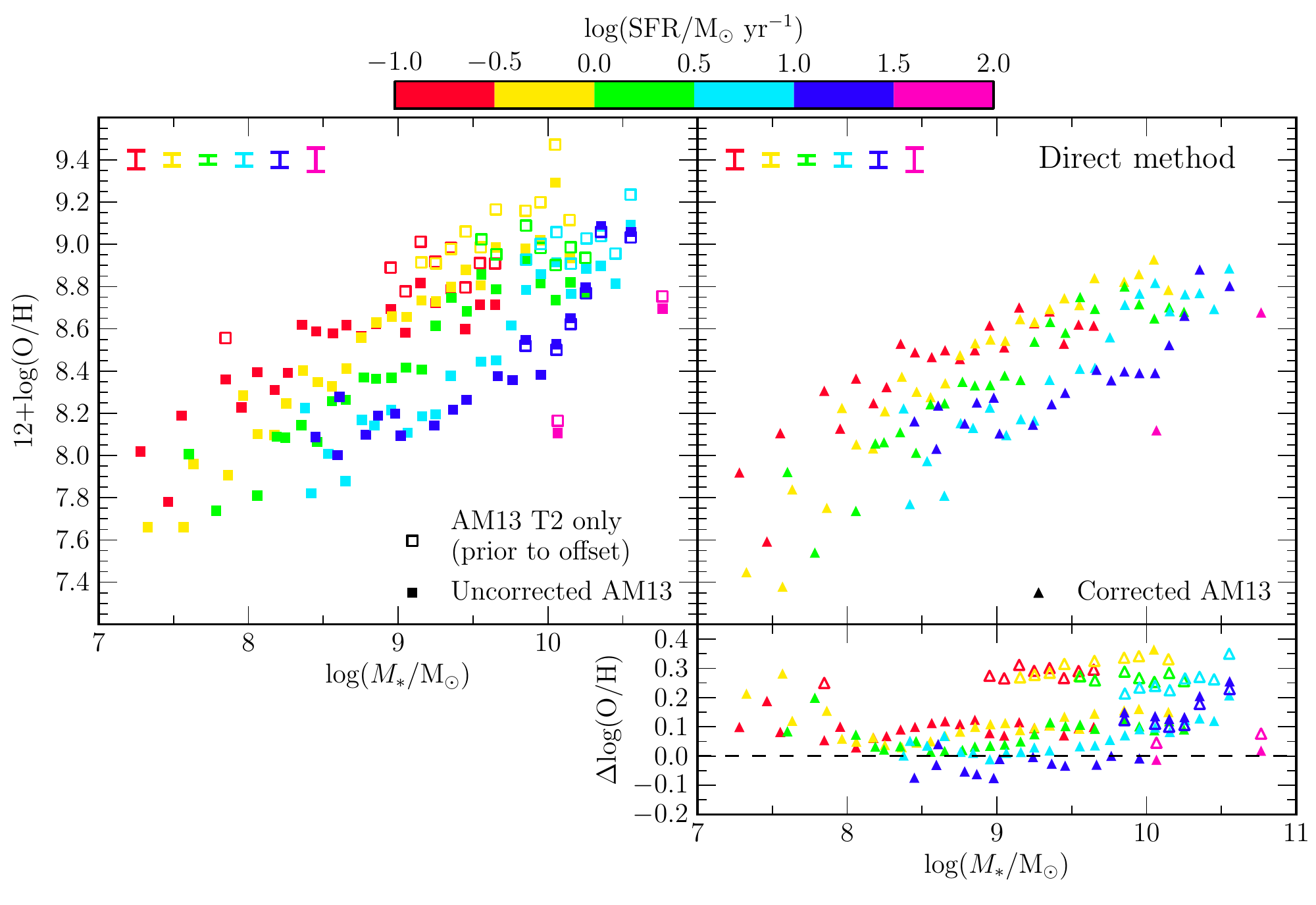}
 \caption{The direct-method FMR for \mstar-SFR stacks from AM13, color-coded by SFR.
  Metallicities in the left panel have not been corrected for the effects of flux-weighting or DIG
 contamination.  Filled squares denote points using the uncorrected AM13 metallicities, while hollow squares
 show 12+log(O/H)$_{\text{T2}}$ assuming the \ttwo-\te\ relation of equation~\ref{eq:t2t3} for those stacks
 with \ttwo\ estimates but no \te\ estimates.  In the right panel, metallicities have been corrected for
 flux-weighting effects and DIG contamination.
  The mean uncertainty in metallicity for each SFR bin is displayed in the upper left corner.
  Filled triangles in the bottom-right panel show the difference
 between the uncorrected and corrected direct-method metallicities,
 where hollow symbols show the difference
 between uncorrected 12+log(O/H)$_{\text{T2}}$ and corrected 12+log(O/H) for those bins with \ttwo\ only.
  After correction, the SFR dependence in the FMR is weaker since there is a positive correlation
 between $\Delta$log(O/H) and SFR at most stellar masses.
}\label{fig:fmrdirect}
\end{figure*}

The FMR as determined using direct-method metallicities will also be subject to biases from
 flux-weighting effects and DIG emission.
  We investigate the effects of flux-weighting and DIG contamination on the FMR using the
 \mstar-SFR stacks of AM13.
  We recalculate the direct-method metallicities of the AM13 \mstar-SFR stacks using the
 methodology in Section~\ref{sec:hiiregions} that includes updated atomic data.
  In order to reproduce the results of AM13 using new atomic data, we calculate the original
 metallicities for those stacks with only \ttwo\ measurements by subtracting the median difference
 between 12+log(O/H)$_{\text{T2}}$ and 12+log(O/H)$_{\text{T2,T3}}$ from 12+log(O/H)$_{\text{T2}}$
 for those stacks with measurements of both temperatures in the same SFR bin.
  This process yields the uncorrected AM13 FMR, shown in the left panel of Figure~\ref{fig:fmrdirect}
 as filled squares color-coded by SFR.  The hollow squares show 12+log(O/H)$_{\text{T2}}$ for bins
 with \ttwo\ only prior to the application of the offset.

We correct the AM13 metallicities of each \mstar-SFR bin for the effects of flux-weighting and DIG contamination.
  The strength of the SFR dependence of the MZR may change after correcting for metallicity biases
 since \fdig\ decreases as SFR increases at fixed \mstar.
  Because \fdig\ depends on SFR, we cannot use the \stackmod model with \fdig=0.55 to correct
 the metallicities in each \mstar-SFR bin, but must instead use a different \fdig\ for each \mstar-SFR bin.
  Accordingly, from the strong-line comparison sample that is selected following
 AM13, we select the subset of galaxies in a particular \mstar-SFR bin and determine the median
 \fdig\ using the method outlined in Section~\ref{sec:stackmodels}.  We then produce a model for each
 \mstar-SFR bin with the inferred \fdig, while all other input parameters are the same as for the
 \stackmod model.  Using these new models, we fit the metallicity biases using equation~\ref{eq:ohpoly}
 and apply these new fits to correct the direct-method metallicity of each \mstar-SFR bin.

The corrected direct-method FMR for AM13 \mstar-SFR stacks is presented in the right
 panel of Figure~\ref{fig:fmrdirect}.  The difference between the uncorrected
 and corrected log(O/H) for each bin is presented in the bottom-right panel, where the hollow
 triangles show $\Delta$log(O/H) between the corrected value and the uncorrected
 12+log(O/H)$_{\text{T2}}$ before applying the offset.
  SFR dependence is still clearly present in the corrected direct-method FMR with higher-SFR
 galaxies having lower metallicities at fixed \mstar, in agreement with other observations of the
 SFR dependence of the local MZR \citep{man10,lar10,and13}.  However, the SFR dependence
 is weaker after correcting for biases in the metallicity estimates.  At fixed \mstar, $\Delta$log(O/H)
 correlates with SFR such that galaxies with lower SFR have larger positive biases, while galaxies
 with high SFR have smaller positive or sometimes slight negative biases.
  This effect weakens the strength of the SFR dependence, occurring because DIG contamination
 causes galaxies to appear more metal rich when direct-method metallicities are employed
 due to increased low-ionization line strength and
 decreased \ttwo.  This bias is strongest in low-SFR galaxies in which DIG emission begins
 to dominate the line fluxes, leading to large corrections at low-SFR and smaller corrections
 as SFR increases.

\begin{figure}
 \includegraphics[width=\columnwidth]{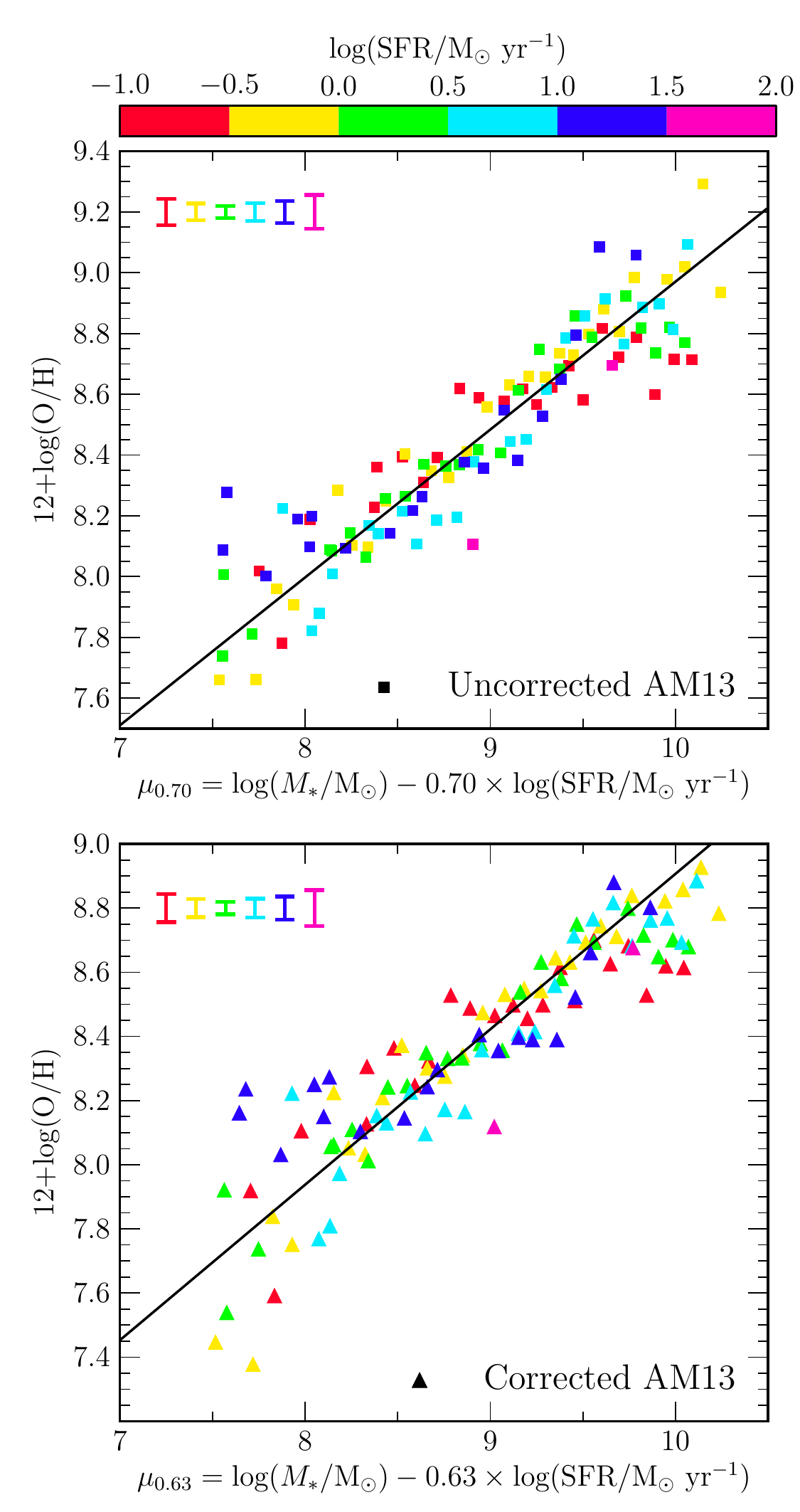}
 \caption{The planar projection of the FMR for AM13 \mstar-SFR stacks.
  We use the FMR parameterization of \citep{man10}, and show the FMR projection using uncorrected (top)
 and corrected (bottom) direct-method metallicities.
  In each panel, the black line shows the best-fit linear relation around which the scatter is minimized
 for the best-fit value of $\alpha$.
  The value of the parameter $\alpha$ that minimizes the scatter around the plane was found to be
 $\alpha=0.70\pm0.015$ when using uncorrected metallicities and $\alpha=0.63\pm0.016$ after correction.
  Correcting for flux-weighting effects and DIG contamination slightly decreases the strength
 of the SFR dependence of the FMR.
  Error bars in the upper left corner show the mean uncertainty in 12+log(O/H) for each SFR bin.
}\label{fig:fmrproj}
\end{figure}

\citet{man10} parameterized the SFR dependence with a planar projection using the parameter $\mu_{\alpha}$
 that is a linear combination of \mstar\ and SFR:
\begin{equation}
\mu_{\alpha}=\text{log}(M_*/\text{M}_{\odot}) - \alpha\times \text{log}(\text{SFR}/\text{M}_{\odot}\text{ yr}^{-1}) .
\end{equation}
We evaluate the SFR strength of the direct-method FMR based on uncorrected and corrected metallicities
 by finding the value of $\alpha$ that minimizes the scatter around a linear fit in each case.
  We find that the uncorrected AM13 metallicities yield $\alpha=0.70\pm0.015$
 (compare to $\alpha=0.66$ reported in AM13),
 while the SFR dependence is slightly weaker after correcting the metallicities, with $\alpha=0.63\pm0.016$.
  The best-fit projections of the uncorrected and corrected direct-method FMRs are presented in the
 top and bottom panels of Figure~\ref{fig:fmrproj}.
  The smaller value of $\alpha$ after correcting the metallicities confirms that DIG contamination
 leads to an overestimation of the strength of the SFR dependence.
  However, this small decrease in $\alpha$ does not bring estimates using the direct-method into agreement
 with those made using strong-line metallicities.  Investigations using strong-line indicators
 find much weaker SFR dependence ranging from $\alpha=0.19$ \citep{yat12} to $\alpha=0.32$
 \citep{man10}.

\subsection{B16 and Cowie et al. 2016 results are primarily caused by DIG contamination}

B16 estimated direct-method metallicites of stacks of SDSS galaxies in bins of \mstar\ and distance
 from the $z\sim0$ \mstar-SSFR relation ($\Delta$SSFR),
 and showed that empirical strong-line metallicity calibrations have a systematic dependence on $\Delta$SSFR.
  In particular, these authors found that galaxies with higher $\Delta$SSFR
 display systematically higher N2, lower O3N2, and higher N2O2 values at fixed direct-method metallicity.
  B16 provided new calibrations that include a $\Delta$SSFR term to account for this variation.
  \citet{cow16} found similar results based on individual $z\sim0$ SDSS galaxies with auroral-line detections,
 instead using H$\beta$ luminosity as the secondary parameter.  These authors
 found that, at fixed direct-method metallicity,
 galaxies with higher H$\beta$ luminosity displayed higher N2, N2O2, and N2S2.
  \citet{cow16} provided new strong-line calibrations including an additional
 H$\beta$ luminosity term, and interpreted the trends as an increase in both N/O and ionization parameter
 as SFR increases.  Since SSFR and H$\beta$ luminosity are strongly correlated, it appears that
 the two studies observed the same phenomenon using different parameterizations.

\begin{figure*}
 \includegraphics[width=\textwidth]{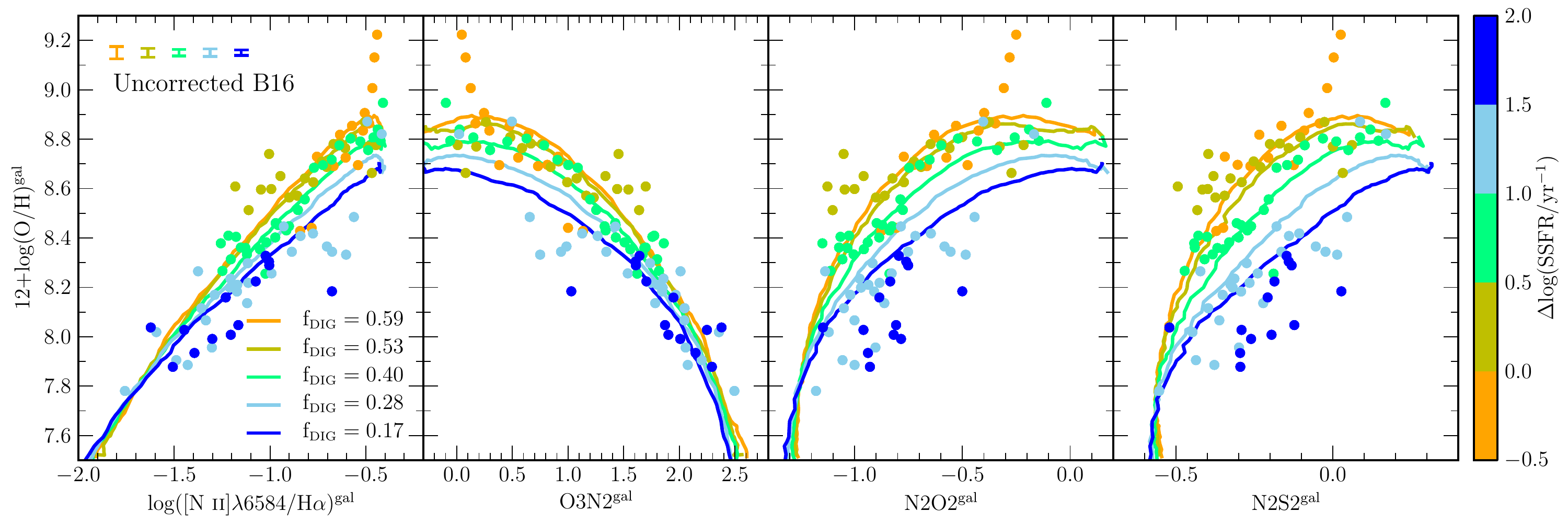}
 \caption{Global galaxy direct-method metallicity as a function of observed strong-line ratios for
 the \mstar-$\Delta$SSFR stacks of B16.  Filled circles denote data from B16 color-coded by $\Delta$log(SSFR),
 where the metallicities have been recalculated using updated atomic data.  Metallicities and line ratios
 have not been corrected for flux-weighting effects or DIG contamination.  Predicted global galaxy
 metallicity and uncorrected line ratios from models matched in \fdig\ to each
 $\Delta$SSFR bin are shown as solid lines of the corresponding color.
  Error bars in the upper left corner display the mean uncertainty in 12+log(O/H) for each $\Delta$log(SSFR) bin.
}\label{fig:b16ratios}
\end{figure*}

In Figure~\ref{fig:b16ratios}, we show direct-method metallicity as a function of the strong-line
 ratios N2a, O3N2, N2O2, and N2S2.  We plot the points from the \mstar-$\Delta$SSFR stacks of
 B16, color-coded by $\Delta$log(SSFR).
  We recalculate uncorrected B16 direct-method metallicities using our methodology and updated atomic data
 presented in Section~\ref{sec:hiiregions}.
  Following AM13 and B16, we estimate the uncorrected metallicities of
 stacks for which only \ttwo\ was measured by adjusting 12+log(O/H)$_{\text{T2}}$ by the median
 offset between 12+log(O/H)$_{\text{T2}}$ and 12+log(O/H)$_{\text{T2,T3}}$ for those stacks with
 both \ttwo\ and \te\ measurements in the same $\Delta$SSFR bin.
  There are no stacks in the $\Delta$log(SSFR$)=-0.25$ bin with \te\ estimates, thus the offset
 for \ttwo-only metallicities cannot be determined in the same way for this bin.  B16 did not
 apply any offset to the metallicities of stacks in this lowest-$\Delta$SSFR bin, instead adopting
 the value of 12+log(O/H)$_{\text{T2}}$ assuming the \ttwo-\te\ relation followed by \hii regions.
  To place the metallicities in the $\Delta$log(SSFR$)=-0.25$ bin onto the same scale as those of the
 other bins, we apply the offset for the closest $\Delta$SSFR bin ($\Delta$log(SSFR$)=0.25$) to
 12+log(O/H)$_{\text{T2}}$.  This solution is robust because, while the offset increases with
 decreasing $\Delta$SSFR, the rate of change of the offset size with $\Delta$SSFR decreases
 with decreasing $\Delta$SSFR.  The two bins closest in $\Delta$SSFR to the $\Delta$log(SSFR$)=-0.25$ bin
 have the smallest difference in offset of only 0.026~dex, so this solution should yield the uncorrected
 metallicities of the $\Delta$log(SSFR$)=-0.25$ stacks within $\lesssim0.02$~dex.
  It is important to note that these metallicities and strong-line ratios are inferred directly from
 the observed line fluxes of each stack, and have not been corrected for any biases.

We expect \fdig\ to correlate with $\Delta$SSFR since \sigha, from which \fdig\ is estimated, correlates strongly
 with SFR and SSFR, but does not show a strong dependence on \mstar.
  It is therefore expected that \fdig\ will change significantly across samples that vary greatly
 in SSFR and SFR, as in B16 and \citet{cow16}.
  If \fdig\ changes significantly between bins of $\Delta$SSFR or H$\beta$ luminosity, then the bias
 arising from DIG contamination will also vary systematically between such bins.
  We investigate the connection between DIG emission and the B16 and \citet{cow16} results by
 determining the median \fdig\ for subsets of the B16 sample.  We begin with the SDSS strong-line comparison
 sample of individual galaxies that is selected in a nearly identical manner to the samples of AM13 and B16.
  We divide the full sample into subsamples in 0.5~dex-wide bins of $\Delta$log(SSFR) using the
 parameterization of the mean $z\sim0$ \mstar-SSFR relation from B16.  For each subsample, we determine
 \sigha\ and \fdig\ for the individual galaxies and use the distribution of \fdig\ values to infer
 the median \fdig\ following the methods described in Section~\ref{sec:stackmodels}.
  We find the median \fdig\ for bins centered on $\Delta\log(\text{SSFR/yr}^{-1})=[-0.25, 0.25, 0.75, 1.25, 1.75]$
 to be $f_{\text{DIG}}^{\text{med}}=[0.59, 0.53, 0.40, 0.28, 0.17]$.
  We create a set of five models that have all model parameters set to the same values as for the \stackmod model
 except for \fdig, which is set to the $f_{\text{DIG}}^{\text{med}}$ value for each $\Delta$SSFR bin.
  In Figure~\ref{fig:b16ratios}, we plot the predicted global galaxy strong-line ratios and
 uncorrected direct-method metallicities for the models matched to each $\Delta$SSFR bin.
  The values plotted for the models are the predicted observed values as would be inferred from
 global galaxy spectra before correcting for any biases.

\begin{figure*}
 \includegraphics[width=\textwidth]{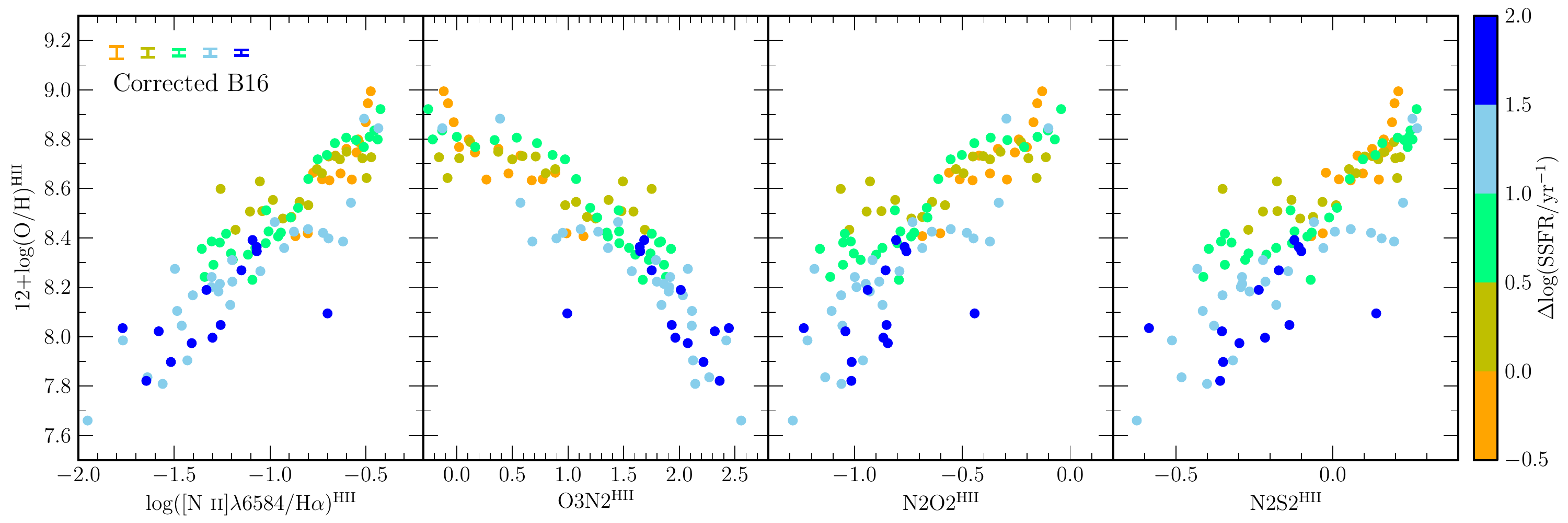}
 \caption{Corrected direct-method metallicity as a function of corrected strong-line ratios for
 the \mstar-$\Delta$SSFR stacks of B16.  The effects of flux-weighting and DIG emission
 have been corrected for using the models shown in Figure~\ref{fig:b16ratios}.
  Correcting for DIG contamination reduces the scatter and systematic dependence on
 $\Delta$log(SSFR) in these relations.
  Mean uncertainties on the metallicity for each $\Delta$SSFR bin are shown in the upper left corner.
}\label{fig:b16corrected}
\end{figure*}

For each line ratio, we find that the predicted global galaxy line ratios and uncorrected metallicites from
 the models are in excellent agreement with the observations of B16,
 although the models somewhat underpredict the deviation
 in N2O2 and N2S2 that is observed in the highest $\Delta$SSFR bins.
  This disagreement at high $\Delta$SSFR may indicate that the \fdig-\sigha\ relation
 of equation~\ref{eq:fdig} overpredicts \fdig\ at high \sigha\ and may require some revision.
  Additionally, lines of constant \fdig\ in the models match lines of constant $\Delta$SSFR in the B16 stacks.
  It is therefore plausible that the systematic trends observed in B16 and \citet{cow16} can be
 explained by variation of \fdig\ with $\Delta$SSFR and H$\beta$ luminosity.
  Following Sections~\ref{sec:strongbias} and~\ref{sec:ohbias}, we fit the strong-line ratio and
 direct-method metallicity biases using equations~\ref{eq:ratpoly} and~\ref{eq:ohpoly}, respectively,
 for each of the models matched to the B16 $\Delta$SSFR bins.  These best-fit polynomials are then used
 to correct the strong-line ratios and direct-method metallicites of the points in each B16 $\Delta$SSFR bin
 for flux-weighting effects and DIG contamination.
  Figure~\ref{fig:b16corrected} shows the corrected direct-method metallicities as a function of
 corrected strong-line ratios for the B16 \mstar-SSFR stacks.  The dependence on $\Delta$SSFR
 of each strong-line ratio at fixed metallicity has decreased or disappeared once biases in both
 properties are accounted for.  This resolution is most apparent in N2O2 and N2S2, which displayed
 the strongest $\Delta$SSFR dependence prior to correction because DIG contamination affects O2
 and S2 more strongly than O3 or N2.
  
We conclude that the majority of the systematic offsets as a function of $\Delta$SSFR and H$\beta$
 luminosity observed by B16 and \citet{cow16} are a result of the decreasing importance of DIG
 emission as star formation intensity increases.  Offsets in strong-line ratios at fixed
 direct-method metallicity occur because both the strong-line ratio and the direct-method metallicity
 are biased, predominantly due to DIG contamination.  The spread in line ratio at fixed metallicity is
 largest at high metallicities (12+log(O/H$)>8.3$), where singly-ionized oxygen is the dominant ionic
 species, because DIG contamination strongly affects O2, \ttwo, and, consequently, the O$^+$/H estimate.
  After accounting for \fdig\ variation as a function of $\Delta$SSFR, any remaining systematic offset
 as a function of $\Delta$SSFR is small and does not require large systematic changes in
 the physical conditions of the \hii region gas to explain.
  Since the B16 results appear to be equivalent to those of
 \citet{cow16}, a large systematic increase in N/O and ionization parameter with increasing
 SFR is not needed to explain the shift in strong-line ratios at fixed direct-method metallicity
 observed by \citet{cow16}.

\subsection{Correcting the \citet{cur17} empirical calibrations}

C17 recently used stacks of SDSS star-forming galaxies in bins of O3 and O2 to construct a set of
 fully-empirical strong-line calibrations for a range of commonly applied line ratios
 using direct-method metallicities.  Utilizing a fully-empirical calibration dataset with a large
 dynamic range in metallicity improves upon past metallicity calibrations based upon galaxy
 spectra, which required the use of photoionization models at high metallicities where auroral lines are
 not detected for individual SDSS galaxies \citep{mai08}.
  While the C17 calibrations can be used over a wider range of metallicities than any other
 empirical calibration to date, both the direct-method metallicities and strong-line ratios used in the
 calibrations are subject
 to biases from flux-weighting effects and DIG contamination as described in this work.
  In order to use the C17 calibrations to estimate the characteristic metallicity of the star-forming
 regions in galaxies by removing DIG contamination or flux-weighted combination effects,
 we recommend first using the observed uncorrected galaxy line ratios to determine the
 uncorrected metallicity, then correcting the metallicity inferred from the C17 calibrations
 using the fit to the bias in 12+log(O/H)$_{\text{T2,T3}}$ (left panel of Figure~\ref{fig:ohbias}).
  The best-fit coefficients are given in Table~\ref{tab:coeffs}.
  This method will yield robust corrected metallicities that are representative of the distribution of
 \hii region metallicities galaxies.

\section{Implications for high-redshift studies}\label{sec:highz}

The MZR is known to evolve with redshift such that galaxies have lower metallicities at fixed
 stellar mass as redshift increases \citep[e.g.,][]{erb06, mai08, ste14, tro14, san15, ono16}.
  High-redshift metallicity studies have relied nearly uniformly on strong-line calibrations
 to estimate metallicity because of the difficulty of detecting faint auroral lines at
 $z>1$.  We have shown how contamination from DIG emission can affect strong-line ratios
 and thus impact strong-line metallicity estimates.  Correcting for these biases can lead to
 significant changes in the inferred shape of the local MZR.  It is of interest to consider
 what effects DIG contamination might have on the interpretation of high-redshift strong-line ratios.

High redshift galaxies have both smaller size \citep{van14} and higher SFR
 \citep{whi14,shi15} at fixed \mstar than $z\sim0$ galaxies on average.
  Thus, typical \sigha\ values are much higher at high redshift than for local galaxies,
 with typical $z\sim2$ galaxies having \sigha\ as high as local starburst galaxies.
  If the relationship between \sigha\ and \fdig\ in equation~\ref{eq:fdig} holds in the same
 form at high redshifts, then DIG emission should only account for a small fraction ($\sim0-20$\%) of line
 emission in high-redshift star-forming galaxies.  Based on the observations and simple model
 of \citet{oey07}, we expect that DIG emission becomes increasingly less important with increasing
 redshift.  Accordingly, correction of high-redshift galaxy line ratios should be performed using
 models that have \fdig=0, such as the \hiimod model.  Additionally, the high SSFR of high-redshift
 galaxies and accompanying strong feedback may efficiently mix metals into the ISM such that
 the width of the \hii region metallicity distribution is small.
  Such a scenario can explain the flat metallicity gradients observed in some high-redshift
 galaxies \citep{jon13,lee16,wan17,ma17}, and is similar to the inferred reason behind
 flatter gradients in lower mass local galaxies \citep{ho15}.
  If efficient metal distribution is a common feature at high redshift,
 then an appropriate model should also have smaller $\sigma_{\text{T}}$ than the \stackmod model.

If DIG emission is negligible at high redshift, it carries implications for the interpretation
 of the evolution of strong-line ratios.  Galaxies at $z\sim1-2$ display systematically different
 emission-line ratios from those typically observed in local galaxies, including an offset towards
 higher [O\iii]/H$\beta$ and/or [N\ii]/H$\alpha$ in the O3N2 diagram
 \citep{sha05,liu08,kew13,ste14,sha15} and higher O3 and [O\iii]/[O\ii] values at fixed \mstar\ than those
 observed in the local universe \citep{nak14,ono16,san16a}.
  Such evolution in galaxy strong-line ratios may imply that local strong-line metallicity calibrations
 do not produce reliable metallicity estimates for high-redshift galaxies.
  By comparing the positions of high-redshift
 galaxies to those of local galaxies in diagnostic emission line diagrams, such evolution in strong-line
 ratios has been explained with evolving physical conditions of the ionized gas in \hii regions, including
  gas density, ionization parameter, N/O abundance ratio, and shape of the ionizing spectrum.
  These interpretations have assumed that all of the emission line flux from both $z\sim0$
 and high-redshift galaxies originates from \hii regions.  If DIG emission is negligible in
 high-redshift star-forming galaxies, then the amount of inferred evolution in gas physical conditions
 such as the ionization parameter or hardness of the ionizing spectrum
 is likely overestimated because DIG emission tends to shift $z\sim0$ global galaxy line ratios towards
 lower-excitation states compared to their constituent \hii regions.  It would then be more appropriate
 to compare high-redshift strong-line ratios to those of individual \hii regions instead of
 SDSS global galaxy spectra, or else first correct SDSS line ratios using the best-fit functions
 in Figure~\ref{fig:strongbias} before inferring evolution of \hii region physical conditions.
  Such corrections are most important in line-ratio spaces that are significantly affected by
 DIG emission, such as the O3S2 and [O\iii]/[O\ii] vs. \mstar\ diagrams.
  Evolution in some ionized gas physical properties is still clearly required because
 DIG contamination (or lack thereof) cannot drive an offset in the O3N2 diagram (see Figure~\ref{fig:allratios})
 for example.

Revealing the true DIG contribution to global galaxy line fluxes in high-redshift galaxies
 requires high-spatial-resolution emission-line maps to disentangle \hii and DIG regions and
 determine their relative importance.  The SDSS-IV MaNGA IFU survey has shown that line ratio
 maps from such a dataset can efficiently identify \hii and DIG regions based on systematic
 changes in strong-line ratios as a function of H$\alpha$ surface brightness \citep{zha17}.
  Similar high-redshift datasets should be able to identify significant DIG emission if
 the spatial resolution is sufficient to begin to resolve \hii regions ($\lesssim1$~kpc).
  Such maps have been obtained for a small number of gravitationally-lensed objects
 \citep[e.g.,][]{jon10,yua11,lee16},
 but understanding the typical DIG contribution at high redshift necessitates larger samples
 spanning a wide range of galaxy properties.
  Confirming the nature of DIG emission at high-redshift is crucial for properly interpreting
 the evolution in galaxy strong-line ratios.

\section{Summary and conclusions}\label{sec:summary}

We have presented a set of empirically-motivated models that treat galaxies as a collection of
 multiple line-emitting regions with different physical properties.
  In addition to line emission from classical \hii regions, these models incorporate
 DIG emission based on observed DIG strong-line ratio excitation sequences for the first time.
  We present the first measurement of DIG region excitation sequences over a range of excitation levels
 using data from the SDSS-IV MaNGA IFU survey \citep{zha17}.
  Our model framework tracks contributions from DIG and \hii regions to both strong and auroral
 optical emission lines.  Previous models of galaxy line emission have treated galaxies as single
 \hii regions with effective physical properties.  Such descriptions of galaxy line emission are not
 sufficient to simultaneously match strong and auroral emission line properties in all line-ratio
 diagrams simulaneously.  Including multiple \hii regions with a range of excitation
 levels is required to reproduce the offset of global galaxy spectra in the \ttwo-\te\ diagram
 \citep[Figures~\ref{fig:t2t3digfrac055} and~\ref{fig:t2t3digfrac040};][]{pil10,pil12,and13}.
  Furthermore, inclusion of DIG emission is necessary to properly reproduce galaxy excitation
 sequences in strong-line ratio diagrams, as evidenced by the distinct excitation sequences of
 \hii regions, SDSS galaxies, and DIG regions in the O3N2, O3S2, and O3O2 diagrams (Figure~\ref{fig:allratios}).

We constructed the \stackmod model in which DIG emission contributes 55\% of the total Balmer emission,
 which provides a good description of typical $z\sim0$ star-forming galaxies as represented by stacks
 of SDSS galaxies from \citet{and13}, \citet{bro16}, and \citet{cur17}.
  We find that the ionic temperature \ttwo\ of DIG regions must be $\sim15$\% lower than \ttwo\ of \hii
 regions at fixed metallicity to match the strong-line ratios of SDSS stacks at fixed \ttwo.
  This result may indicate that DIG region electron temperature systematically deviates from the electron
 temperature of associated \hii regions, but may also represent a systematic effect in the process of
 combining line emission from multiple regions to form a global galaxy spectrum.  Observations
 of auroral lines from DIG regions are needed to investigate this effect.
  When following this assumption about DIG \ttwo, the \stackmod model is in excellent agreement
 with SDSS stacks simultaneously in diagrams involving strong-line ratios, electron temperatures, and
 direct-method oxygen abundances.

We used the \stackmod model to characterize biases in strong-line ratios, electron temperatures,
 and direct-method oxygen abundances as inferred from global galaxy spectra.
  Contamination of the global galaxy spectrum by DIG emission is the primary driver of biases in the
 \stackmod model.  DIG contamination tends to inflate the strength of low-ionization lines and lower
 the ionic temperature \ttwo, making global galaxy spectra appear more metal-rich than is true of the
 metallicity distribution of star-forming regions within each galaxy.  We quantified
 biases in these properties as the difference between the value inferred from a global galaxy spectrum
 and the median value of the \hii region distribution of that property within each galaxy.
  We provided polynomial fits to the bias in each property (Table~\ref{tab:coeffs}) that can be
 subtracted from global galaxy values to correct for the effects of flux-weighting and DIG
 contamination.  The corrections presented in Section~\ref{sec:biases} are appropriate for samples
 of galaxies that are representative of the local star-forming population.  A recipe for correcting
 individual galaxies or unrepresentative samples is given in Appendix~\ref{appendix} in which the
 \stackmod model is generalized to have any value of \fdig.

We applied these corrections to investigate observational biases in the $z\sim0$ MZR and FMR.
  Nearly all metallicity calibrations are based on \hii regions, including the direct-method
 and both empirical and theoretical strong-line metallicity calibrations.  It is thus imperative
 that emission-line ratios of global galaxy spectra are corrected to be representative of the
 underlying \hii region distribution before using calibrations based on \hii regions to
 estimate metallicity.
  After correcting for flux-weighting effects and DIG contamination, we found that the $z\sim0$
 direct-method MZR has $\sim0.1$~dex lower normalization and a slightly steeper low-mass slope
 ($\gamma=0.75$) compared to the uncorrected MZR ($\gamma=0.67$).  These changes in the MZR shape
 have a significance of $2\sigma$.
  The direct-method FMR displays slightly weaker SFR dependence after correction since DIG
 tends to make low-SFR galaxies appear more metal-rich, artificially strengthening the trend with SFR.
  We also investigated the effects of DIG contamination and flux-weighting on the local MZR
 as determined using multiple strong-line calibrations.  DIG contamination can substantially
 affect the inferred shape of the MZR, flattening the low-mass slope and changing the normalization of
 theoretical calibrations in particular.
  Future studies of metallicity scaling relations can use the corrections given in this work to obtain
 robust galaxy metallicity estimates that are placed on a scale that can be compared directly to
 gas-phase metallicities reported by chemical evolution models.

We showed that the systematic trends in strong-line ratios at fixed direct-method metallicity
 with SSFR and H$\beta$ luminosity observed by \citet{bro16} and \citet{cow16} can be explained
 almost entirely by a decreasing \fdig\ with increasing SSFR.  The importance of DIG is naturally expected to
 decrease with increasing star-formation intensity as classical \hii regions occupy a larger volume
 of the ionized ISM and dominated line emission \citep{oey07}.
  This result demonstrates the importance in correcting for DIG contamination before inferring correlations
 of \hii region physical properties with galaxy properties.

Our results have implications for the inferred evolution of \hii region physical properties with redshift.
  If the trend between \fdig\ and \sigha\ holds out to high redshifts, we expect that DIG emission is
 negligible in typical high-redshift galaxies that are more highly star-forming \citep{whi14}
 and more compact \citep{van14} than their
 $z\sim0$ counterparts at fixed \mstar.  Inferring evolution in \hii region properties by comparing positions of
 high-redshift galaxies to those of global galaxy spectra in strong-line ratio diagrams will likely
 overestimate the magnitude of evolution in, e.g., metallicity and ionization parameter.  DIG contamination
 increases low-ionization line ratios in local star-forming galaxies, making them appear to have a lower
 level of excitation
 than the ionized gas in their constituent \hii regions.  Such an effect can artificially augment the
 offset between $z\sim0$ and $z>1$ star-forming regions in strong-line ratio diagrams, leading to incorrect
 assumptions about the evolution of ionized gas properties.
  A more robust comparison can be achieved by correcting $z\sim0$ global galaxy observations for DIG
 contamination prior to comparing to high-redshift samples.

We stress that, in cases where it is desirable to measure properties that are characteristic of the \hii regions
 within galaxies, deriving properties directly from observed line ratios of global galaxy spectra will not
 yield the desired result, but instead will be systematically biased.
  This is true of any dataset where the spectroscopic aperture (e.g., fiber, slit, etc.) contains light
 from multiple \hii regions and the diffuse gas that exists between \hii regions.
  Models of galaxy line-emission must incorporate both multiple emitting regions with a spread
 in properties and DIG emission in order to accurately match the emission-line properties of
 real galaxies.  The increasing number of spatially-resolved spectroscopic surveys of local
 galaxies (e.g., MaNGA, SAMI, CALIFA) will allow for an accurate determination of the strong-line
 properties of DIG regions.  Observational constraints are still needed on the auroral-line properties
 and electron temperatures of diffuse gas.  We encourage future studies modeling the line emission of
 star-forming galaxies to avoid treating galaxies as single emitting regions with a set of effective
 properties, and instead design models that reflect the substructure and diversity observed in
 the ISM of real galaxies.

\acknowledgements We acknowledge the importance of the First Carnegie Symposium in Honor of Leonard Searle
 for presentations and discussions without which this work would not have been possible.
  We acknowledge support from the NSF AAG grant AST-1312780 and
 grant NNX16AF54G from the NASA ADAP program.
  RY acknowledges support by NSF AAG grant AST-1715898.
  Funding for the Sloan Digital Sky Survey IV has been provided by the Alfred P. Sloan Foundation, the U.S. Department of Energy Office of Science, and the Participating Institutions. SDSS-IV acknowledges
support and resources from the Center for High-Performance Computing at
the University of Utah. The SDSS web site is www.sdss.org.
SDSS-IV is managed by the Astrophysical Research Consortium for the 
Participating Institutions of the SDSS Collaboration including the 
Brazilian Participation Group, the Carnegie Institution for Science, 
Carnegie Mellon University, the Chilean Participation Group, the French Participation Group, Harvard-Smithsonian Center for Astrophysics, 
Instituto de Astrof\'isica de Canarias, The Johns Hopkins University, 
Kavli Institute for the Physics and Mathematics of the Universe (IPMU) / 
University of Tokyo, Lawrence Berkeley National Laboratory, 
Leibniz Institut f\"ur Astrophysik Potsdam (AIP),  
Max-Planck-Institut f\"ur Astronomie (MPIA Heidelberg), 
Max-Planck-Institut f\"ur Astrophysik (MPA Garching), 
Max-Planck-Institut f\"ur Extraterrestrische Physik (MPE), 
National Astronomical Observatories of China, New Mexico State University, 
New York University, University of Notre Dame, 
Observat\'ario Nacional / MCTI, The Ohio State University, 
Pennsylvania State University, Shanghai Astronomical Observatory, 
United Kingdom Participation Group,
Universidad Nacional Aut\'onoma de M\'exico, University of Arizona, 
University of Colorado Boulder, University of Oxford, University of Portsmouth, 
University of Utah, University of Virginia, University of Washington, University of Wisconsin, 
Vanderbilt University, and Yale University.

\appendix

\section{Appendix A: A recipe for correcting individual galaxies or unrepresentative samples}\label{appendix}

In Section~\ref{sec:biases}, we presented polynomial functions that represent the median bias in
 strong-line ratios, electron temperatures, and direct-method metallicities when these properties
 are inferred directly from global galaxy spectra.  However, the \stackmod model with \fdig=0.55,
 upon which the corrections in Section~\ref{sec:biases} are based, is only appropriate to apply
 to a sample of galaxies that is representative of the typical $z\sim0$ star-forming population
 or to individual galaxies that fall near the mean relations.  In this appendix,
 we present generalized results for a set of models in which
 \fdig\ is varied from 0.0 (equivalent to the \hiimod model) to 0.8, and supply
 a recipe to follow when applying these generalized results.

The relative contribution of DIG to Balmer emission, \fdig, is inferred using \sigha\ and
 equation~\ref{eq:fdig}.  Thus, the inferred \fdig\ depends on the star formation properties of
 the galaxy, since \sigha\ will increase with increasing SFR.  If a galaxy falls near the
 mean $z\sim0$ \mstar-SFR relation, or if the mean of a sample of galaxies lies near the mean
 local relation, then the corrections given in Section~\ref{sec:biases} may be applied.
  However, it is often of interest to study unrepresentative or extreme objects.
  For example, the sample of individual $z\sim0$ SDSS galaxies with auroral-line detections
 have higher SFR at fixed \mstar\ than is typical of the local star-forming population, and
 thus requires a lower median \fdig\ value as demonstrated in Section~\ref{sec:auroralmodels}.
  Investigating SFR dependence of local scaling relations requires dividing the local galaxy
 population into subsamples that are unrepresentative by construction, as in studies of
 the $z\sim0$ FMR \citep{man10,lar10,and13}.  Extreme local galaxies are also of interest
 because they may provide local analogs of the ISM conditions in high-redshift galaxies
 \citep[e.g.,][]{bro14,bia16}.  We therefore provide results for models spanning a wide range
 in \fdig\ so that flux-weighting effects and DIG contamination may be corrected for
 in individual galaxies and samples with a wide range in SFR and SSFR.

\begin{figure}
 \includegraphics[width=\columnwidth]{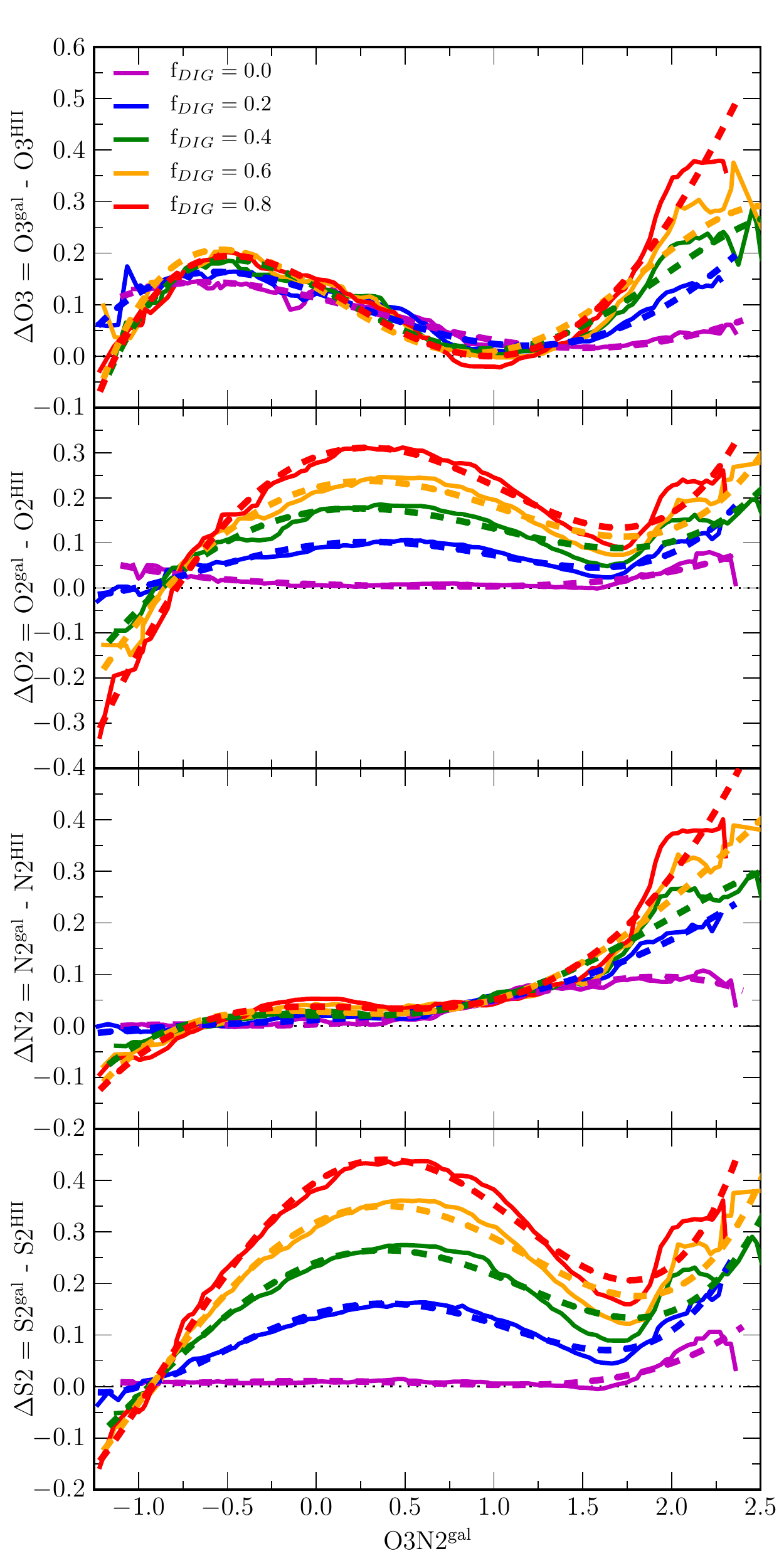}
 \caption{The difference between the global galaxy line ratio and median line ratio of the \hii region
 distribution, $\Delta$X, as a function of O3N2\gal\ for the strong-line ratios X=O3, O2, N2, and S2.
  Solid lines show the running median of 2500 mock galaxy realizations in bins of O3N2\gal\
 for models with \fdig=0.0 to 0.8.
  In each panel, the dashed lines display the best-fit fourth-order polynomial to the bias in the
 global galaxy line ratio, $\Delta$X, for the model of the corresponding color.
  The best-fit coefficients are presented in Table~\ref{tab:appendixstrongcoeffs}.
}\label{fig:appendixstrongbias}
\end{figure}

\begin{figure}
 \includegraphics[width=\columnwidth]{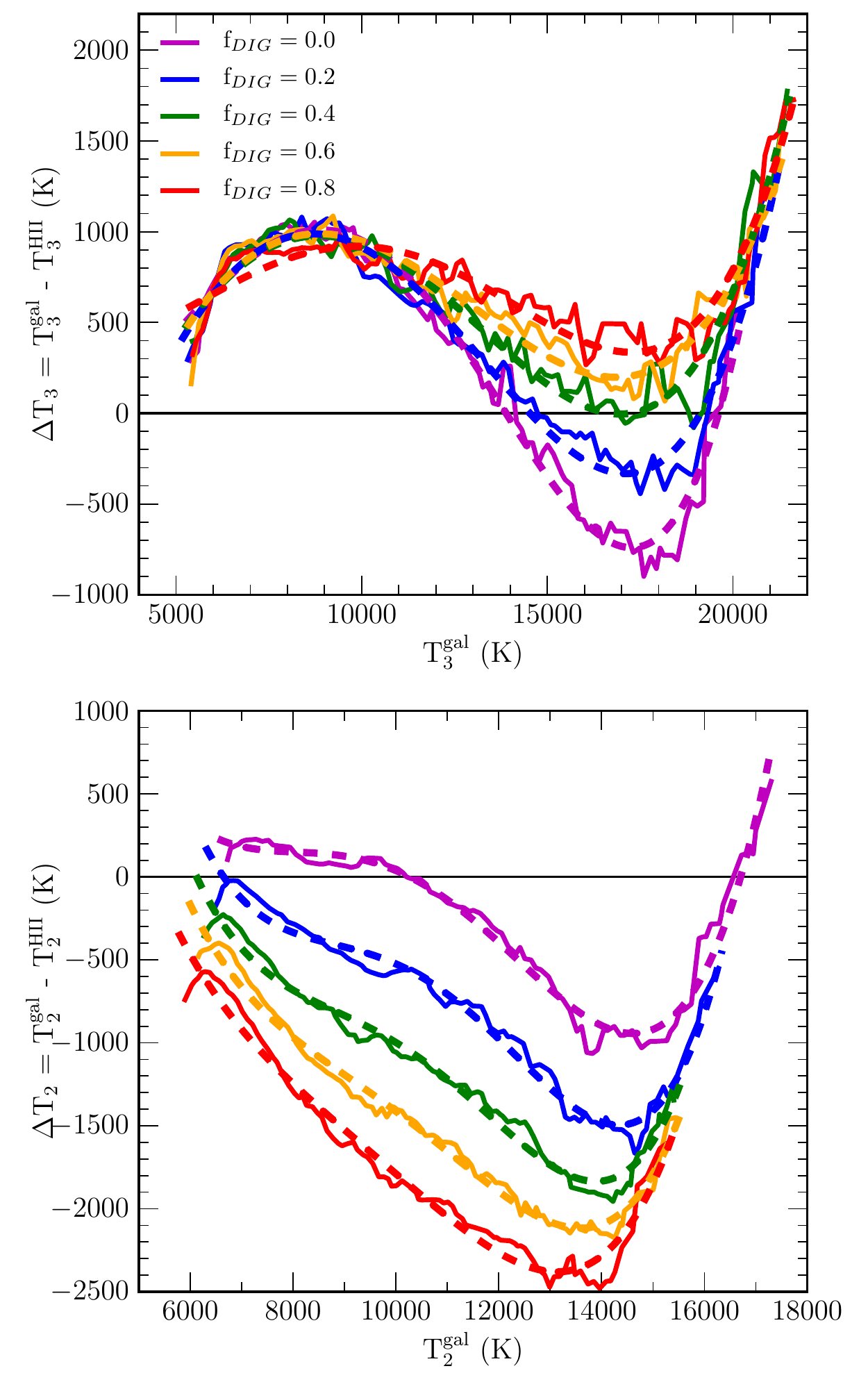}
 \caption{
The difference between the electron temperature inferred from the global galaxy spectrum
 and the median electron temperature of
 the \hii region distribution as a function of electron temperature.  Results for \te\ are shown
 in the top panel, while the bias in \ttwo\ is presented in the bottom panel, for models with
 DIG contribution ranging from \fdig=0.0 to 0.8.
  Best-fit fourth-order polynomials for each model are presented as dashed lines of the corresponding color.
  The best-fit coefficients are given in Table~\ref{tab:appendixtempcoeffs}.
}\label{fig:appendixtempbias}
\end{figure}

\begin{figure*}
 \includegraphics[width=\textwidth]{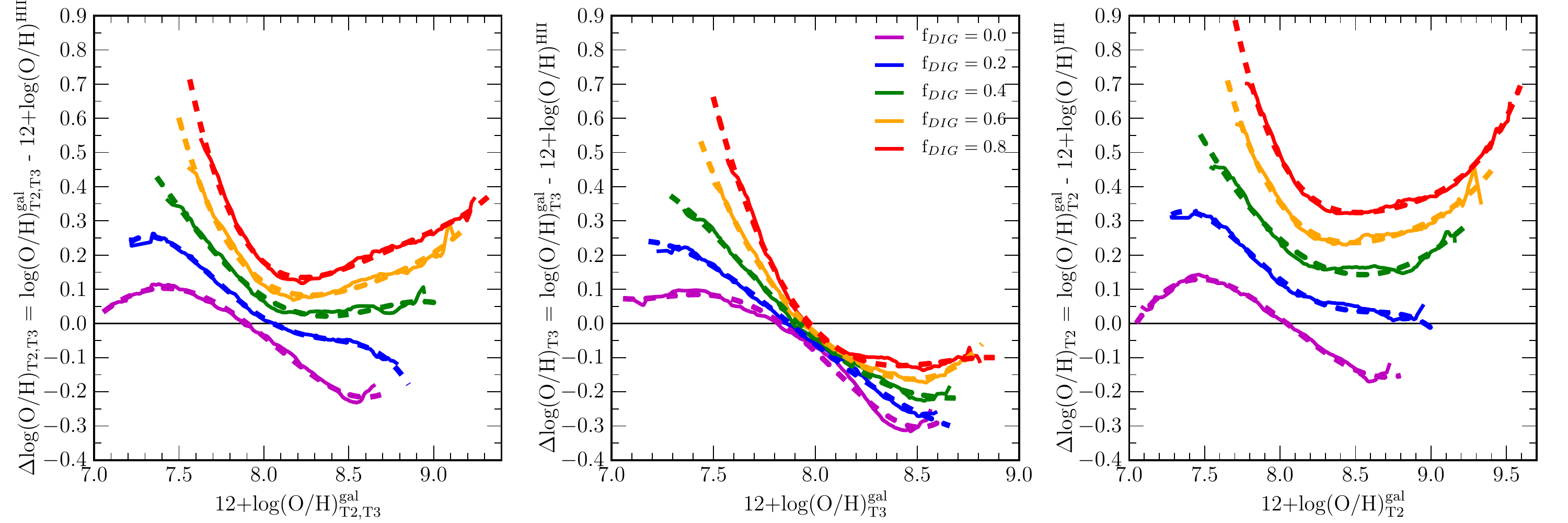}
 \caption{
The difference between the global galaxy direct-method metallicity, inferred from the observed galaxy spectrum,
 and the median metallicity of the \hii region distribution for models with \fdig=0.0 to 0.8.
  We show the bias in global galaxy metallicity for three cases: both \te\ and \ttwo\ are measured from
 the galaxy spectrum (left panel); only \te\ is measured directly and \ttwo\ is estimated using the
 \ttwo-\te\ relation of equation~\ref{eq:t2t3} (middle panel); and only \ttwo\ is measured directly
 and \te\ is estimated using equation~\ref{eq:t2t3} (right panel).
  Dashed lines show the best-fit fourth-order polynomials for the model with the corresponding color.
  The best-fit coefficients are presented in Table~\ref{tab:appendixohcoeffs}.
}\label{fig:appendixohbias}
\end{figure*}

Following the methodology presented in Section~\ref{sec:framework}, we create five models with
 the same input parameters as for the \stackmod model (N$_{\text{HII}}$=25, $\sigma_{\text{T}}$=0.07~dex,
 \fdig=0.55), except we vary the value of \fdig\ from 0.0 to 0.8 in increments of 0.2.
  For this set of models, the bias in properties inferred from global galaxy spectra relative
 to the median properties of the distribution of \hii regions in each galaxy is shown in
 Figure~\ref{fig:appendixstrongbias} for strong-line ratios, Figure~\ref{fig:appendixtempbias}
 for electron temperatures, and Figure~\ref{fig:appendixohbias} for direct-method oxygen abundances.
  As before, we display the bias in direct-method oxygen abundance for three scenarios in which
 (1) both \te\ and \ttwo\ are estimated directly from the galaxy spectrum, (2) only \te\ is known
 and \ttwo\ is estimated from equation~\ref{eq:t2t3}, and (3) only \ttwo\ is known and \te\ is
 estimated from equation~\ref{eq:t2t3}.  For each model, we fit the bias in each property with
 the fourth-order polynomials of equations~\ref{eq:ratpoly}-\ref{eq:ohpoly}.
  The best-fit coefficients for the strong-line bias are presented in Table~\ref{tab:appendixstrongcoeffs},
 while the best-fit coefficients for the electron temperature and direct-method oxygen abundance biases
 are given in Tables~\ref{tab:appendixtempcoeffs} and~\ref{tab:appendixohcoeffs}, respectively.

\begin{table}
 \centering
 \caption{Global galaxy bias coefficients for strong-line ratios (equation~\ref{eq:ratpoly})
 }\label{tab:appendixstrongcoeffs}
 \renewcommand{\arraystretch}{1.2}
 \begin{tabular}{ c | l l l l l }
   \hline\hline
   \multicolumn{6}{c}{$\Delta$O3} \\
   \fdig & $c_0$ & $c_1$ & $c_2$ & $c_3$ & $c_4$ \\[0pt]
   \hline
   0.0  &  0.113 & -0.0816 & -0.0299 & 0.0341 & -0.00443 \\
   0.2  &  0.124 & -0.115 & -0.0417 & 0.0663 & -0.00948 \\
   0.4  &  0.134 & -0.175 & -0.0571 & 0.145 & -0.0343 \\
   0.6  &  0.133 & -0.211 & -0.0364 & 0.167 & -0.0437 \\
   0.8  &  0.139 & -0.188 & -0.0769 & 0.150 & -0.0241 \\
   \hline\hline
   \multicolumn{6}{c}{$\Delta$O2} \\
   \fdig & $c_0$ & $c_1$ & $c_2$ & $c_3$ & $c_4$ \\[0pt]
   \hline
   0.0  &  0.00828 & -0.0132 & 0.0112 & -0.00632 & 0.00377 \\
   0.2  &  0.0950 & 0.0472 & -0.0756 & -0.0115 & 0.0176 \\
   0.4  &  0.169 & 0.0591 & -0.134 & 0.0308 & 0.00655 \\
   0.6  &  0.223 & 0.0989 & -0.183 & 0.0288 & 0.0132 \\
   0.8  &  0.291 & 0.143 & -0.269 & 0.0413 & 0.0211 \\
   \hline\hline
   \multicolumn{6}{c}{$\Delta$N2} \\
   \fdig & $c_0$ & $c_1$ & $c_2$ & $c_3$ & $c_4$ \\[0pt]
   \hline
   0.0  &  0.00307 & 0.0145 & 0.0295 & 0.00878 & -0.00786 \\
   0.2  &  0.0115 & 0.0180 & 0.0113 & 0.0102 & -0.000491 \\
   0.4  &  0.0204 & -0.00620 & 0.00106 & 0.0503 & -0.0128 \\
   0.6  &  0.0280 & -0.0123 & -0.0147 & 0.0618 & -0.0120 \\
   0.8  &  0.0376 & -0.00259 & -0.0388 & 0.0565 & -0.002357 \\
   \hline\hline
   \multicolumn{6}{c}{$\Delta$S2} \\
   \fdig & $c_0$ & $c_1$ & $c_2$ & $c_3$ & $c_4$ \\[0pt]
   \hline
   0.0  &  0.0105 & -0.000192 & -0.0110 & -0.00180 & 0.00586 \\
   0.2  &  0.142 & 0.0960 & -0.112 & -0.0343 & 0.0313 \\
   0.4  &  0.240 & 0.125 & -0.166 & -0.00721 & 0.0235 \\
   0.6  &  0.318 & 0.166 & -0.219 & -0.00621 & 0.0289 \\
   0.8  &  0.397 & 0.223 & -0.287 & -0.0283 & 0.0479 \\
   \hline
 \end{tabular}
\end{table}

\begin{table}
 \centering
 \caption{Global galaxy bias coefficients for electron temperatures (equation~\ref{eq:temppoly})
 }\label{tab:appendixtempcoeffs}
 \renewcommand{\arraystretch}{1.2}
 \begin{tabular}{ c | l l l l l }
   \hline\hline
   \multicolumn{6}{c}{$\Delta$T$_3$} \\
   \fdig & $c_0$ & $c_1$ & $c_2$ & $c_3$ & $c_4$ \\[0pt]
   \hline
   0.0  & 407.9   & -4,953   & 16,610 & -15,140 & 4,031   \\
   0.2  & -2,444 &  7,038  & -1,307   & -3,995    & 1,623 \\
   0.4  & -1,949  &  5,794  & -795.2   & -3,512    & 1,394 \\
   0.6  & -2,796 &  9,447  & -6,403   & 120.1     &  580.6  \\
   0.8  & 854.9    & -3,791 &  9,842  & -7,991    & 2,005 \\
   \hline\hline
   \multicolumn{6}{c}{$\Delta$T$_2$} \\
   \fdig & $c_0$ & $c_1$ & $c_2$ & $c_3$ & $c_4$ \\[0pt]
   \hline
   0.0  & 14,810 & -63,090  &  99,850 & -68,230  & 16,690  \\
   0.2  & 28,460 & -114,500 & 169,400 & -109,900 & 25,970  \\
   0.4  & 25,560 & -103,700 & 154,300 & -101,800 & 24,640  \\
   0.6  & 18,980 & -76,430  & 112,600 & -75,170  & 18,620  \\
   0.8  & 13,370 & -54,490  &  80,700 & -55,920  & 14,550  \\
   \hline
 \end{tabular}
\end{table}

\begin{table}
 \centering
 \caption{Global galaxy bias coefficients for direct-method oxygen abundances (equation~\ref{eq:ohpoly})
 }\label{tab:appendixohcoeffs}
 \renewcommand{\arraystretch}{1.2}
 \begin{tabular}{ c | l l l l l }
   \hline\hline
   \multicolumn{6}{c}{$\Delta$log(O/H) (\te\ and \ttwo)\tablenotemark{a}} \\
   \fdig & $c_0$ & $c_1$ & $c_2$ & $c_3$ & $c_4$ \\[0pt]
   \hline
   0.0  &  -0.0382 & -0.402 & -0.0948 & 0.381 &  0.123 \\
   0.2  &  0.0150 & -0.299 & 0.408 & 0.100 & -0.565 \\
   0.4  &  0.0776 & -0.320 & 0.472 & 0.0335 & -0.201 \\
   0.6  &  0.121 & -0.327 & 0.847 & -0.692 & 0.257 \\
   0.8  &  0.178 & -0.431 & 1.26 & -1.10 & 0.358 \\
   \hline\hline
   \multicolumn{6}{c}{$\Delta$log(O/H) (\te\ only)\tablenotemark{b}} \\
   \fdig & $c_0$ & $c_1$ & $c_2$ & $c_3$ & $c_4$ \\[0pt]
   \hline
   0.0  & -0.0825 & -0.550 & -0.241 & 0.662 & 0.514  \\
   0.2  & -0.0604 & -0.473 & 0.0717 & 0.192 & -0.0671  \\
   0.4  & -0.0424 & -0.518 & 0.320 &  0.202 & -0.164  \\
   0.6  & -0.0287 & -0.545 & 0.586 & -0.211 & 0.310  \\
   0.8  & -0.0177 & -0.617 & 1.13 & -0.650 &  0.0385  \\
   \hline\hline
   \multicolumn{6}{c}{$\Delta$log(O/H) (\ttwo\ only)\tablenotemark{c}} \\
   \fdig & $c_0$ & $c_1$ & $c_2$ & $c_3$ & $c_4$ \\[0pt]
   \hline
   0.0  & 0.0152 & -0.341 & -0.0727 & 0.306 & -0.0113  \\
   0.2  & 0.129 & -0.346 & 0.266 &  0.283 & -0.346  \\
   0.4  & 0.246 & -0.387 & 0.364 & 0.00148 & -0.0165  \\
   0.6  & 0.352 & -0.590 & 0.999 & -0.662 & 0.203  \\
   0.8  & 0.485 & -0.813 & 1.44 & -1.08 & 0.347  \\
   \hline
 \end{tabular}
 \tablenotetext{1}{The direct-method 12+log(O/H) case where both \te\ and \ttwo\ are directly determined from the galaxy spectrum.}
 \tablenotetext{2}{The case where only \te\ is estimated directly, while \ttwo\ is inferred using equation~\ref{eq:t2t3}.}
 \tablenotetext{3}{The case where only \ttwo\ is estimated directly, while \te\ is inferred using equation~\ref{eq:t2t3}.}
\end{table}

We recommend using the following procedure to apply corrections to individual galaxies
 or samples that are unrepresentative of the $z\sim0$ star-forming population.
  First, estimate \fdig\ for each galaxy or the median \fdig\ of the sample using \sigha\
 and equation~\ref{eq:fdig}.  Identify the models presented in this appendix that bracket
 this \fdig\ value.  Interpolate between the best-fit polynomials of these bracketing
 models to obtain corrections for strong-line ratios, electron temperatures, or direct-method
 metallicities appropriate for the galaxy or sample of galaxies.  Subtract the interpolated
 correction values for a given property from the values of that property as inferred from
 the global galaxy spectrum in order to correct for flux-weighting effects and DIG contamination.
  This procedure should yield robust corrections to individual galaxies or samples of galaxies that
 do not follow the mean local \mstar-SFR relation.
  When inferring metallicities from strong-line calibrations, we recommend first correcting the
 simple strong-line ratios O3, O2, N2, and S2, then constructing the corrected metallicity
 indicator (e.g., R23, O3N2) from these corrected simple ratios before using strong-line
 calibrations (empirical or theoretical) based on \hii regions to estimate metallicity.

\section{Appendix B: How do changes in $\sigma_{\text{T}}$ affect predicted line ratios and electron temperatures?}\label{appendix2}

In our models of $z\sim0$ galaxies, we assume that \ttwo\ of DIG regions is 15\% lower than
 \ttwo\ of \hii regions at fixed metallicity.  This assumption was motivated by offsets between
 observations of $z\sim0$ galaxies and the \fdig=0.55 model in the strong-line ratio vs. \ttwo\ diagrams
 (Fig.~\ref{fig:vstdigfrac055}) when we assumed that DIG and \hii region \ttwo\ were the same at fixed
 metallicity.
  Our assumption regarding lower DIG \ttwo\ additionally brought the magnitude of the predicted offset between
 \hii regions and $z\sim0$ galaxies in the \ttwo-\te\ diagram into agreement with observations
 (Fig.~\ref{fig:t2t3digfrac055}).
  Given that there are no direct observational constraints of the electron temperature of DIG other than
 along one line-of-sight in the Milky Way \citep{rey01},
 and the ionizing spectrum and gas physical conditions differ for DIG and \hii regions,
 our assumption regarding DIG \ttwo\ is not unreasonable.  Nevertheless, it is worthwhile
 to consider whether the \ttwo\ discrepancies can be resolved under a different set of assumptions.

The adopted value of the width of the input \hii region \te\ distribution, $\sigma_{\text{T}}$,
 can significantly affect the electron temperatures inferred from mock global galaxy spectra.
  This effect has been demonstrated by \citet{pil12no}, who were able to reproduce the offset
 between \hii regions and $z\sim0$ galaxies in the \ttwo-\te\ diagram by modeling galaxies as
 ensembles of \hii regions with a range of metallicities (equivalent to a range of \te).
  These authors found that the magnitude of the \ttwo-\te\ offset increased as the range
 of metallicities of the combined \hii regions increased.  In our model framework, a wider range in metallicity
 is equivalent
 to increasing the value of $\sigma_{\text{T}}$.  We adopted a value of $\sigma_{\text{T}}$=0.07~dex
 based on the observed \te\ distributions of \hii regions in nearby spiral galaxies \citep{ber15,cro15,cro16}.
  The results of \citet{pil12no} suggest that adopting a larger value of $\sigma_{\text{T}}$ could potentially
 resolve the discrepancy between models and observations in the \ttwo-\te\ diagram without assuming different
 \hii and DIG \ttwo\ at fixed metallicity.  We investigate the effects of adopting different values of
 $\sigma_{\text{T}}$ on the predicted strong-line ratios and electron temperatures in order to determine
 whether different values of $\sigma_{\text{T}}$ offer a viable solution to the discrepancies present in
 diagrams involving \ttwo.

We produce a set of models that have the same input parameters except for $\sigma_{\text{T}}$, which is varied.
  We consider five values of the width of the log-normal \te\ distribution:
 $\sigma_{\text{T}}$=[0.04, 0.07, 0.1, 0.15, 0.2].
  Other parameters are set to the values adopted in our fiducial models (N$_{\text{HII}}=25$, \fdig=0.55,
 log(T$_{\text{cent}}/$K$)=3.7$~to~4.3).  In these models, we do not assume a lower DIG \ttwo\ at fixed metallicity,
 but instead assume that \ttwo\ of \hii and DIG regions is equal at fixed metallicity.

\begin{figure}
 \includegraphics[width=\columnwidth]{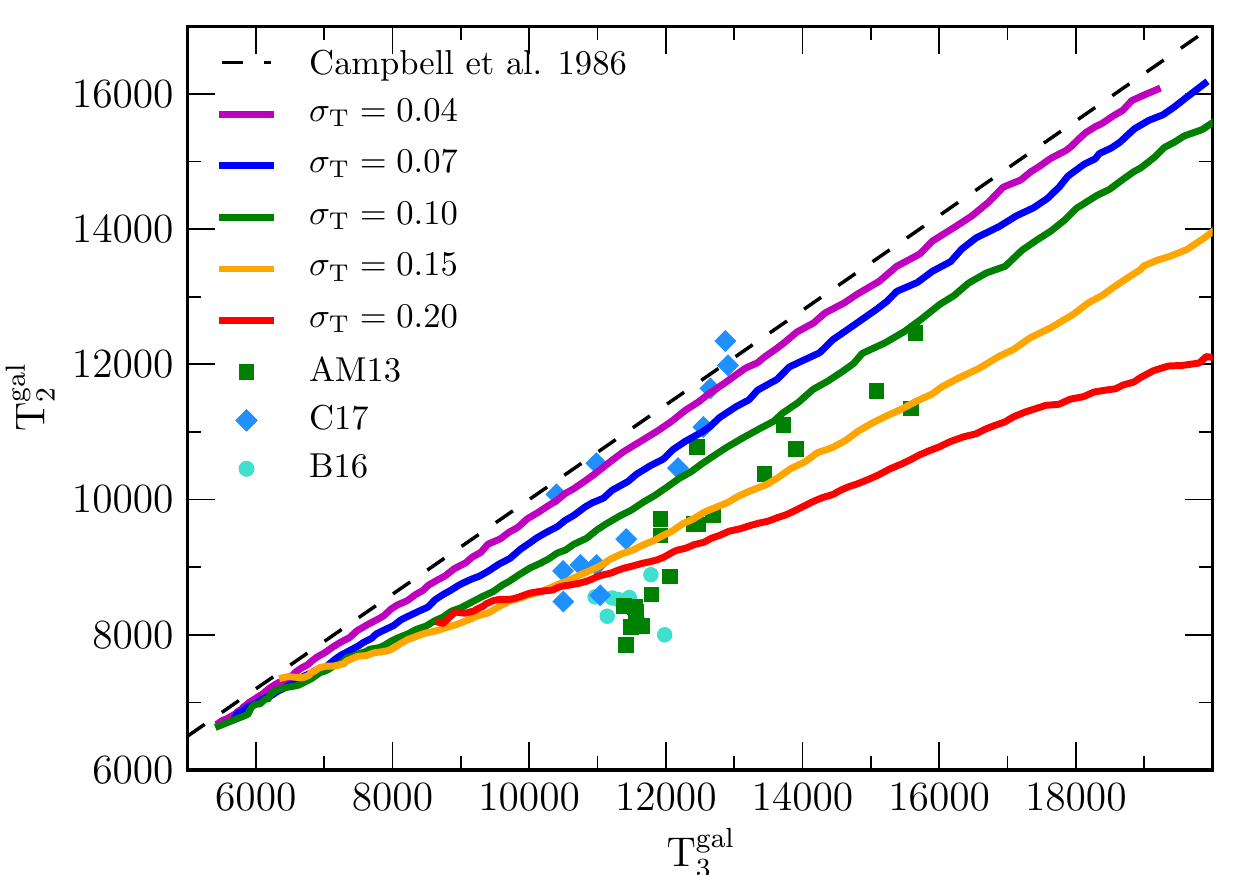}
 \caption{The global galaxy ionic temperature diagram of \ttwogal\ vs. \tegal, including models for which
 only $\sigma_{\text{T}}$ is varied.
  The solid colored lines show predictions of models with \fdig=0.55 and different values of $\sigma_{\text{T}}$,
 under the assumption that DIG and \hii region \ttwo\ is the same at fixed metallicity.
  The dashed black line shows the \hii region \ttwo-\te\ relation of \citet{cam86} given in
 equation~\ref{eq:t2t3}.
  The colored points indicate stacks of $z\sim0$ SDSS galaxies with auroral-line measurements.
}\label{fig:t2t3sigmat}
\end{figure}

\begin{figure*}
 \centering
 \includegraphics[width=\textwidth]{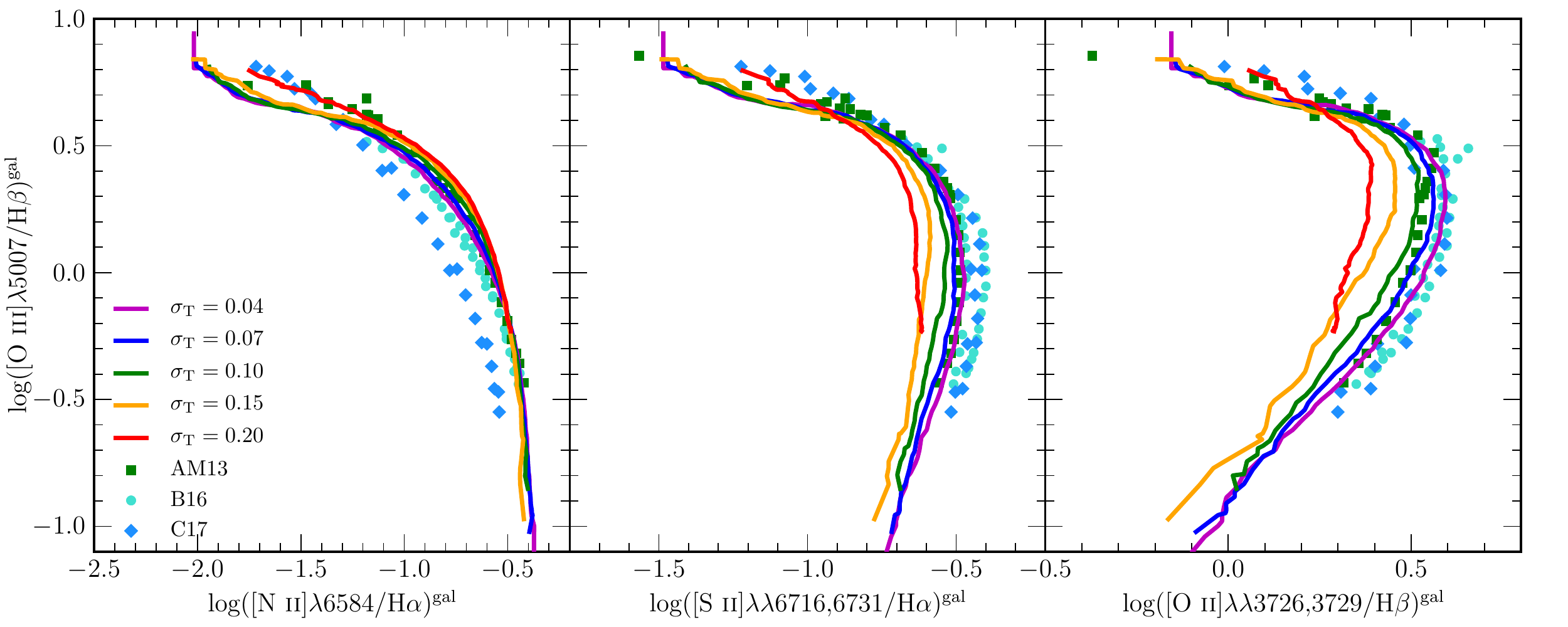}
 \caption{The O3N2 (left), O3S2 (middle), and O3O2 (right) strong-line ratio
 diagrams for stacks of SDSS galaxies and models with varying $\sigma_{\text{T}}$.
  Colored lines and points are the same as in Figure~\ref{fig:t2t3sigmat}.
}\label{fig:bptsigmat}
\end{figure*}

The models with varied $\sigma_{\text{T}}$ are shown in the \ttwo-\te\ diagram in Figure~\ref{fig:t2t3sigmat}.
  When working under the assumption that \hii region and DIG \ttwo\ are equal at fixed metallicity,
 $\sigma_{\text{T}}$$\approx$0.15~dex is required to match $z\sim0$ observations in the \ttwo-\te\ diagram.
  This value of $\sigma_{\text{T}}$ is roughly twice the value observed for \hii region distributions
 in local spiral galaxies \citep{ber15,cro15,cro16}.  We show predictions from the same set of models in the
 strong-line ratio O3N2, O3S2, and O3O2 diagrams in Figure~\ref{fig:bptsigmat}. 
  Increasing $\sigma_{\text{T}}$ results in lower [S\ii]/H$\alpha$ and [O\ii]/H$\beta$ at fixed [O\iii]/H$\beta$.
  Changes to the global galaxy strong-line excitation sequences come about because increasing the range of \hii region
 \te\ also increases the range of \hii and DIG region strong-line ratios (Fig.~\ref{fig:hiiratios}).  Combining light from
 \hii and DIG regions with a wider range of strong-line ratios results in different average excitation sequences because
 the relation between each strong-line ratio and \te\ is different.

The $\sigma_{\text{T}}$=0.15 model significantly underpredicts [S\ii]/H$\alpha$ and [O\ii]/H$\beta$
 at fixed [O\iii]/H$\beta$ for $z\sim0$ galaxies.
  While this discrepancy could potentially be resolved by adopting both a larger
 $\sigma_{\text{T}}$ and larger \fdig, reconciling the $\sigma_{\text{T}}$=0.15 model with observations
 in the O3S2 and O3O2 diagrams would require \fdig$\gtrsim$0.8.  Such a high fraction of Balmer emission originating
 from DIG is in conflict with narrowband H$\alpha$ studies of nearby galaxies which place the DIG fraction at
 $30-60$\% \citep{zur00,oey07}.  While adopting a larger $\sigma_{\text{T}}$ than our fiducial value of 0.07~dex
 can reproduce the \ttwo-\te\ offset without additional assumptions regarding DIG \ttwo, this assumption also
 results in strong-line ratios that disagree significantly with observations of galaxies.
  We conclude that our assumed value of $\sigma_{\text{T}}$=0.07~dex is reasonable alongside the assumption
 that DIG \ttwo\ is lower than that of \hii regions at fixed metallicity.
  Increasing the adopted values of
 $\sigma_{\text{T}}$ and \fdig\ while assuming equal DIG and \hii region \ttwo\ at fixed metallicity
 cannot provide a solution to the \ttwo\ discrepancies in
 Figures~\ref{fig:vstdigfrac055} and~\ref{fig:t2t3digfrac055}.

\bibliography{modelpaper}

\end{document}